\documentclass{aastex}
\usepackage{spr-astr-addons}
\usepackage{url}
\usepackage{graphicx}

\RequirePackage{color}

\begin{document}

\title{Calculation of the relativistic Bulk Viscosity, Shear Viscosity and Shear-Stress Viscosity of Accretion Disks around the Rotating Black Holes.}
\slugcomment{Not to appear in Nonlearned J., 45.}
\shorttitle{Relativistic Viscosity}
\shortauthors{M. Moeen Moghaddas}

\author{M. Moeen Moghaddas\altaffilmark{1}}
\affil{Mahboobe Moeen Moghaddas \email{mahboobemoen@gmail.com}}

\altaffiltext{1}{Department of Sciences, Kosar University of Bojnord, Bojnord, Iran.}


\begin{abstract}
In this paper we study the relativistic bulk tensor, shear tensor and the shear stress viscosity of the relativistic accretion disks around the rotating black holes. we calculate all non-zero components of the bulk tensor, shear tensor and shear stress viscosity in the equatorial plan in the BLF. We obtain the relations of the relativistic bulk tensor, shear tensor and shear stress viscosity in the terms of components of four velocity and  derivatives, Christoffel symbols and metric components. To see the behaviorr of the components of relativistic bulk tensor, shear tensor and shear stress viscosity, we introduce a radial form for the radial component of the four velocity in the LNRF then all components of the bulk tensor, shear tensor and shear stress viscosity are derived in BLF. Figures of non-zero components of the bulk tensor, shear tensor and shear stress viscosity are shown for some states. The importance and influence of the relativistic bulk viscosity in the accretion disks around the rotating black holes are studied in this paper.

\end{abstract}

\keywords{relativistic bulk tensor, relativistic shear tensor, relativistic shear stress viscosity, black holes’ accretion disks, relativistic disks.}

\section{Introduction} \label{sec:intro}

The accretion disks around the rotating black holes were studied by some of authors. Because of strong gravity of black holes we must study those disks with the relativistic method. An important parameter to energy distribution of the accretion disks around the black holes is shear stress viscosity. 
\cite{ac} used a standard and nonrelativistic form for viscosity to study ADAFs around the Kerr metric. \cite{pa} calculated the $r\phi$ component of the shear tensor to study the transonic viscous accretion disks in the Kerr metric. \cite{gp} studied the thin ADAF disks around the Kerr black holes and they calculated the shear tensor by using the nonrelativistic and relativistic casual viscosity. To calculate shear tensor the non-relativistic and relativistic causal viscosity were used by \cite{ta} to studied the stationary hydrodynamic equations of the three states of the accretion disks around the Kerr black holes. \cite{mg} calculated the shear tensor with introducing a zero radial component for the four velocity to solve the hydrodynamic equations of the optically thin, steady state, accretion disks around the Kerr black holes analytically. \cite{mo} calculate the bulk and shear tensor for an accretion disks around the non-rotating black holes with introducing a simple model for radial velocity and shows that bulk viscosity in those disks may be important. 

It is not clear that how the shear stress viscosity affects the relativistic accretion disks. Shear stress viscosity include two types of viscosity which are bulk and shear viscosity. In the previous papers the most of authors have considered zero bulk viscosity especially in accretion disks around the rotating black holes. So we try to calculate all the non-zero components of bulk and shear viscosity. 
In this paper, we study the relativistic shear stress viscosity of the stationary, axisymmetric, accretion disks in the equatorial plan around the rotating black holes. Metric, units and reference frames are in $\S$\ref{sec:2}. The relativistic bulk viscosity are derived in $\S$\ref{sec:3}. All non-zero components of the relativistic shear viscosity are acquired in $\S$\ref{sec:4}. In $\S$\ref{sec:5} we introduce a radial model for the radial component of the four velocity in the LNRF and other components of the four velocity in the BLF are derived. Components of the shear and bulk tensors in some $n$ are calculated in $\S$\ref{sec:6}, figures of those components are in this section. Non-zero components and figures of shear stress viscosity are in $\S$\ref{sec:7}. Influence of the bulk tensor, summery and conclusion are seen in $\S$\ref{sec:8}. 

 \section{Metric, Units and Reference Frames }
 \label{sec:2}
\subsection{Background Metric and Units}
We use the Boyer-Lindquist coordinates of the rotating black hole space time. In the Boyer-Lindquist coordinates, the Kerr metric is
\begin{equation}\label{1}
ds^{2}=g_{\alpha
\beta}dx^{\alpha}dx^{\beta}=-\alpha^{2}dt^{2}+\gamma_{ij}(dx^{i}+\beta^{i}dt)(dx^{j}+\beta^{j}dt),
\end{equation}
where $i,j=r,\theta$ and $\phi$. The non-zero components of the lapse
function $\alpha$, the shift vector $\beta^{i}$ and the spatial matrix
$\gamma_{ij}$ are given in the geometric units as
\begin{eqnarray}\label{2}
&\alpha&=\sqrt{\frac{\Sigma\Delta}{A}},\qquad \beta^{\phi}=-\omega,\qquad \gamma_{rr}=\frac{\Sigma}{\Delta},\nonumber\\
&\gamma_{\theta\theta}&=\Sigma,\qquad \gamma_{\phi\phi}=\frac{Asin^{2}\theta}{\Sigma}.
\end{eqnarray}
Here, we use geometric mass $m=GM/c^{2}$,
$\Sigma=r^{2}+a^{2}\cos^{2}\theta$, $\Delta=r^{2}-2Mr+a^{2}$ and
$A=\Sigma\Delta+2mr(r^{2}+a^{2})$. The position of outer and inner
horizons, $r\pm$, are calculated by inserting $\Delta=0$ to get
$r\pm=m\pm(m^2-a^2)^{1/2}$. The angular velocity of the frame
dragging due to the black hole rotation is
$\omega=-g_{t\phi}/g_{\phi\phi}=\frac{2mar}{A}$, where M is the
black hole mass, G is the gravitational constant and c is the speed
of light and the angular momentum of the black hole, J, is described
as
\begin{equation}\label{3}
a=Jc/GM^{2},
\end{equation}
where $-1<a<1$.

Similar to \cite{gp}, for basic scaling we set $G=M=c=1$. The non-zero components of metric, $g_{\mu\nu}$, and its inverse, $g^{\mu\nu}$ are
\begin{eqnarray}\label{4}
&& g_{tt}=-(1-\frac{2mr}{\Sigma}),\qquad g_{rr}=\frac{\Sigma}{\Delta},\qquad g_{\theta\theta}=\Sigma,\nonumber\\
&& g_{\phi\phi}=\frac{Asin^{2}\theta}{\Sigma},\qquad g_{t\phi}=-\frac{2marsin^{2}\theta}{\Sigma},
\end{eqnarray}
\begin{eqnarray}\label{5}
&& g^{tt}=-(\frac{A}{\Delta\Sigma}),\qquad g^{rr}=\frac{\Delta}{\Sigma},\qquad g^{\theta\theta}=\frac{1}{\Sigma},  \nonumber\\
&&g^{\phi\phi}=\frac{1}{\Delta sin^{2}\theta}(1-\frac{2mr}{\Sigma}),\qquad g^{t\phi}=-\frac{2mar}{\Sigma\Delta}.
\end{eqnarray}
\subsection{Reference Frames}
In our study, we use two reference frames. The first one is the
Boyer-Lindquist reference frame (BLF) based on the Boyer-Lindquist coordinates
describing the metric, in which, our calculations are done. The
second one is the locally non-rotating reference frame (LNRF) which
is formed by observers with a future-directed unit vector orthogonal
to $t=constant$. By using the BLF, the LNRF
observer is moving with the angular velocity of frame dragging
($\omega$) arises from the black hole rotating. 

The physical quantities measured in the LNRF are described by using the
hat such as $u^{\hat{\mu}}$ and $u_{\hat{\mu}}$. The transformation
matrixes of the LNRF and the BLF are given in Appendix \ref{apa}.

We study the stationary, equatorially symmetric and axisymmetric accretion fluid in the equatorial plan, therefore in the four velocity $u^{\mu}=(u^{t},u^{r},u^{\theta},u^{\phi})$ we use $u^{\theta}=0$.
\section{Relativistic Bulk Viscosity}
\label{sec:3}
In the relativistic Navier-Stokes flow, the relativistic bulk viscosity is (\cite{mt}):
\begin{equation}\label{6}
B^{\mu\nu}=-\zeta b^{\mu\nu},
\end{equation} 
where $\zeta$ is the coefficient of the bulk viscosity and the bulk tensor $b^{\mu\nu}$ is
\begin{equation}\label{7}
b^{\mu\nu}=\Theta h^{\mu\nu},
\end{equation}
$h^{\mu\nu}=g^{\mu\nu}+u^{\mu}u^{\nu}$ is the projection tensor and $\Theta=u^{\gamma}_{;\gamma}$ is the expansion of the fluid world line which is(\cite{mo})
\begin{equation}\label{8}
\Theta=u^{\gamma}_{;\gamma}=\frac{\partial u^{\gamma}}{\partial x^{\gamma}}+\Gamma_{\gamma\nu}^{\nu}u^{\gamma},
\end{equation}
also $\Gamma_{\gamma\nu}^{\nu}$ is
\begin{equation}\label{9}
\Gamma_{\gamma\nu}^{\nu}=\frac{1}{\sqrt{\left|g\right|}}\frac{\partial \sqrt{\left|g\right|}}{\partial x^{\gamma}}.
\end{equation}
For the Kerr metric $\sqrt{\left|g\right|}=r^{2}$. We study the stationary flow in the equatorial plan, therefore $\frac{\partial}{\partial r}$ is the only non-zero partial derivative, so the expansion of fluid world line in the equatorial plan and in the Kerr metric is
\begin{equation}\label{10}
\Theta  =\frac{\partial u^{\gamma}}{\partial x^{\gamma}}+\frac{2}{r}u^{r}=u^{r}_{,r}+\frac{2u^{r}}{r}
\end{equation}  
We can calculate the ten non-zero components of bulk tensor in the Kerr metric as 
\begin{eqnarray}\label{11}
&& b^{tt}=(u^{r}_{,r}+\frac{2u^{r}}{r})h^{tt}=(u^{r}_{,r}+\frac{2u^{r}}{r})(g^{tt}+(u^{t})^{2}),
\nonumber\\ &&  b^{rr}=(u^{r}_{,r}+\frac{2u^{r}}{r})h^{rr}=(u^{r}_{,r}+\frac{2u^{r}}{r})(g^{rr}+(u^{r})^{2}),
\nonumber\\ && b^{\theta\theta}=(u^{r}_{,r}+\frac{2u^{r}}{r})h^{\theta\theta}=(u^{r}_{,r}+\frac{2u^{r}}{r})g^{\theta\theta},
\nonumber\\ && b^{\phi\phi}=(u^{r}_{,r}+\frac{2u^{r}}{r})h^{\phi\phi}=(u^{r}_{,r}+\frac{2u^{r}}{r})(g^{\phi\phi}+(u^{\phi})^{2}),
\nonumber\\ && b^{tr}=b^{rt}=(u^{r}_{,r}+\frac{2u^{r}}{r})h^{rt}=(u^{r}_{,r}+\frac{2u^{r}}{r})(u^{r}u^{t})
\nonumber\\ && b^{t\phi}=b^{\phi t}=(u^{r}_{,r}+\frac{2u^{r}}{r})h^{t\phi}=(u^{r}_{,r}+\frac{2u^{r}}{r})(g^{t\phi}+u^{t}u^{\phi}), 
\nonumber\\ && b^{r\phi}=b^{\phi r}=(u^{r}_{,r}+\frac{2u^{r}}{r})h^{r\phi}=b^{\phi r}=(u^{r}_{,r}+\frac{2u^{r}}{r})(u^{r}u^{\phi}).\nonumber\\
\end{eqnarray} 
\section{Relativistic shear viscosity}
\label{sec:4}
In the relativistic Navier-Stokes flow, the relativistic shear viscosity ($S^{\mu\nu}$) is (\cite{mt})
\begin{equation}\label{12}
S^{\mu\nu}=-2\lambda\sigma^{\mu\nu}, 
\end{equation}
where $\lambda$ is the coefficient of the dynamical viscosity and the shear tensor ($\sigma^{\mu\nu}$) of the fluid is
\begin{equation}\label{13}
\sigma^{\mu\nu}=g^{\mu\alpha}g^{\nu\beta}\sigma_{\alpha\beta},
\end{equation}
where the shear rate, $\sigma_{\alpha\beta}$ is: (\cite{ta})
\begin{eqnarray}\label{14}
\sigma_{\alpha\beta}&&=\frac{1}{2}(u_{\alpha;\gamma}h^{\gamma}_{\beta}+u_{\beta;\gamma}h^{\gamma}_{\alpha})-\frac{1}{3}\Theta
h_{\mu\nu}\nonumber\\
&&=\frac{1}{2}(u_{\mu;\nu}+u_{\nu;\mu}+a_{\mu}u_{\nu}+a_{\nu}u_{\mu})-\frac{1}{3}\Theta
h_{\mu\nu},
\end{eqnarray}
where $a_{\mu}=u_{\mu;\gamma}u^{\gamma}$ is the four acceleration. 
therefore the shear tensor is
\begin{eqnarray}\label{15}
\sigma^{\mu\nu}&&= g^{\mu\alpha}g^{\nu\beta}\left(\frac{1}{2}(u_{\alpha;\gamma}h^{\gamma}_{\beta}+u_{\beta;\gamma}h^{\gamma}_{\alpha})-\frac{1}{3}\Theta
h_{\mu\nu}\right)
\nonumber\\&&=\frac{1}{2}(u^{\mu}_{;\gamma}h^{\gamma\nu}+u^{\nu}_{;\gamma}h^{\gamma\mu})-\frac{1}{3}\Theta
h^{\mu\nu}
\nonumber\\&&=\frac{1}{2}[(u^{\mu}_{,\gamma}\delta_{\gamma r}+\Gamma^{\mu}_{\gamma \lambda}u^{\lambda})h^{\gamma\nu}+(u^{\nu}_{,\gamma}\delta_{\gamma r}
\nonumber\\&&+\Gamma^{\nu}_{\gamma \lambda}u^{\lambda})h^{\gamma\mu}]-\frac{1}{3}\Theta h^{\mu\nu}.
\end{eqnarray}
All non-zero components of the relativistic shear tensor are calculated completely in Appendix \ref{apc}, we use the non-zero the Christoffel symbols of Appendix \ref{apb} then we have
\begin{eqnarray}
\label{16}
&&\sigma^{tt}= (u^{t}_{,r}+\Gamma ^{t}_{rt}u^{t}+\Gamma ^{t}_{r\phi}u^{\phi})h^{rt}+\Gamma ^{t}_{tr}u^{r}h^{tt}+\Gamma ^{t}_{\phi r}u^{r}h^{t\phi}
\nonumber\\ &&\qquad -\frac{1}{3}(u^{r}_{,r}+\frac{2u^{r}}{r})h^{tt},
\nonumber\\&&\sigma^{tr}=\sigma^{rt}= \frac{1}{2}[(u^{t}_{,r}+\Gamma ^{t}_{rt}u^{t}+\Gamma ^{t}_{r\phi}u^{\phi})h^{rr}+\Gamma ^{t}_{tr}u^{r}h^{rt}
\nonumber\\ &&\qquad +\Gamma ^{t}_{\phi r}u^{r}h^{r\phi}+(u^{r}_{,r}+\Gamma ^{r}_{rr}u^{r})h^{rt}+(\Gamma ^{r}_{tt}u^{t}+\Gamma ^{r}_{t\phi}u^{\phi})h^{tt}
\nonumber\\ &&\qquad +(\Gamma ^{r}_{\phi t}u^{t}+\Gamma ^{r}_{\phi\phi}u^{\phi})h^{t\phi}]-\frac{1}{3}(u^{r}_{,r}+\frac{2u^{r}}{r})h^{rt},
\nonumber\\&&\sigma^{t\phi}=\sigma^{\phi t}= \frac{1}{2}[(u^{t}_{,r}+\Gamma ^{t}_{rt}u^{t}+\Gamma ^{t}_{r\phi}u^{\phi})h^{r\phi}+\Gamma ^{t}_{tr}u^{r}h^{t\phi}
\nonumber\\ &&\qquad +\Gamma ^{t}_{\phi r}u^{r}h^{\phi\phi}+(u^{\phi}_{,r}+\Gamma ^{\phi}_{rt}u^{t}+\Gamma ^{\phi}_{r\phi}u^{\phi})h^{rt}+\Gamma ^{\phi}_{tr}u^{r}h^{tt}
\nonumber\\ &&\qquad +\Gamma ^{\phi}_{\phi r}u^{r}h^{t\phi}]-\frac{1}{3}(u^{r}_{,r}+\frac{2u^{r}}{r})h^{t\phi},
\nonumber\\&&\sigma^{rr}= (u^{r}_{,r}+\Gamma ^{r}_{rr}u^{r})h^{rr}+(\Gamma ^{r}_{tt}u^{t}+\Gamma ^{r}_{t\phi}u^{\phi})h^{rt}+(\Gamma ^{r}_{\phi t}u^{t}
\nonumber\\ &&\qquad +\Gamma ^{r}_{\phi\phi}u^{\phi})h^{r\phi}-\frac{1}{3}(u^{r}_{,r}+\frac{2u^{r}}{r})h^{rr},
\nonumber\\&&\sigma^{r\phi}=\sigma^{\phi r}= \frac{1}{2}[(u^{r}_{,r}+\Gamma ^{r}_{rr}u^{r})h^{r\phi}+(\Gamma ^{r}_{tt}u^{t}+\Gamma ^{r}_{t\phi}u^{\phi})h^{t\phi}
\nonumber\\ &&\qquad +(\Gamma ^{r}_{\phi t}u^{t}+\Gamma ^{r}_{\phi\phi}u^{\phi})h^{\phi\phi}+(u^{\phi}_{,r}+\Gamma ^{\phi}_{rt}u^{t}+\Gamma ^{\phi}_{r\phi}u^{\phi})h^{rr}
\nonumber\\ &&\qquad +\Gamma ^{\phi}_{tr}u^{r}h^{rt}+\Gamma ^{\phi}_{\phi r}u^{r}h^{r\phi}]-\frac{1}{3}(u^{r}_{,r}+\frac{2u^{r}}{r})h^{r\phi},
\nonumber\\&&\sigma^{\theta\theta}= -\frac{1}{3}(u^{r}_{,r}+\frac{2u^{r}}{r})h^{\theta\theta},
\nonumber\\&&\sigma^{\phi\phi}= (u^{\phi}_{,r}+\Gamma ^{\phi}_{rt}u^{t}+\Gamma ^{\phi}_{r\phi}u^{\phi})h^{r\phi}+\Gamma ^{\phi}_{tr}u^{r}h^{t\phi}+\Gamma ^{\phi}_{\phi r}u^{r}h^{\phi\phi}
\nonumber\\ &&\qquad -\frac{1}{3}(u^{r}_{,r}+\frac{2u^{r}}{r})h^{\phi\phi}.
\end{eqnarray}
\section{Radial model of radial component of four velocity}
\label{sec:5}
 To see the behaviour of the bulk tensor and shear tensor and influences of the bulk tensor in the shaer stress viscosity, we use a simple model for the four velocity. In \cite{mo} a simple model for the radial component of the four velocity in the Schwarzschild metric was used. Similary we introduce a simple and radial model for the radial component of the four velocity in the LNRF as
\begin{equation}\label{17} 
u^{\hat{r}}=-\frac{\beta}{r^{n}},
\end{equation}
$\beta$ and $n$ are positive and constant and minus sign is for the direction of $u^{\hat{r}}$ which is toward of the black hole. With the transformations equations we can calculate the radial component of the four velocity in the BLF as:
 \begin{equation}\label{18}
u^{r}=e^{\nu}_{\hat{\mu}}\delta^{r}_{\nu}u^{\hat{\mu}},
\end{equation} 
$e^{\nu}_{\hat{\mu}}$ are the Components of connecting between the BLF and the LNRF which are in the appendix \ref{apb}. $e^{r}_{\hat{\mu}}$ is non-zero in $e^{r}_{\hat{r}}$, therefore the radial component of the four velocity in the BLF is
\begin{equation}\label{19}
u^{r}=e^{r}_{\hat{r}} u^{\hat{r}}=\sqrt{\frac{\Delta}{\Sigma}}u^{\hat{r}}=-\frac{\beta\sqrt{r^{2}-2r+a^{2}}}{r^{n+1}}.
\end{equation} 
If we study the Keplerian accretion disks we have
\begin{equation}\label{20}
\Omega=\frac{u^{\phi}}{u^{t}}=\Omega_{k}^{\pm}=\pm\frac{1}{r^{\frac{3}{2}}\pm a}.
\end{equation}
With $u^{\mu}u_{\mu}=-1$ we have 
\begin{eqnarray}\label{21}
&&-1=u^{t}u_{t}+u^{r}u_{r}+u^{\phi}u_{\phi}\nonumber\\
&&\Rightarrow -1=g_{tt}u^{t}(g_{tt}u^{t}+g_{t\phi}u^{\phi})+g_{rr}(u^{r})^{2}\nonumber\\
&&+u^{\phi}(g_{\phi\phi}u^{\phi}+g_{t\phi}u^{t})\nonumber\\
&&\Rightarrow -1=g_{tt}(u^{t})^{2}+g_{rr}(u^{r})^{2}+g_{\phi\phi}(\Omega)^{2}(u^{t})^{2}\nonumber\\
&&+2g_{t\phi}\Omega(u^{t})^{2}.
\end{eqnarray}
By useing  equations (\ref{20}) and (\ref{19}) in equation (\ref{21}), we have
\begin{equation}\label{22}
u^{t}=\frac{r^{\frac{3}{2}}+a}{r^{n}}\sqrt{\frac{r^{2n+1}+r\beta^{2}}{r^{4}+2ar^{\frac{5}{2}}-3r^{3}}}.
\end{equation}

The time dilation is a relativistic influence which shows the coordinate time ($t$) is larger than the proper time ($\tau$). The time dilation is derived by putting $u^{r}=u^{\theta}=u^{\phi}=0$ also the time dilation is minimum value of the $u^{t}$ is derived as
\begin{equation}\label{23}
u^{t0}=u^{t}_{min}=\sqrt{\left|1/g_{tt}\right|}.
\end{equation}
We must check $u^{t}\geq u^{t0}$ which $u^{t0}=u^{t}_{min}=\sqrt{\left|1/g_{tt}\right|}$ (\cite{mo}). 
\begin{eqnarray}\label{24}
&&\frac{r^{\frac{3}{2}}+a}{r^{n}}\sqrt{\frac{r^{2n+1}+r\beta^{2}}{r^{4}+2ar^{\frac{5}{2}}-3r^{3}}}\geq \sqrt{\frac{r}{r-2}}
\nonumber\\ &&\Rightarrow \frac{(r^{\frac{3}{2}}+a)^{2}(r^{2n+1}+r\beta^{2})}{r^{2n}(r^{4}+2ar^{\frac{5}{2}}-3r^{3})}\geq \frac{r}{r-2}
\nonumber\\ &&\Rightarrow r^{2n+4}-4ar^\frac{3}{2}r^{2n+1}+r^{5}\beta^{2}-2r^{4}\beta^{2}+2ar^\frac{3}{2}r^{2}\beta^{2}
\nonumber\\ &&-4ar^{\frac{5}{2}}\beta^{2}+r^{2}a^{2}\beta^{2}-2ra^{2}\beta^{2}+a^{2}r^{2n+2}-2a^{2}r^{2n+1}\geq 0.\nonumber\\
\end{eqnarray}
All sentences are positive except the last three  sentences which are negative. But in $r\geq 2$, the $r^{2n+4}\geq 4ar^\frac{3}{2}r^{2n+1}$, $r^{5}\beta^{2}\geq 2r^{4}\beta^{2}$, $2ar^\frac{3}{2}r^{2}\beta^{2}\geq 4ar^{\frac{5}{2}}\beta^{2}$, $r^{2}a^{2}\beta^{2}\geq 2ra^{2}\beta^{2}$ and $a^{2}r^{2n+2}\geq 2a^{2}r^{2n+1}$expressions are true, therefore after horizon $u^{t}\geq u^{t0}$ is true. So, in the Keplarian accretion disks the four-velocity in the BLF is
\begin{eqnarray}\label{25}
&&u^{\mu}=(\frac{r^{\frac{3}{2}}+a}{r^{n}}\sqrt{\frac{r^{2n+1}+r\beta^{2}}{r^{4}+2ar^{\frac{5}{2}}-3r^{3}}},\nonumber\\&&-\frac{\beta\sqrt{r^{2}-2r+a^{2}}}{r^{n+1}}, 0,\frac{1}{r^{n}}\sqrt{\frac{r^{2n+1}+r\beta^{2}}{r^{4}+2ar^{\frac{5}{2}}-3r^{3}}}).\nonumber\\
\end{eqnarray}
\begin{figure}
\centerline{\includegraphics[scale=.22]{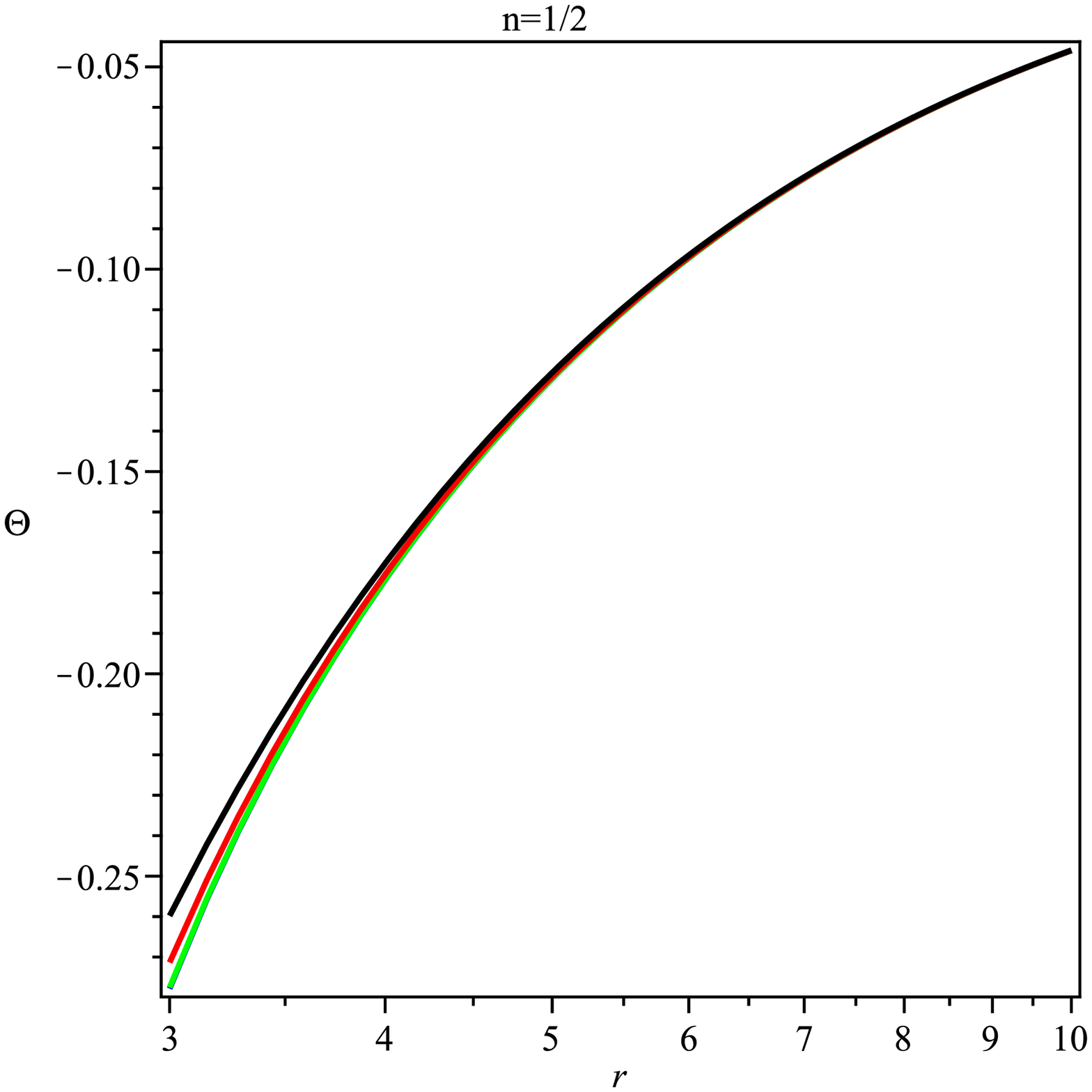}\includegraphics[scale=.22]{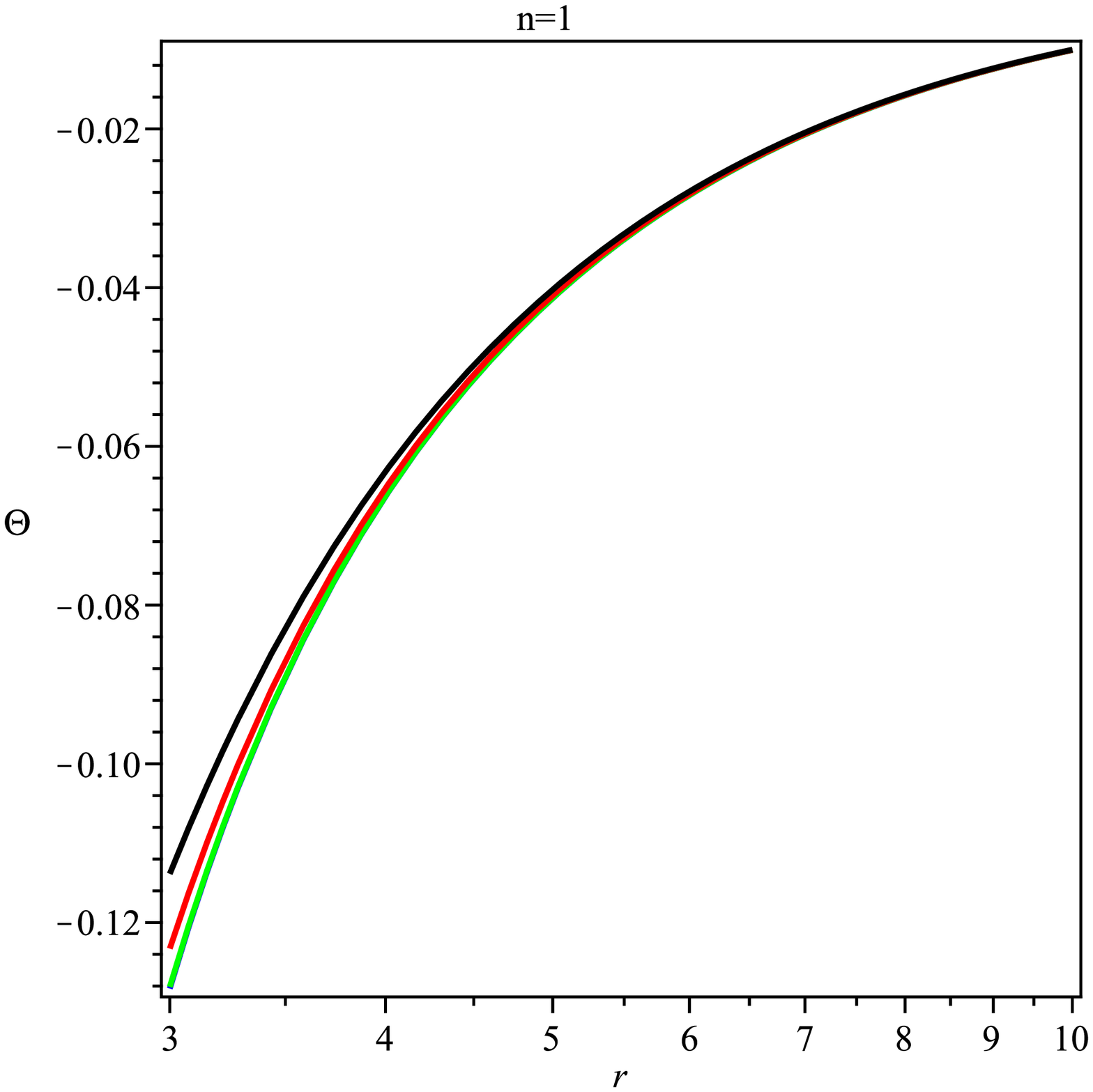}}
\centerline{\includegraphics[scale=.22]{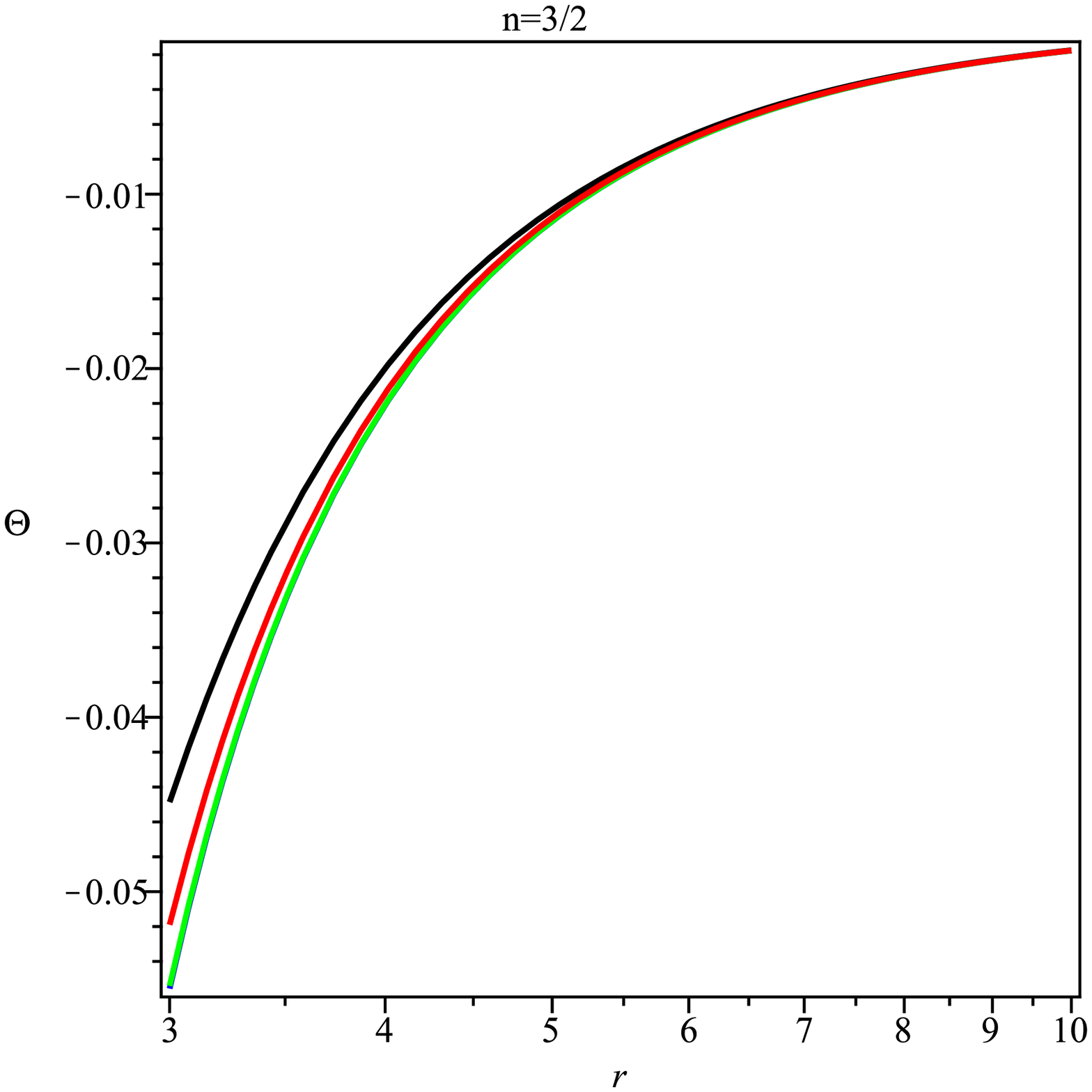}\includegraphics[scale=.22]{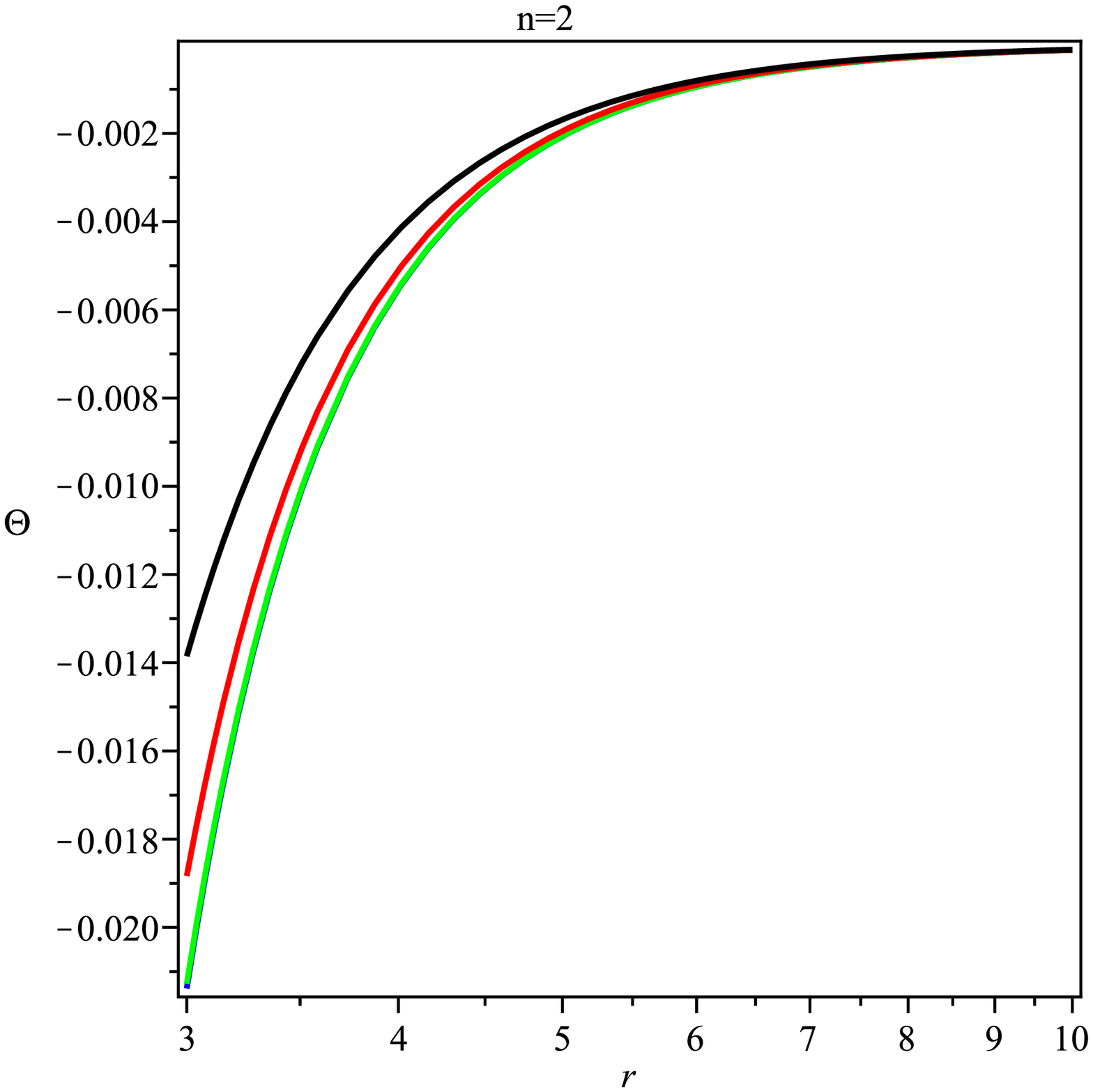}}
\centerline{\includegraphics[scale=.22]{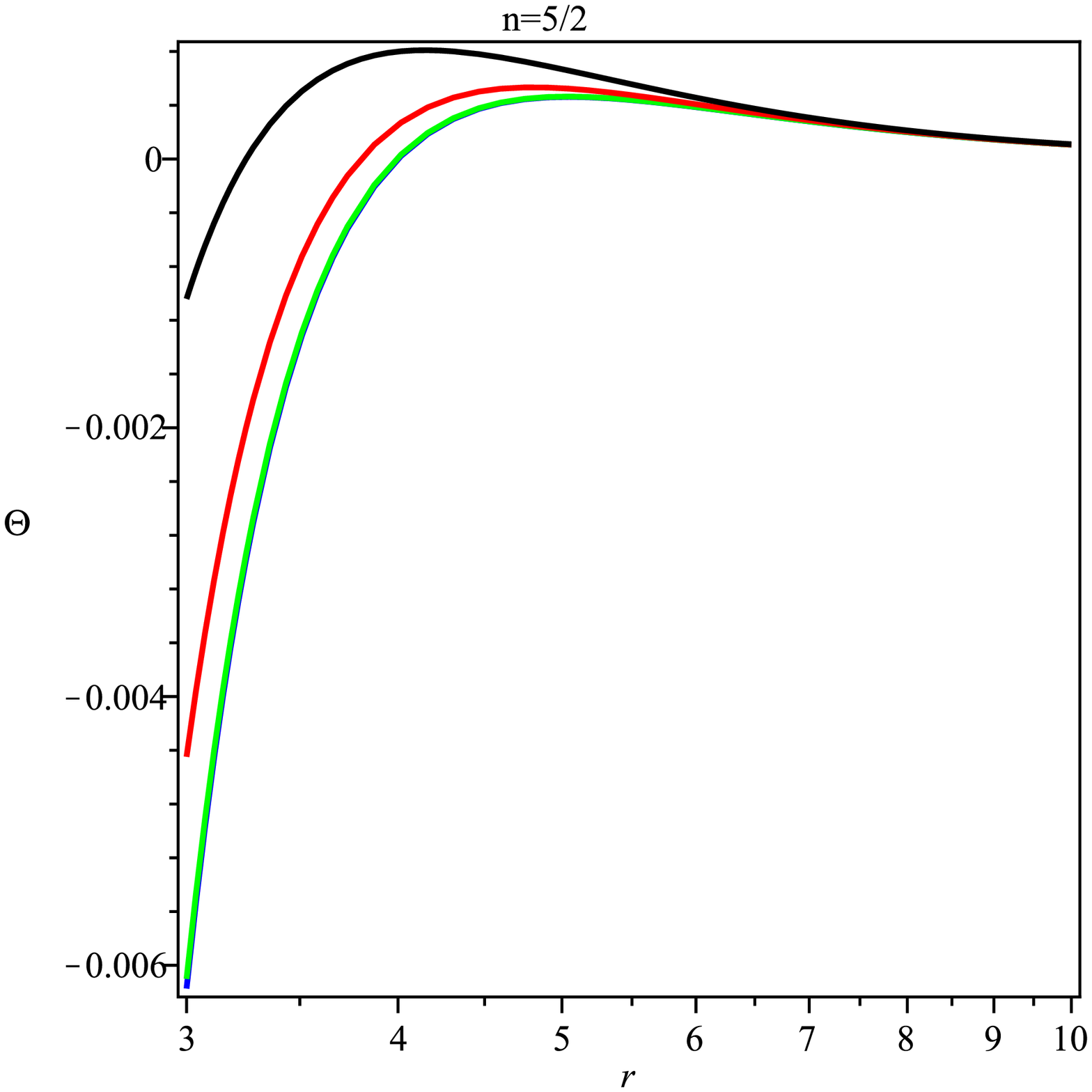}\includegraphics[scale=.22]{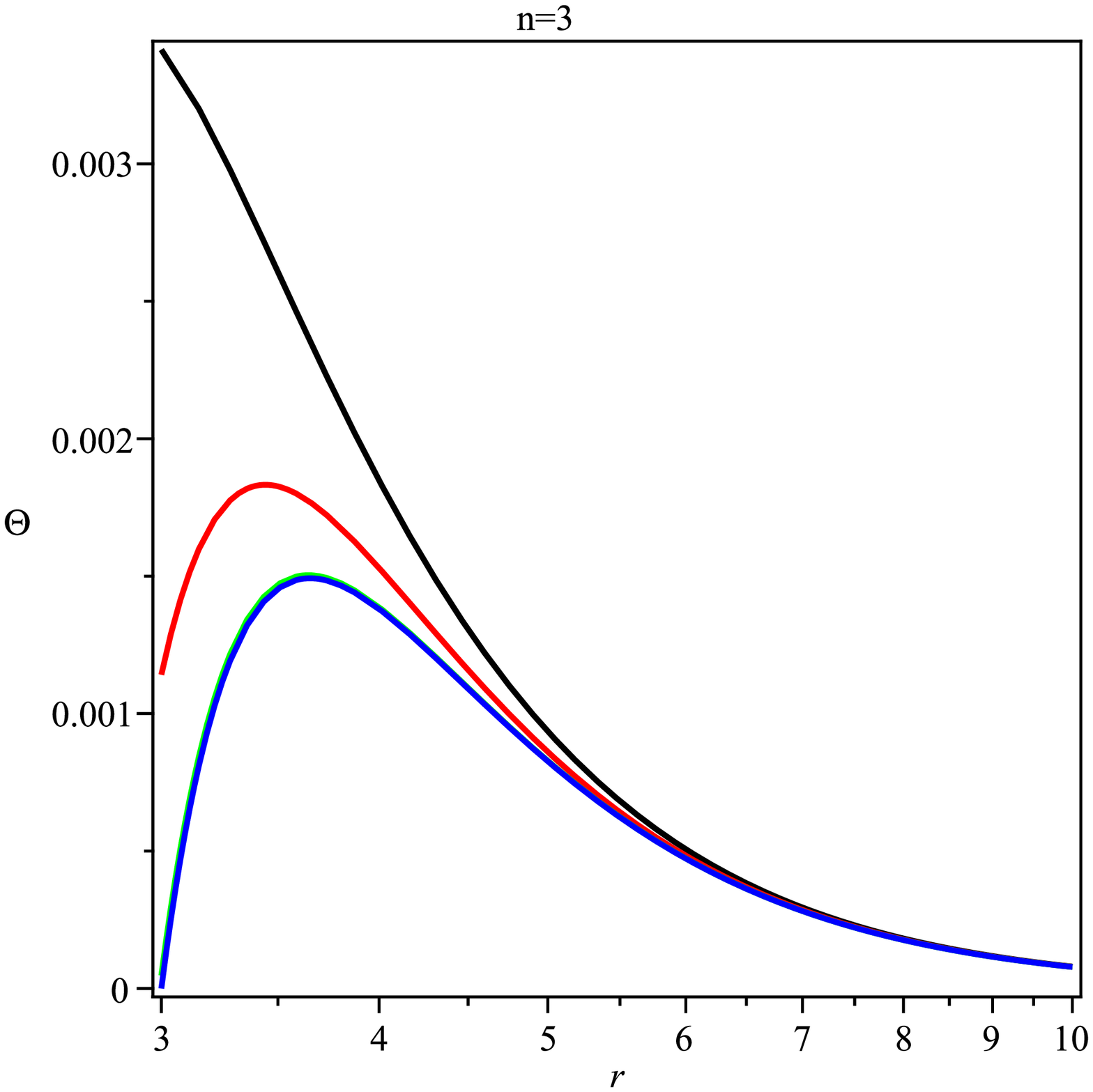}}
 \caption{the projection tensor for $\beta=1$ and $n=\frac{1}{2}, 1, \frac{3}{2}, 2, \frac{5}{2} and 3$. $a=.9$ in black, $a=.5$ in red, $a=.1$ in green and $a=0$ in blue.}
\label{figure1}	
\end{figure}
\section{Shear and bulk tensors}
\label{sec:6}
We use the radial model for calculating the four velocity also the components of shear and bulk tensors in different $n$ are calculated
\subsection{$n=\frac{1}{2}$}
\begin{figure}
\vspace{\fill}
\centerline{\includegraphics[scale=.22]{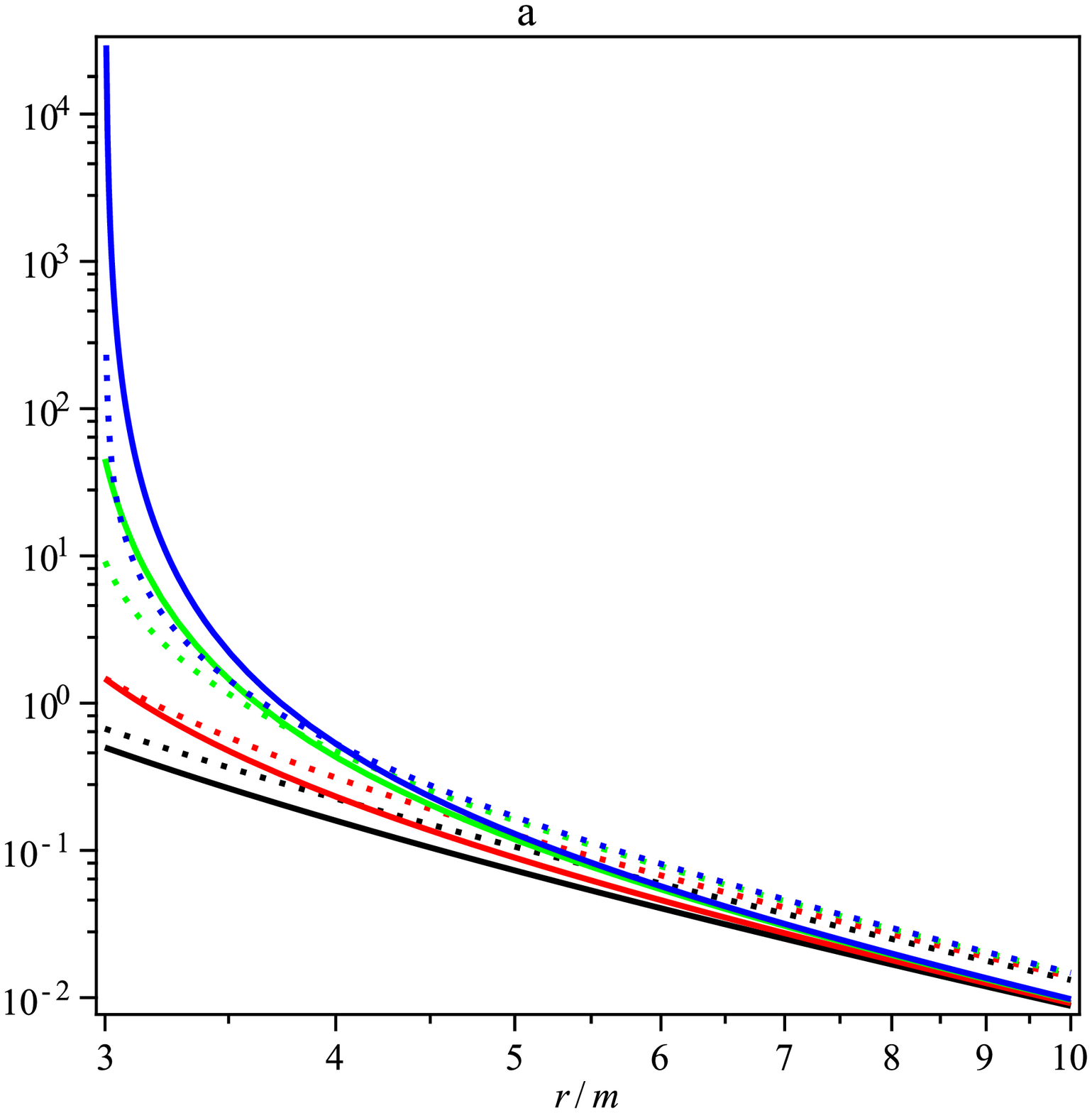}\includegraphics[scale=.22]{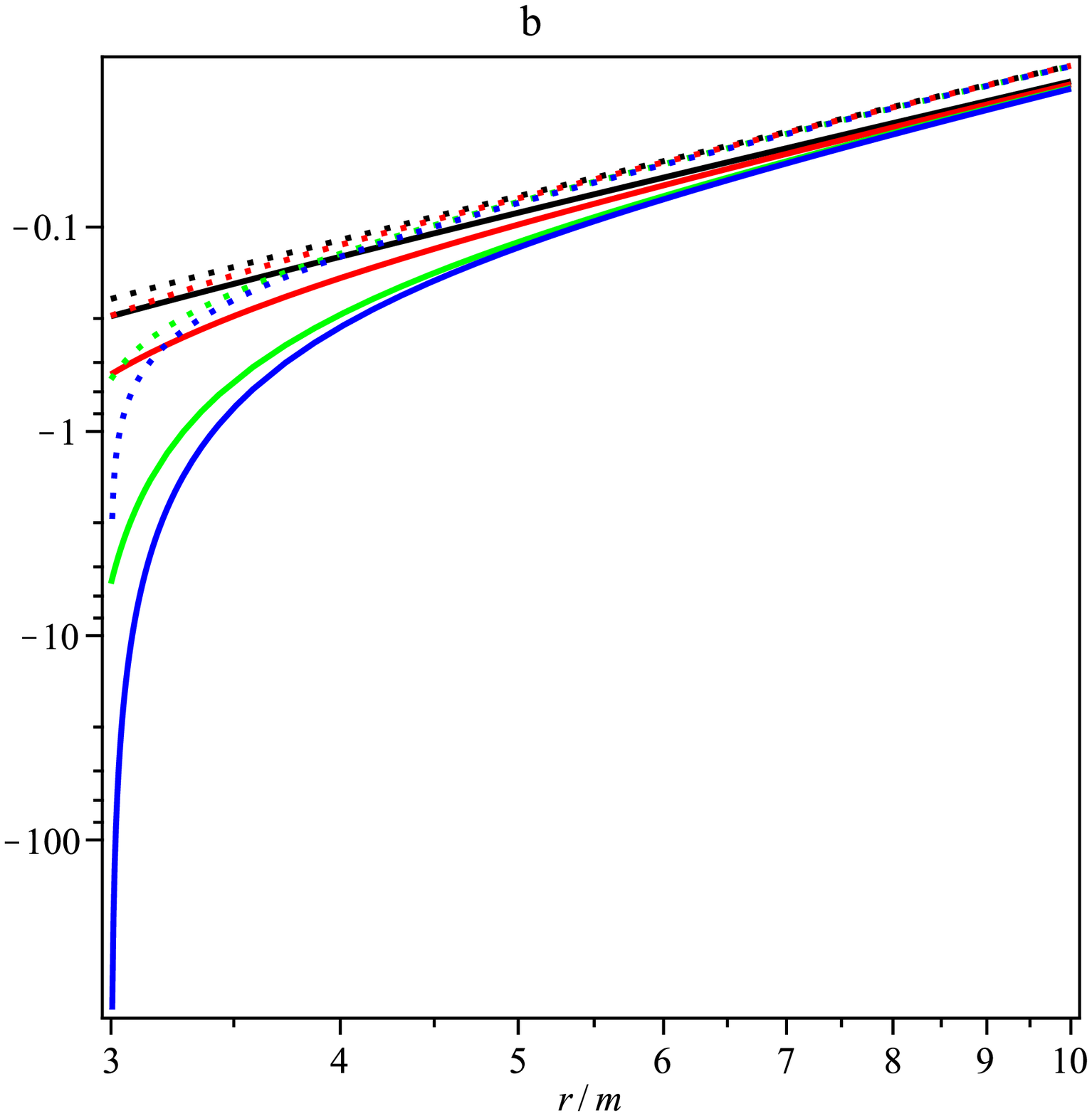}}
\centerline{\includegraphics[scale=.22]{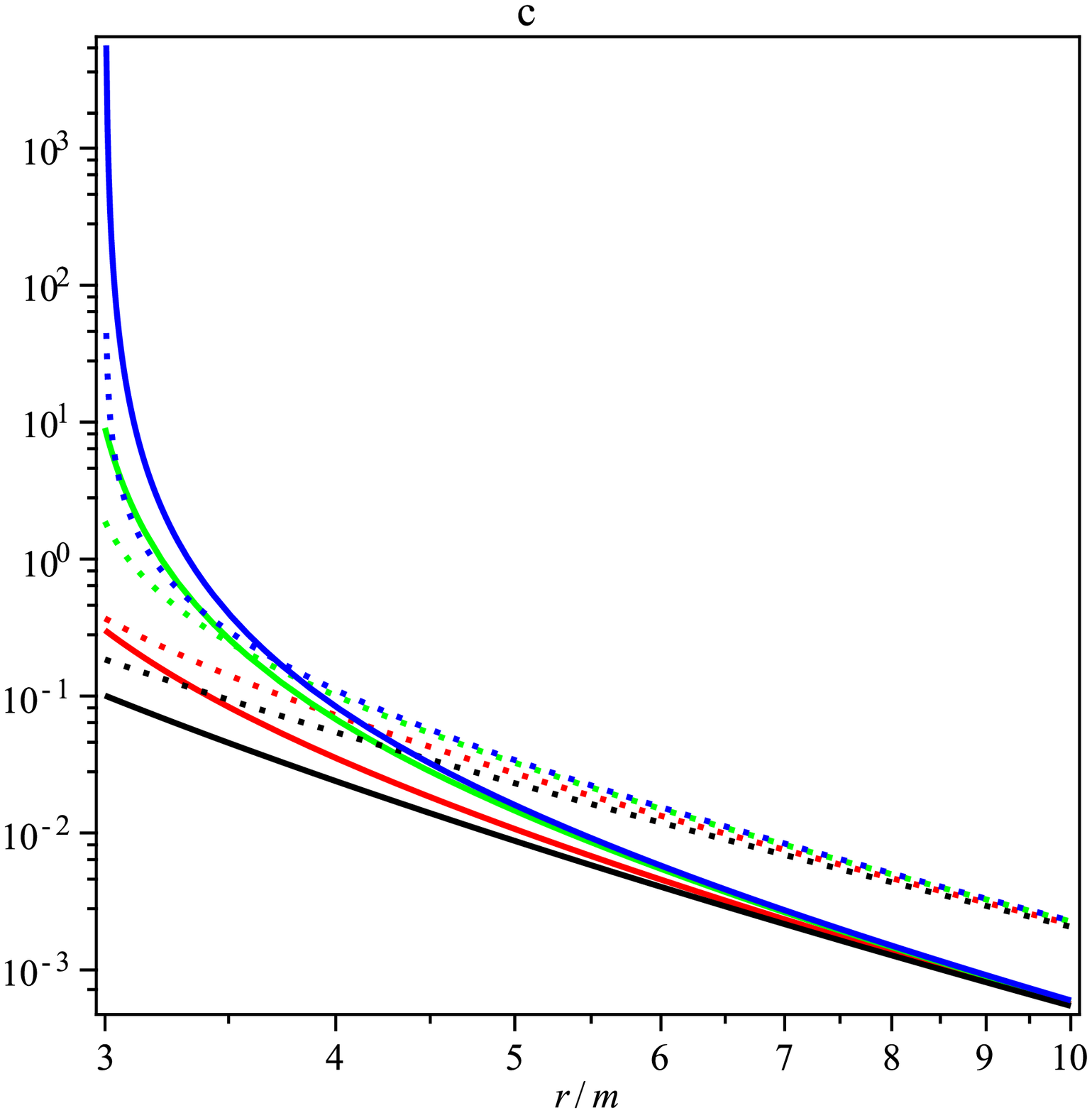}\includegraphics[scale=.22]{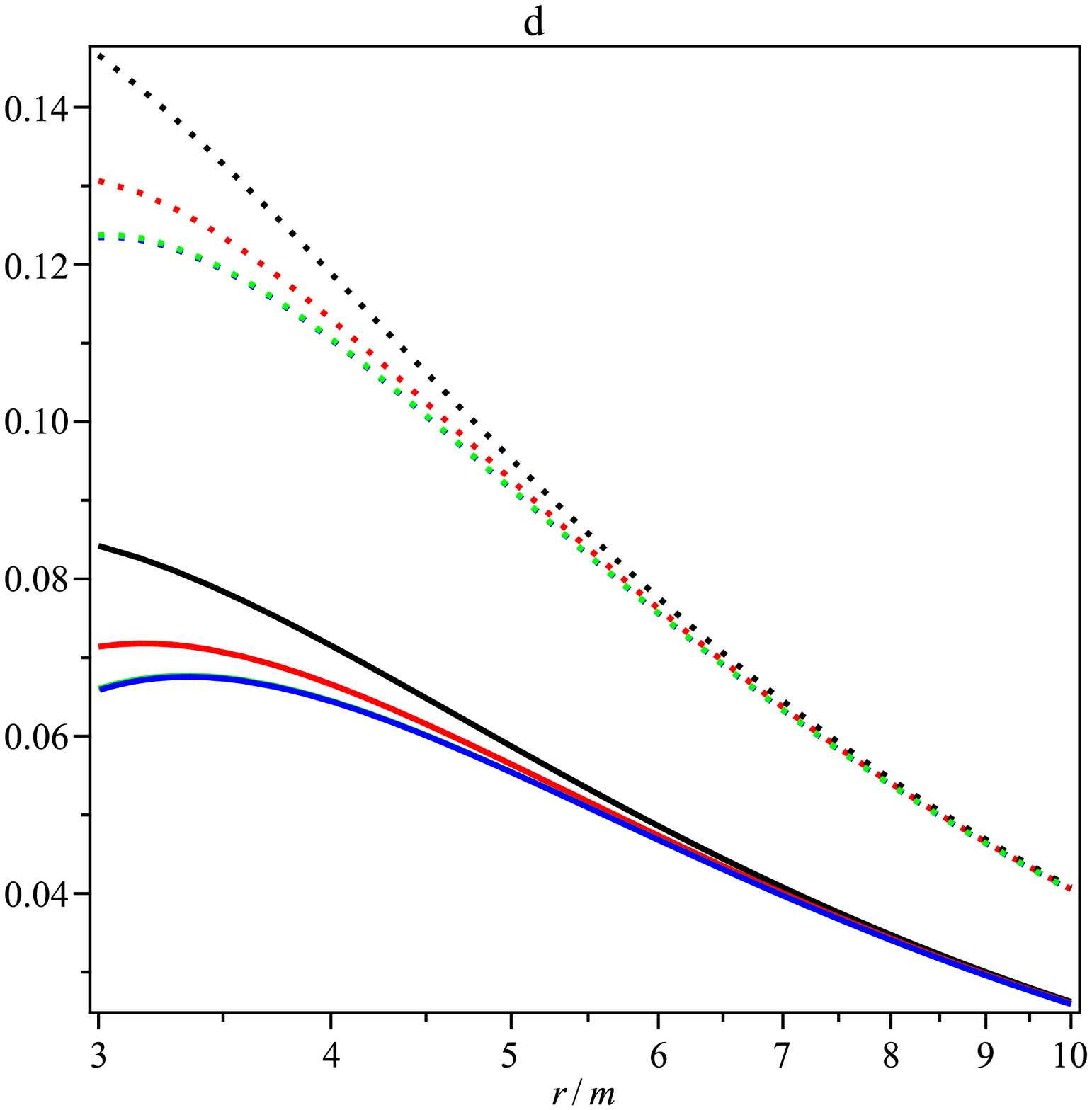}}
\centerline{\includegraphics[scale=.22]{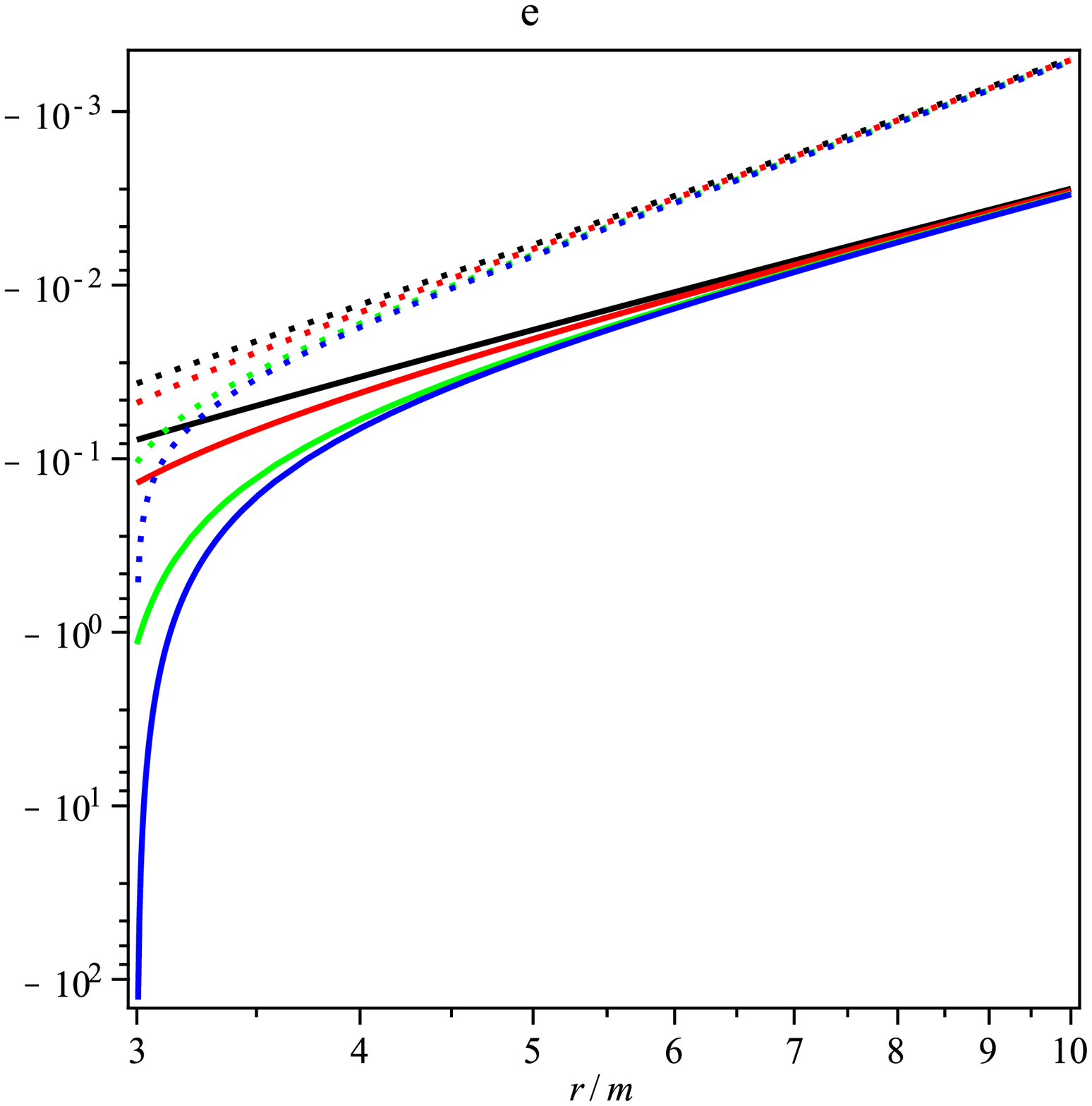} \includegraphics[scale=.22]{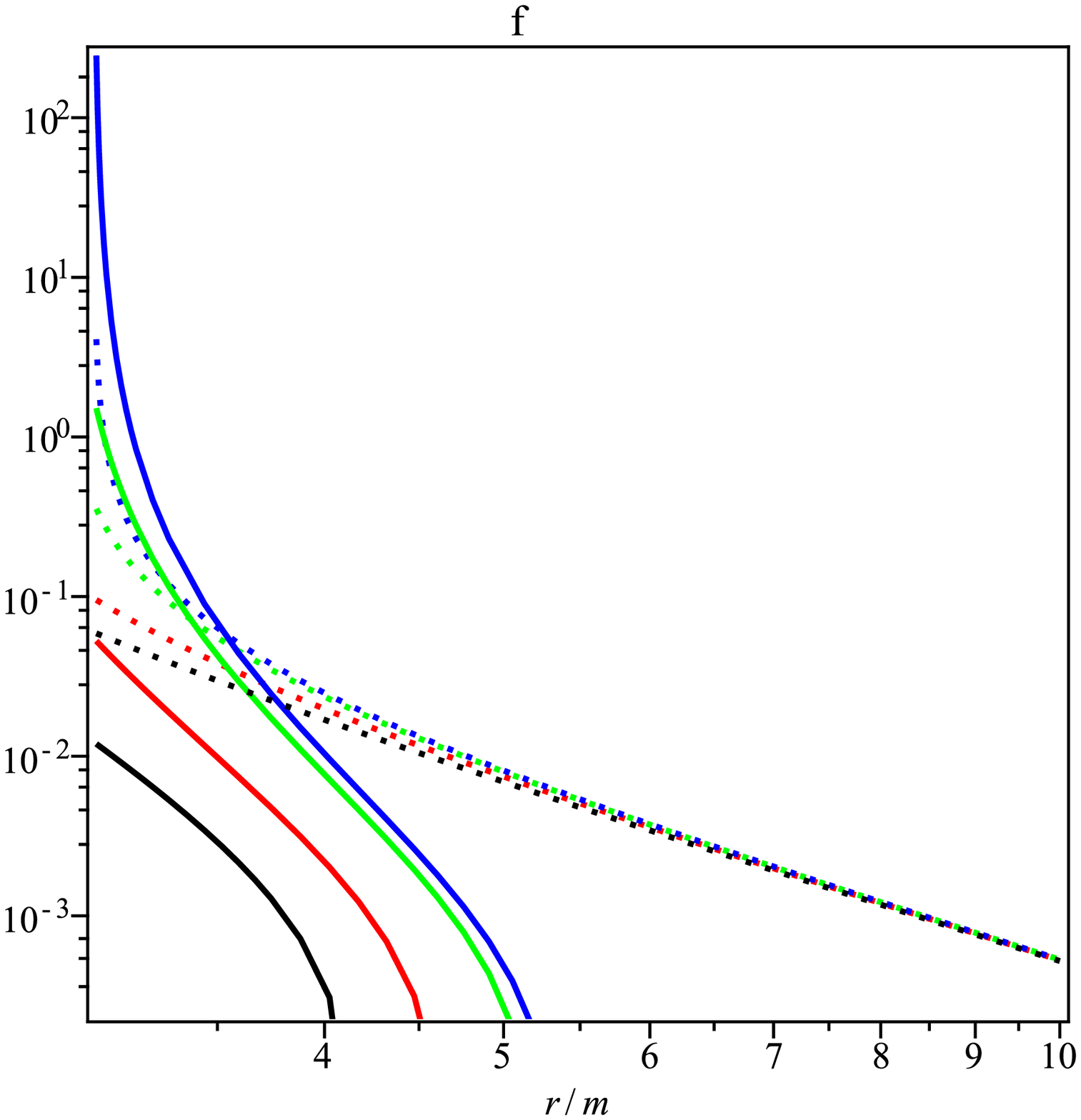}}
\centerline{\includegraphics[scale=.22]{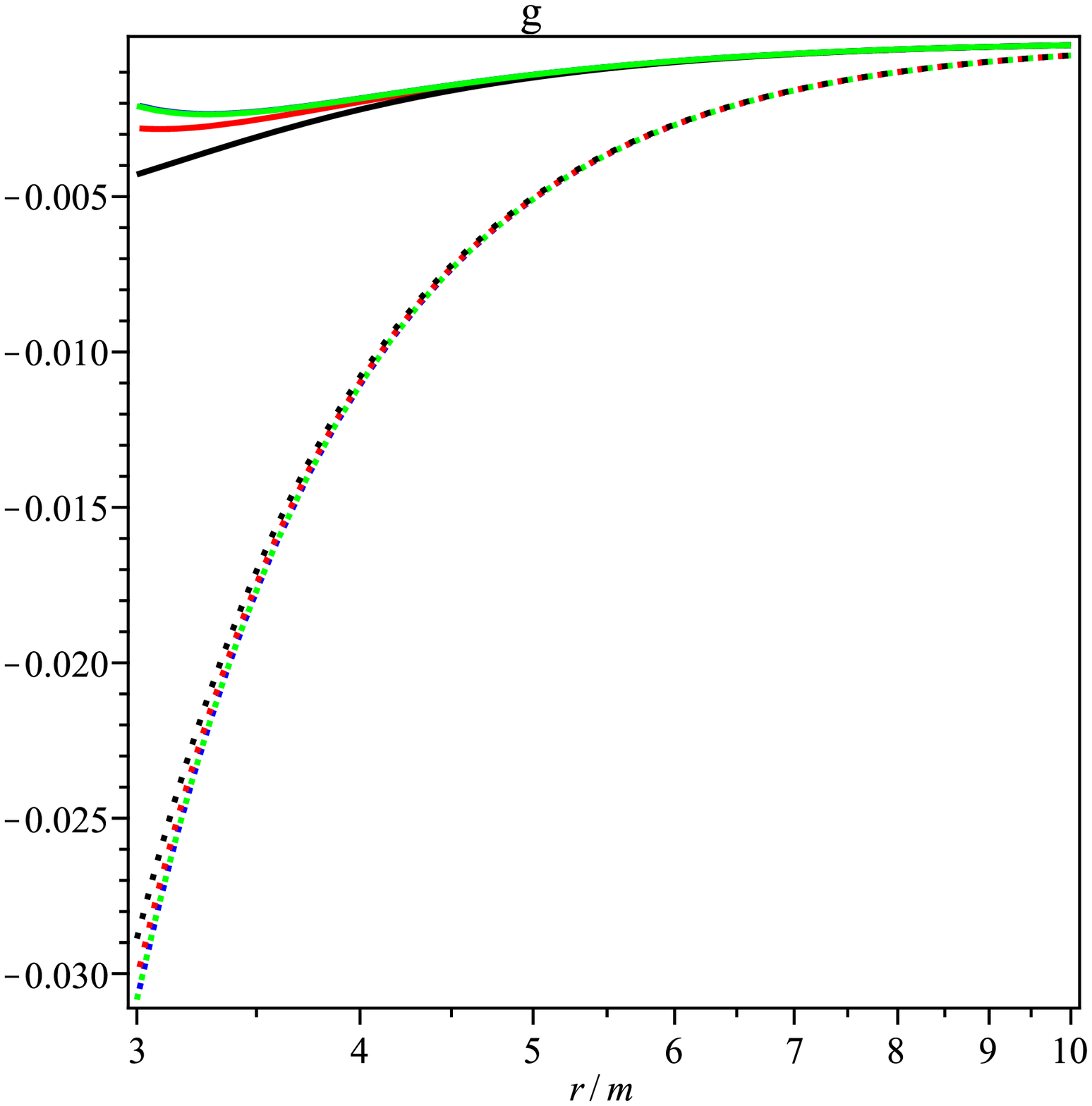}}
\caption{The non-zero components of the shear and the bulk tensors with $\beta=1$ and $n=\frac{1}{2}$. $a=.9$ in black, $a=.5$ in red, $a=.1$ in green and $a=0$ in blue. a: Solid is $\sigma^{tt}$ and dotted is $-b^{tt}$, b: Solid is $\sigma^{rt}$ and dotted is $-b^{rt}$, c: Solid is $\sigma^{t\phi}$ and dotted is $-b^{t\phi}$, d: Solid is $\sigma^{rr}$ and dotted is $-b^{rr}$, e: Solid is $\sigma^{r\phi}$ and dotted is $-b^{r\phi}$, f: Solid is $\sigma^{\phi\phi}$ and dotted is $-b^{\phi\phi}$ and g: Solid is $\sigma^{\theta\theta}$ and dotted is $b^{\theta\theta}$.}
\label{figure2}
\end{figure}

The four velocity in $n=\frac{1}{2}$ can be derived with equation (\ref{25}) as	
\begin{eqnarray}\label{26}
&&u^{\mu}=(\frac{\sqrt{r^{2}+r\beta^{2}}(r^{\frac{3}{2}}+a)}{\sqrt{r}\sqrt{r^{4}+2ar^{\frac{5}{2}}-3r^{3}}},\nonumber\\&&-\frac{\beta\sqrt{r^{2}-2r+a^{2}}}{r^{\frac{3}{2}}},0,\frac{\sqrt{r^2+r\beta^2}}{\sqrt{r}\sqrt{r^{4}+2ar^{\frac{5}{2}}-3r^{3}}}).
\end{eqnarray}
$\Theta=-\frac{\beta(-4r+3r^{2}+a^{2})}{2r^{\frac{5}{2}}\sqrt{r^{2}-2r+a^{2}}}$ is the expansion of the fluid world line and the components of the shear and the bulk tensors are shown in figure \ref{figure2}.
\subsection{$n=1$}
\begin{figure}
\vspace{\fill}
\centerline{\includegraphics[scale=.22]{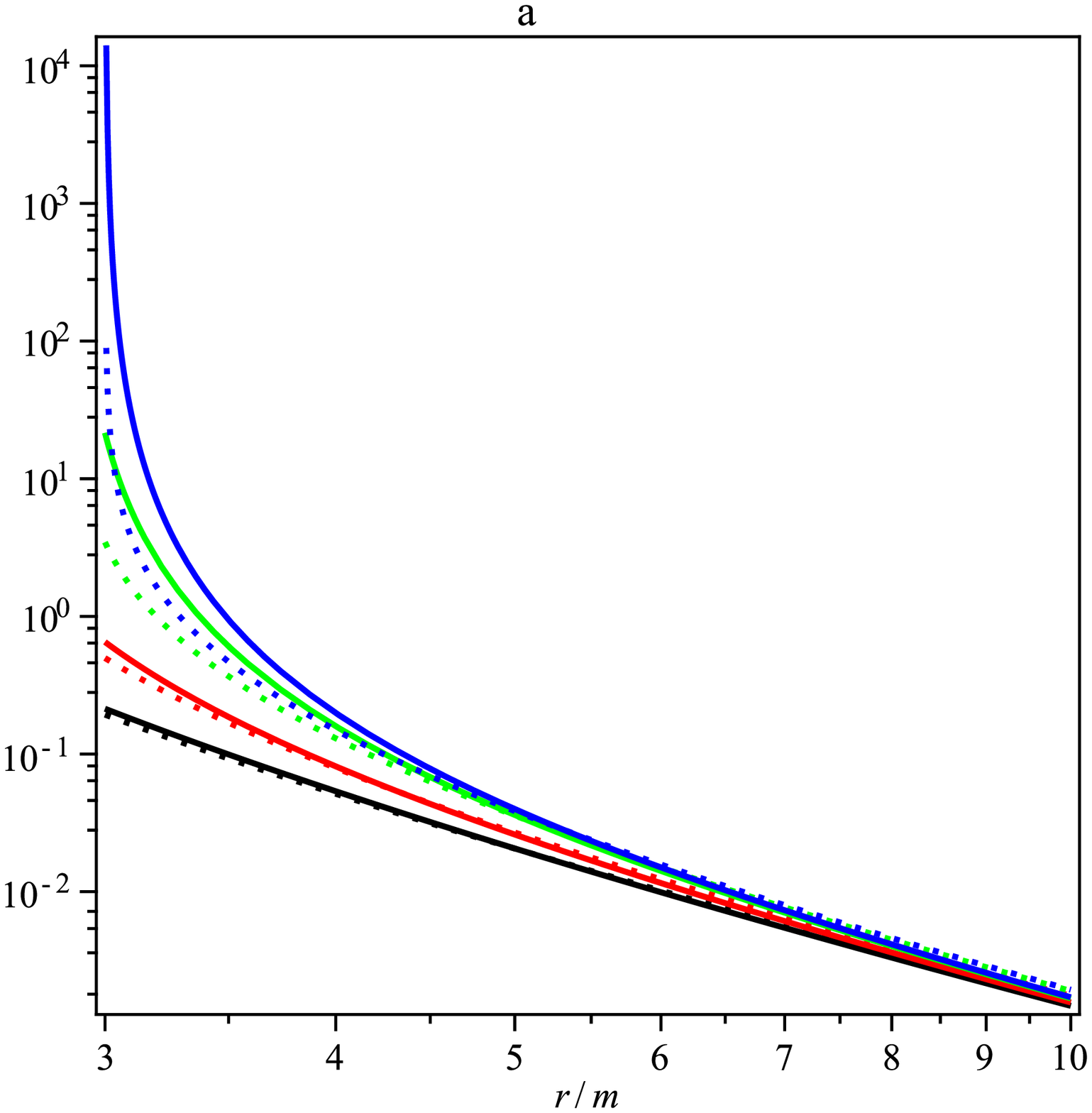}\includegraphics[scale=.22]{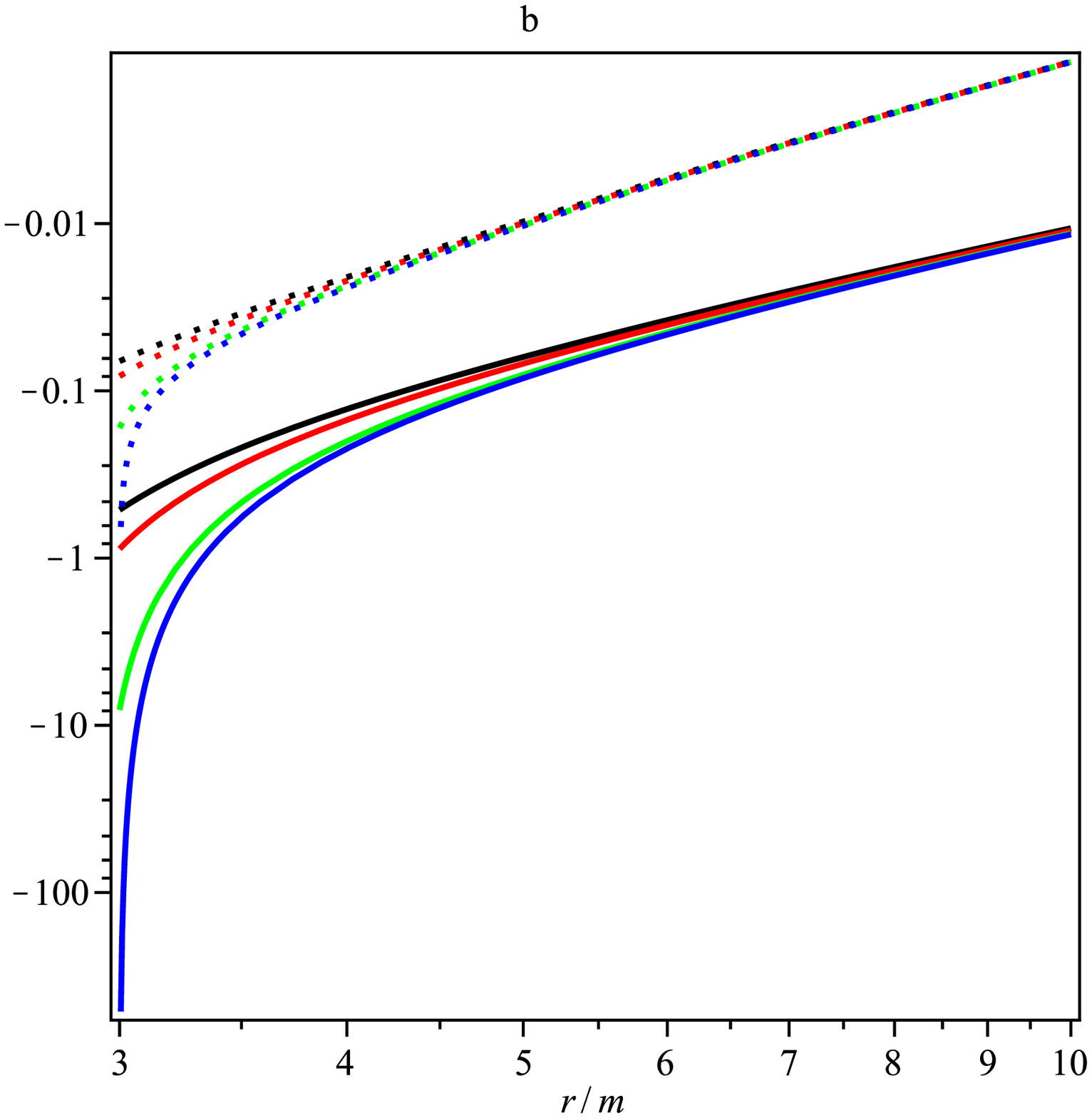}}
\centerline{\includegraphics[scale=.22]{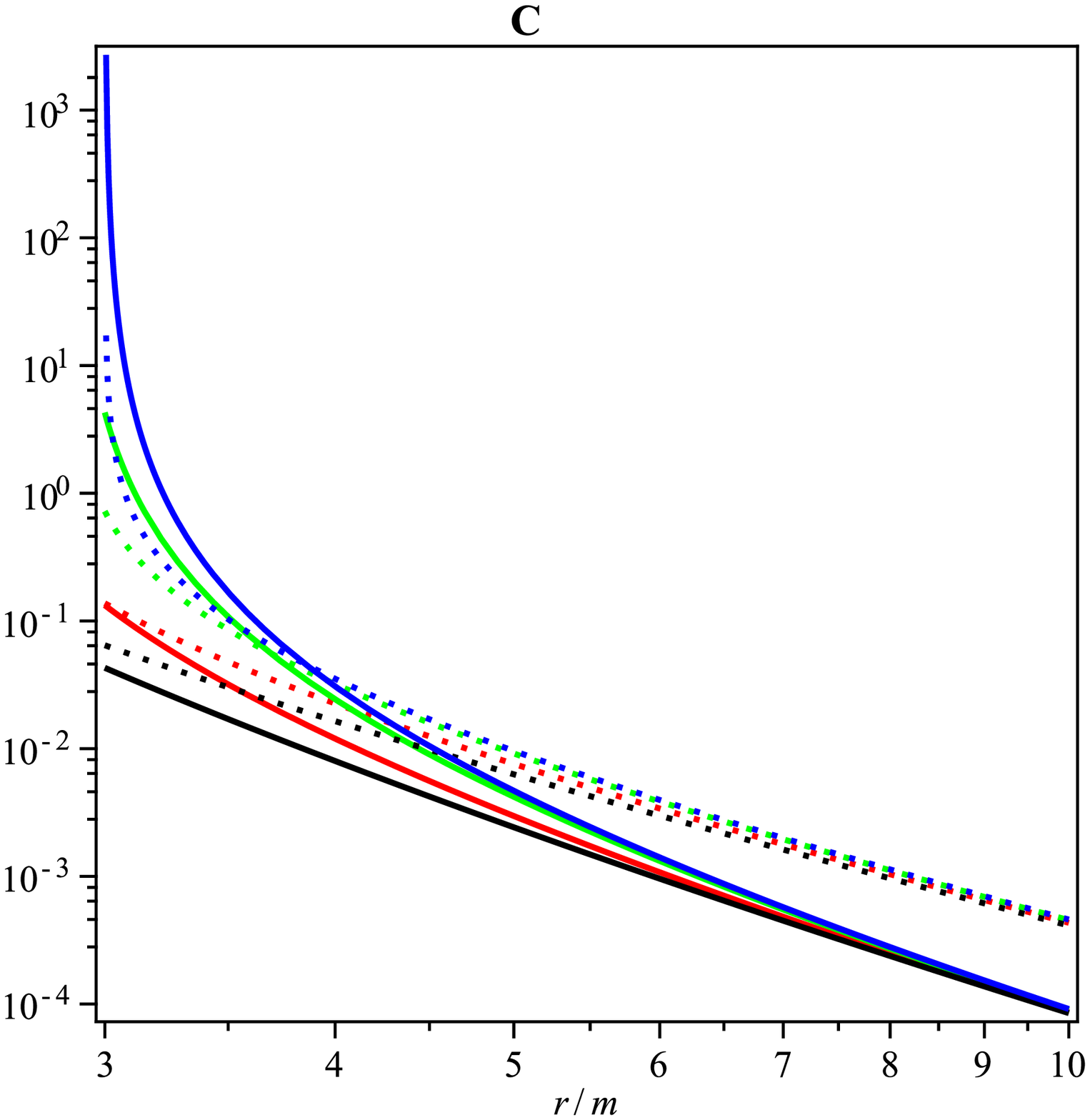}\includegraphics[scale=.22]{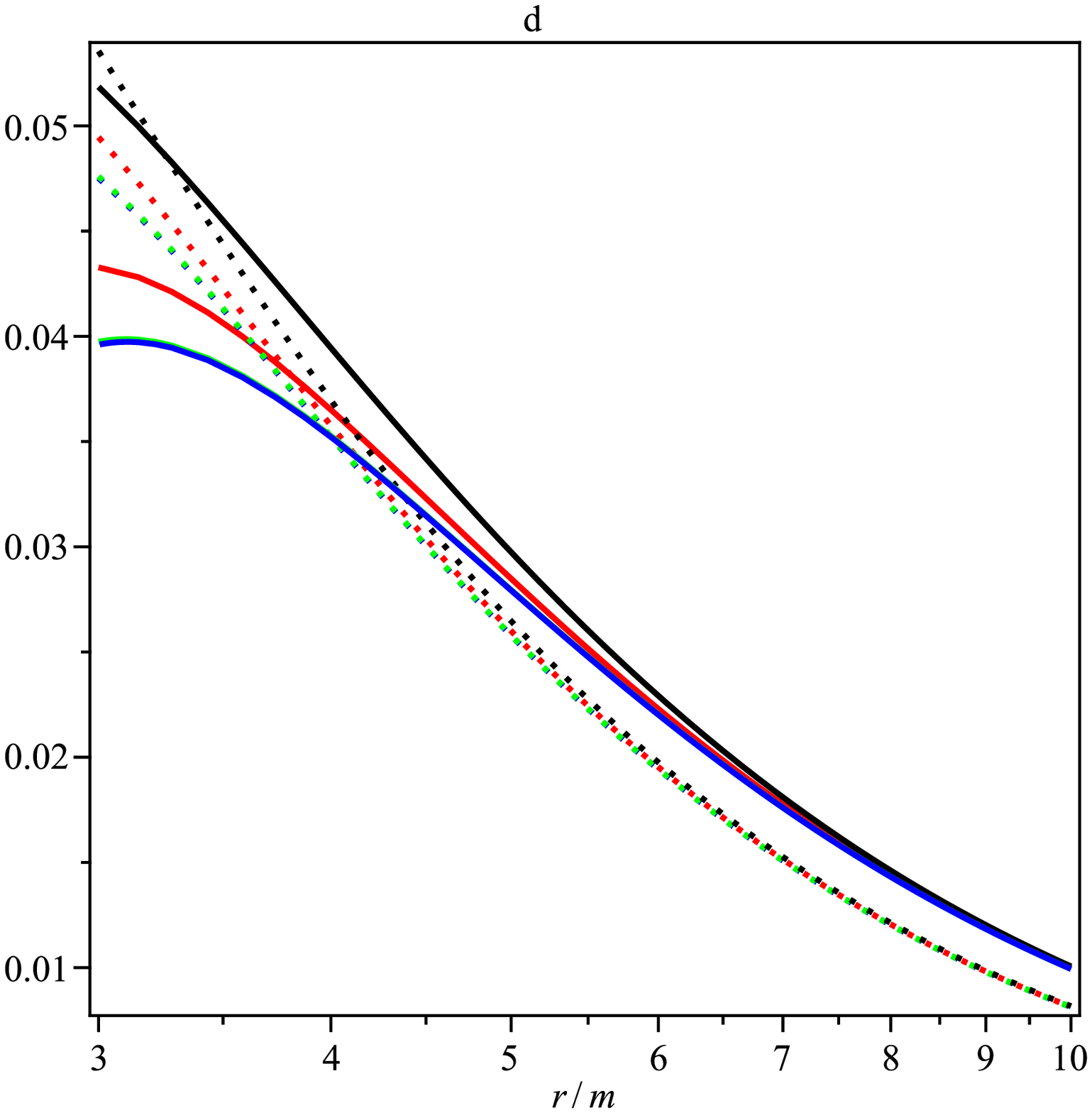}}
\centerline{\includegraphics[scale=.22]{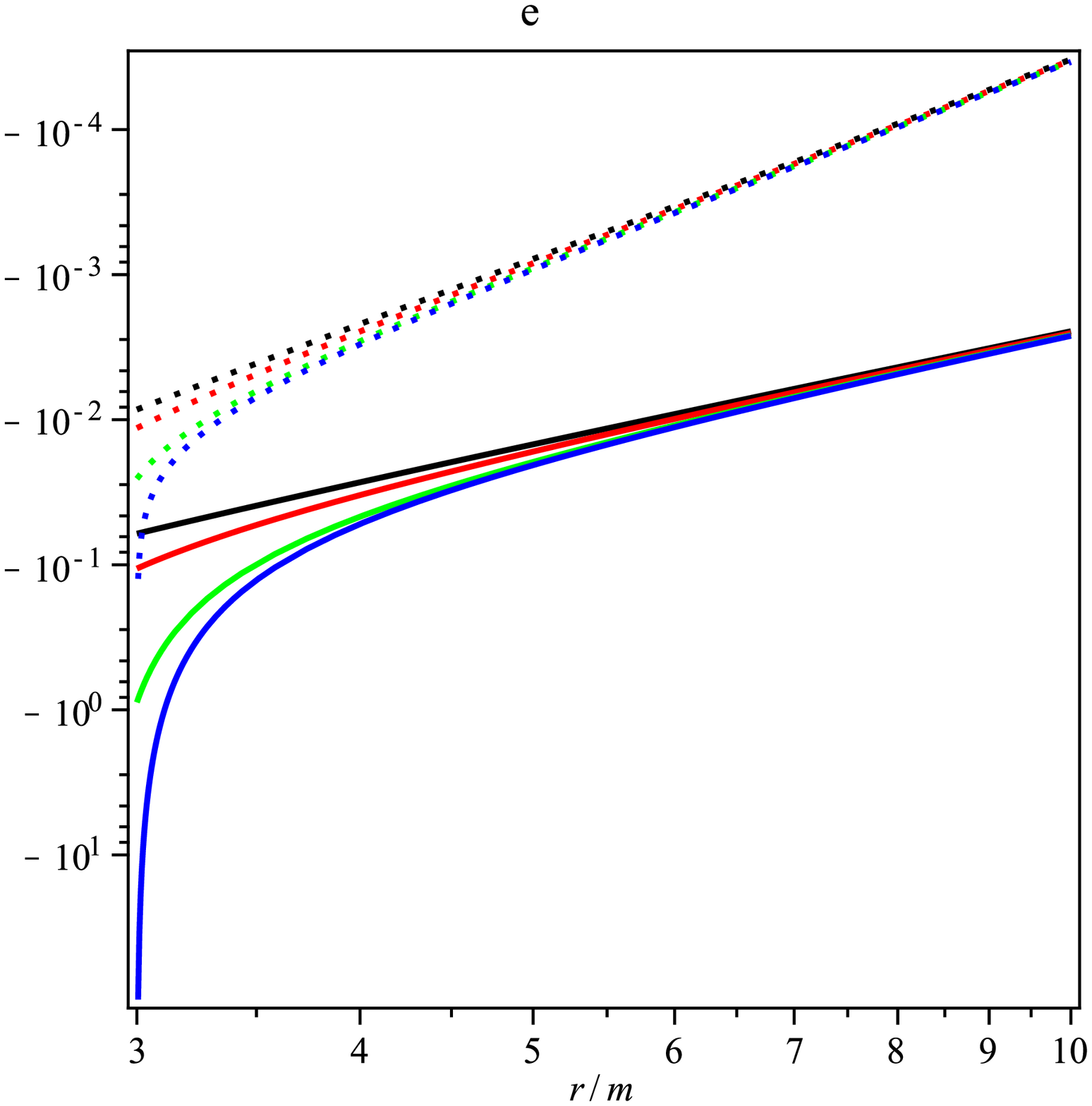} \includegraphics[scale=.22]{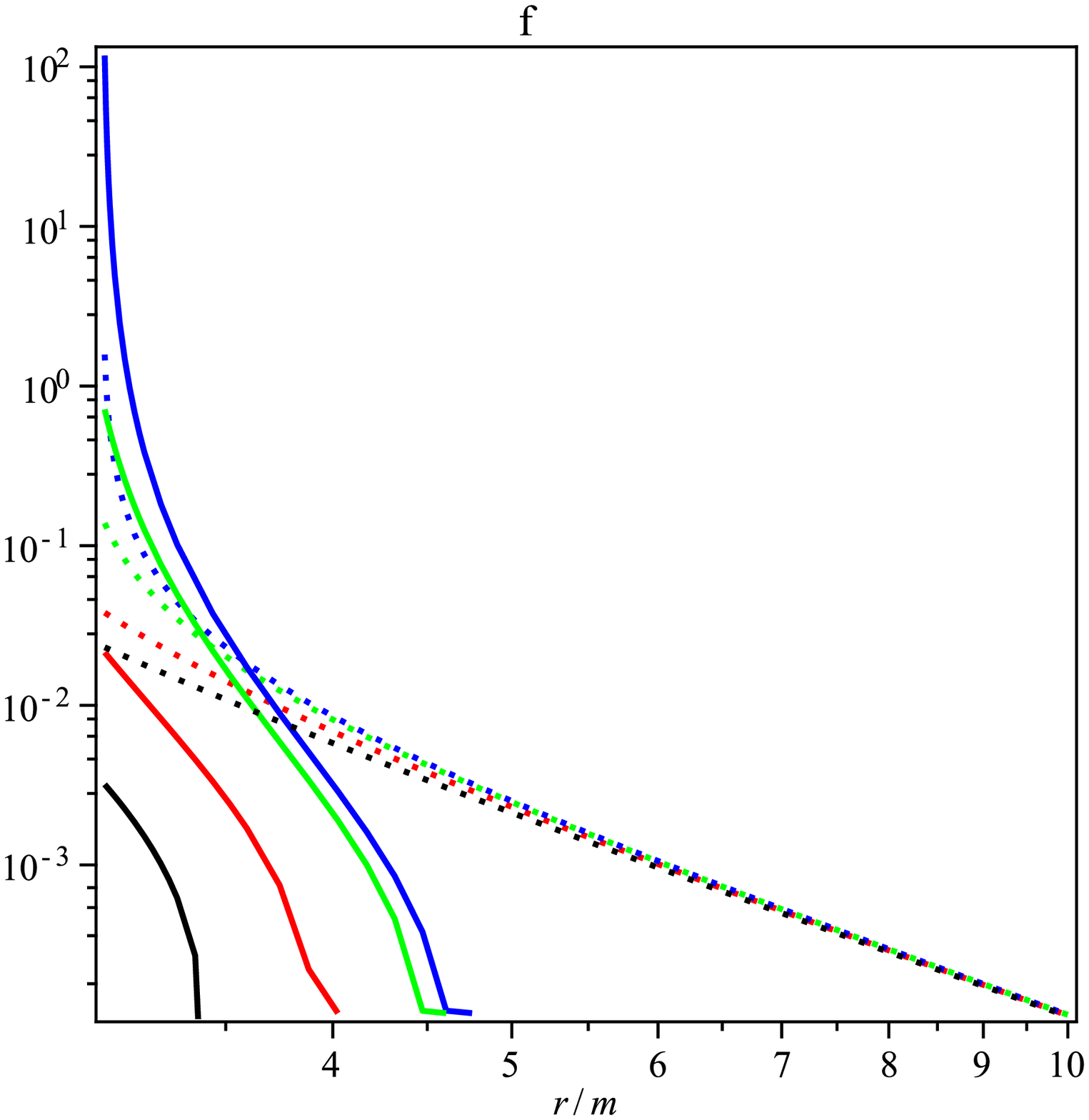}}
\centerline{\includegraphics[scale=.22]{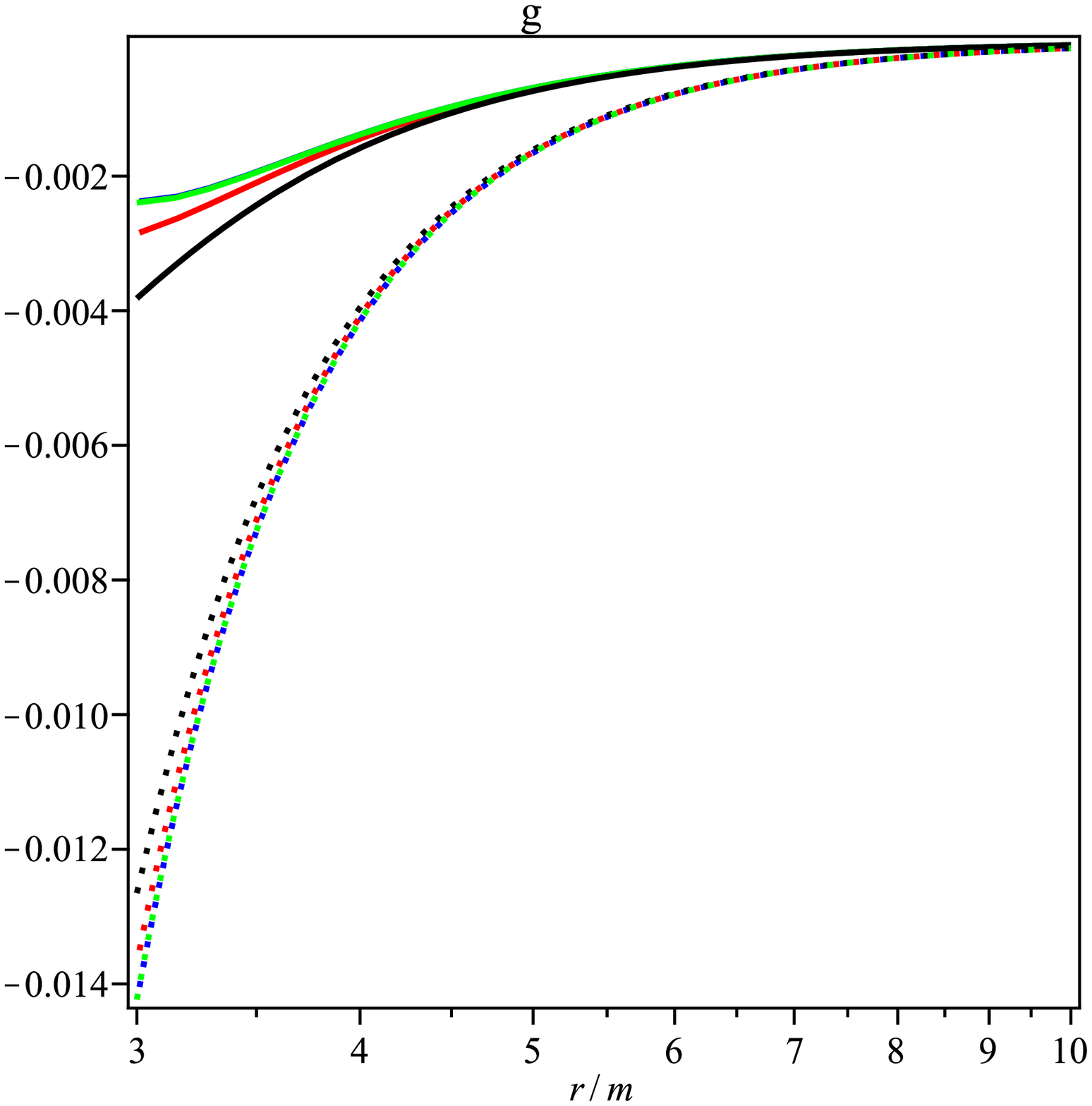}}
 \caption{The non-zero components of the shear and the bulk tensors with $\beta=1$ and $n=1$. $a=.9$ in black, $a=.5$ in red, $a=.1$ in green and $a=0$ in blue. a: Solid is $\sigma^{tt}$ and dotted is $-b^{tt}$, b: Solid is $\sigma^{rt}$ and dotted is $-b^{rt}$, c: Solid is $\sigma^{t\phi}$ and dotted is $-b^{t\phi}$, d: Solid is $\sigma^{rr}$ and dotted is $-b^{rr}$, e: Solid is $\sigma^{r\phi}$ and dotted is $-b^{r\phi}$, f: Solid is $\sigma^{\phi\phi}$ and dotted is $-b^{\phi\phi}$ and g: Solid is $\sigma^{\theta\theta}$ and dotted is $b^{\theta\theta}$.}
\label{figure3}
\end{figure}

In $n=1$ the four velocity is derived with equation (\ref{25}) as	
\begin{eqnarray}\label{27}
&&u^{\mu}=(\frac{\sqrt{r^{3}+r\beta^{2}}(r^{\frac{3}{2}}+a)}{r\sqrt{r^{4}+2ar^{\frac{5}{2}}-3r^{3}}},\nonumber\\&&-\frac{\beta\sqrt{r^{2}-2r+a^{2}}}{r^{2}},0,\frac{\sqrt{r^{3}+r\beta^{2}}}{r\sqrt{r^{4}+2ar^{\frac{5}{2}}-3r^{3}}}).
\end{eqnarray} 
$\Theta=-\frac{\beta(r-1)}{r^{2}\sqrt{r^{2}-2r+a^{2}}}$ is the expansion of the fluid world line and the components of the shear and the bulk tensors are shown in figure \ref{figure3}.
\subsection{$n=\frac{3}{2}$}
\begin{figure}
\vspace{\fill}
\centerline{\includegraphics[scale=.22]{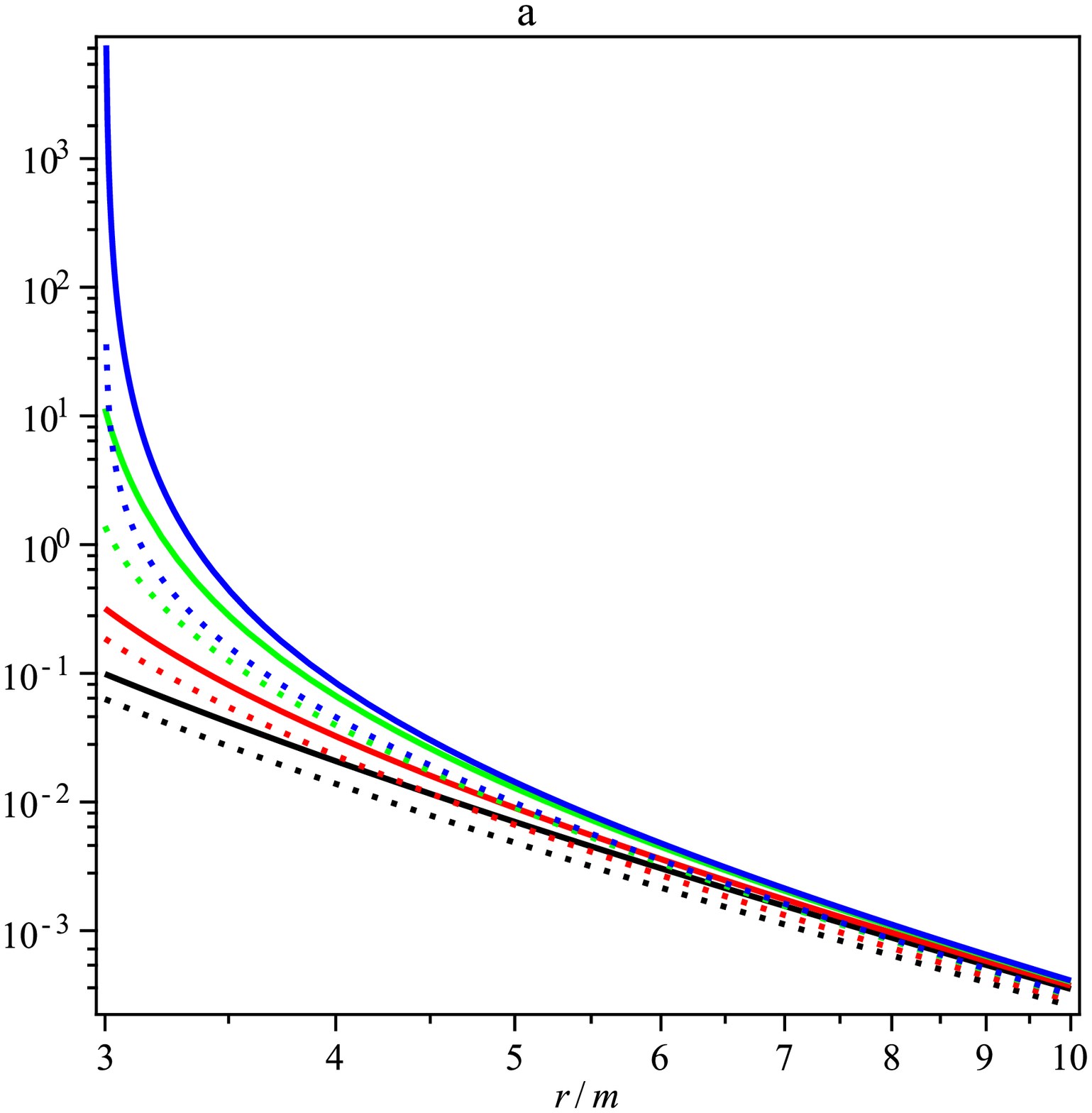}\includegraphics[scale=.22]{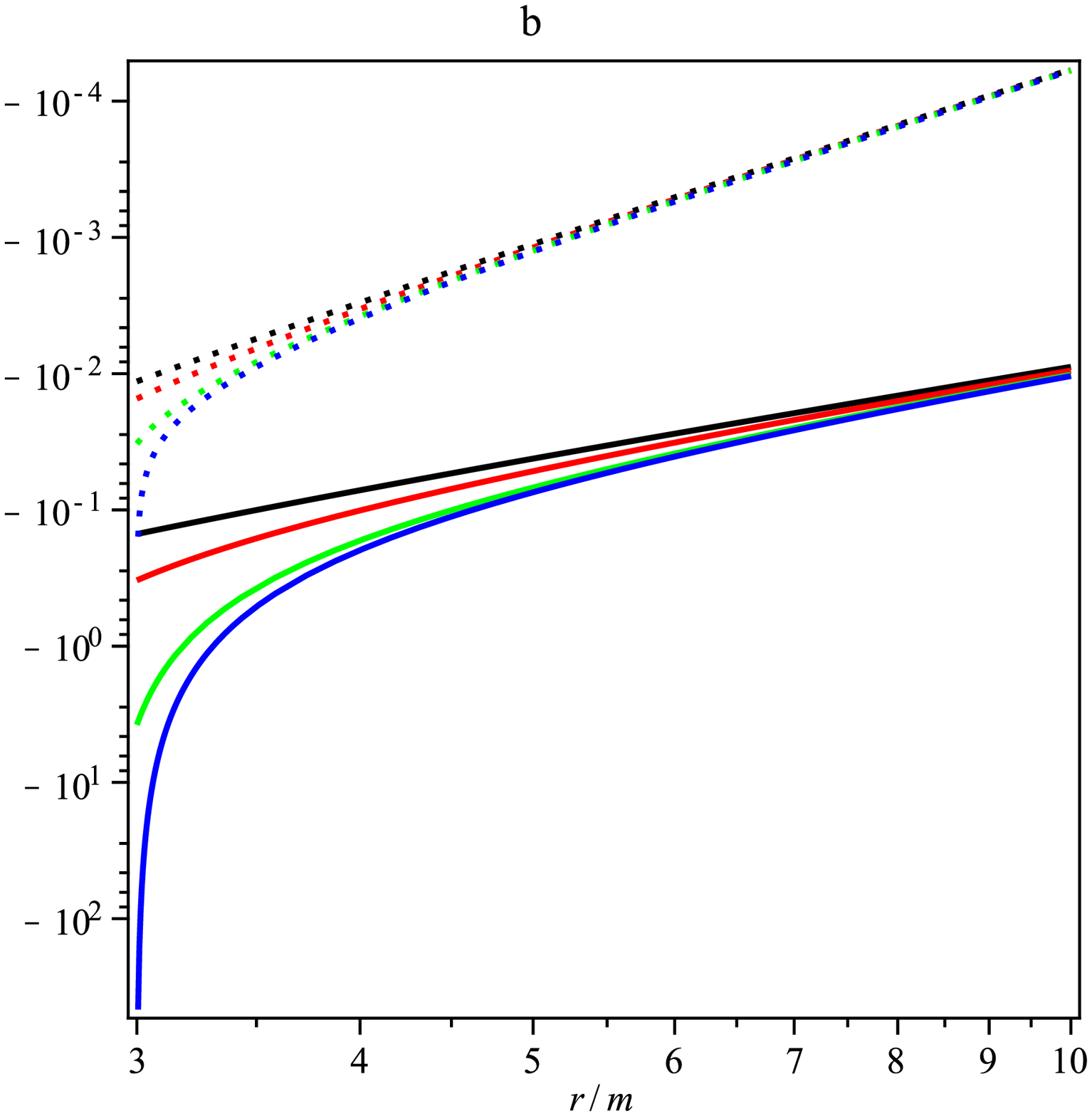}}
\centerline{\includegraphics[scale=.22]{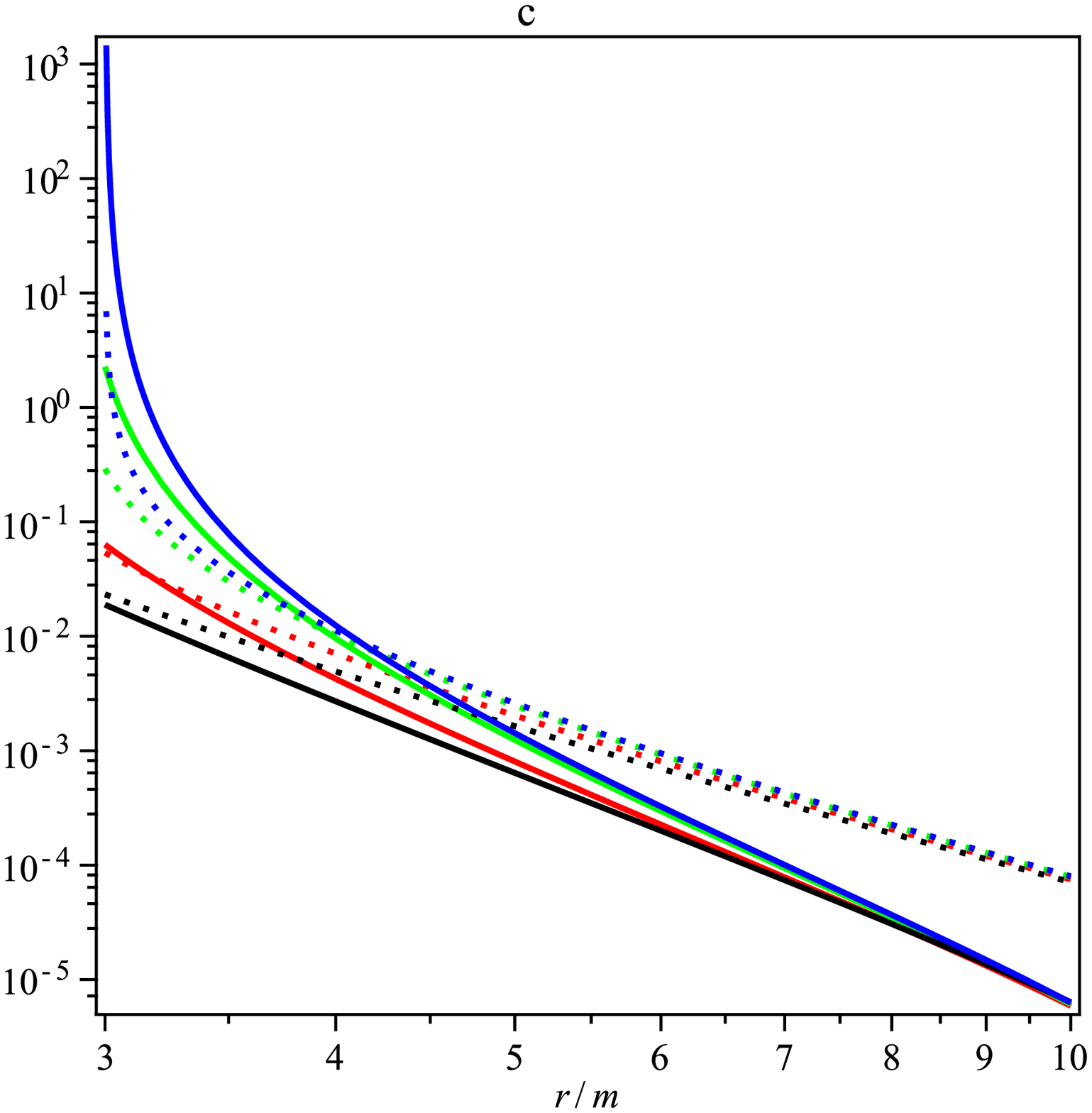}\includegraphics[scale=.22]{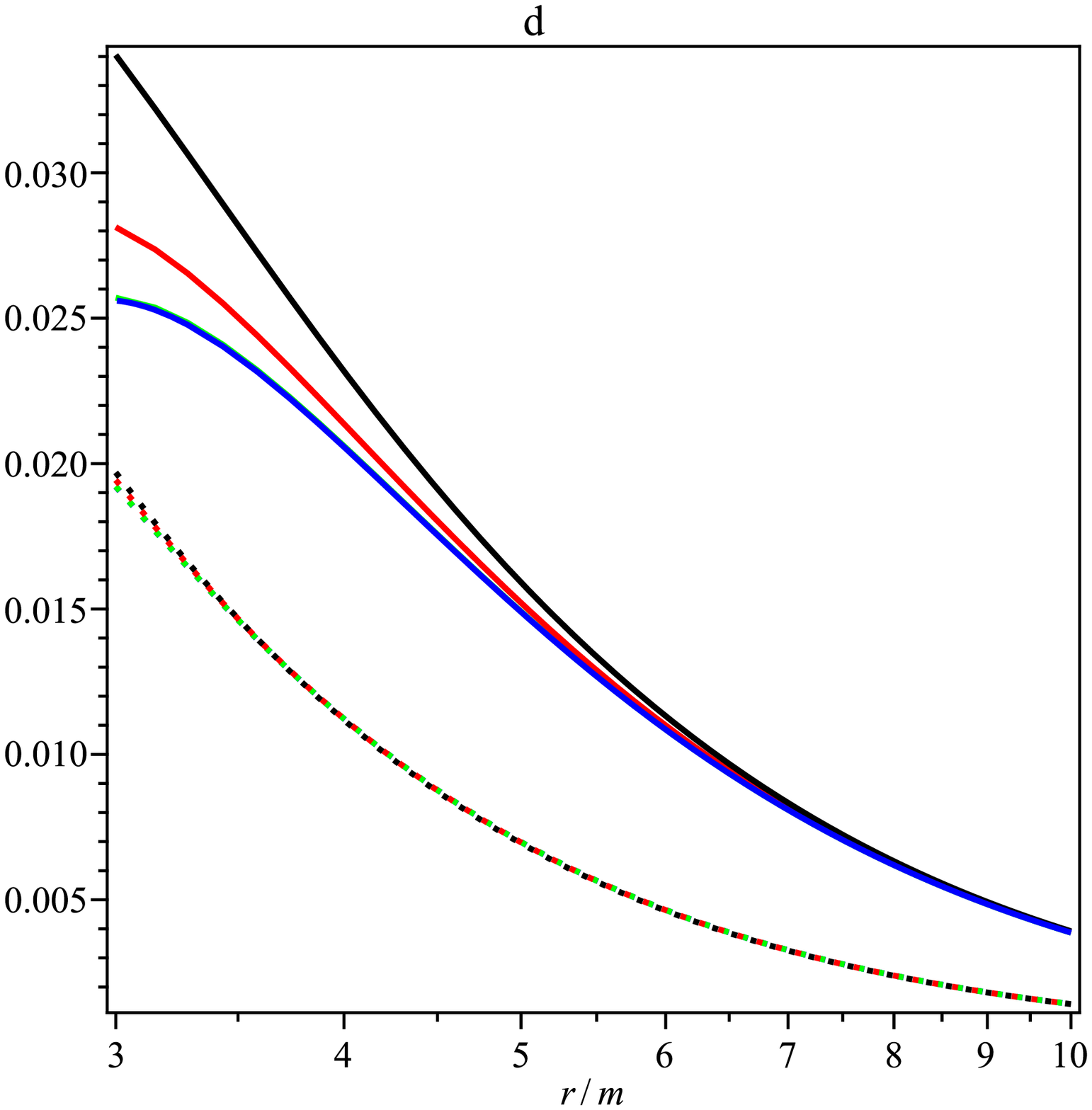}}
\centerline{\includegraphics[scale=.22]{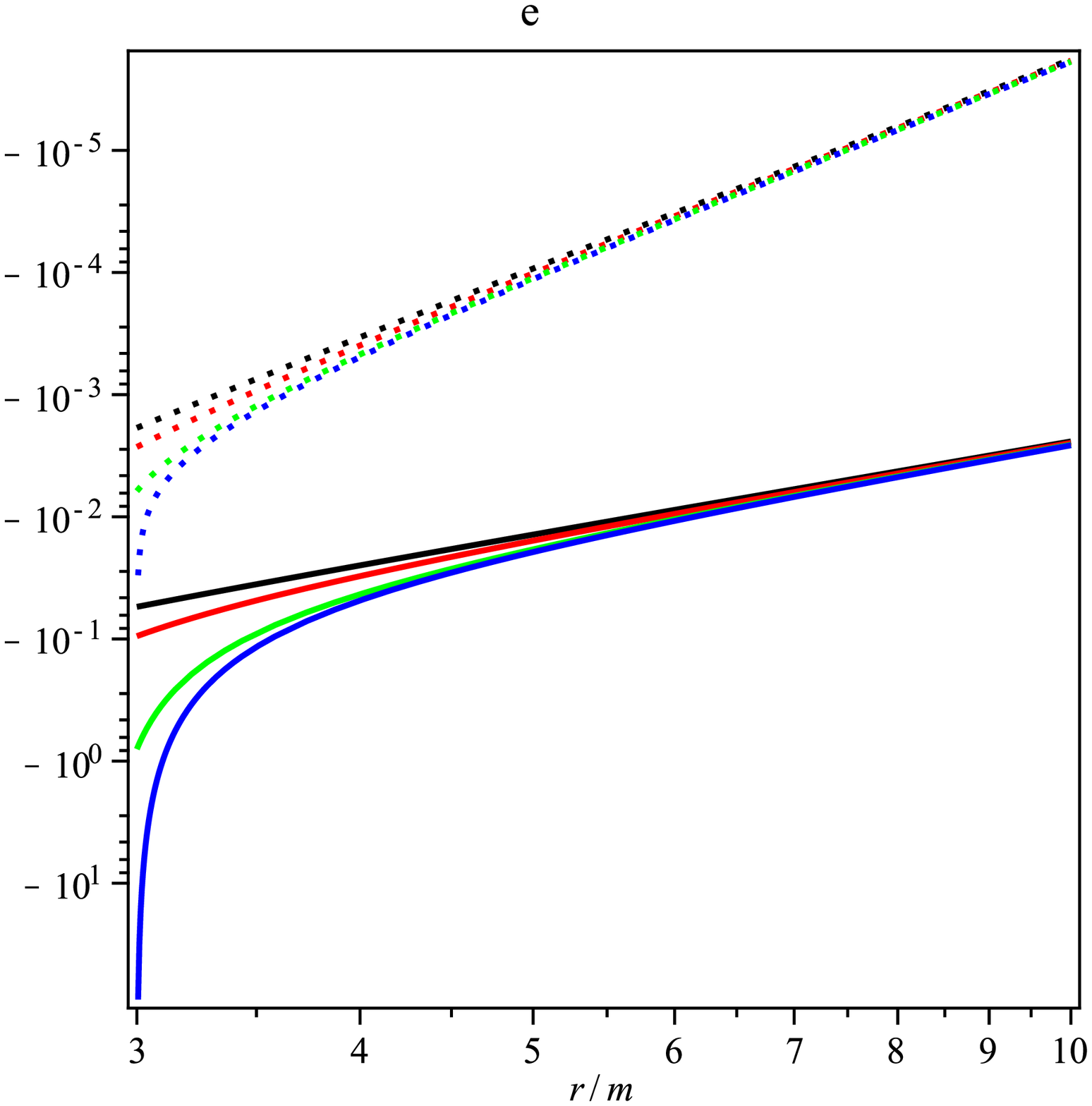} \includegraphics[scale=.22]{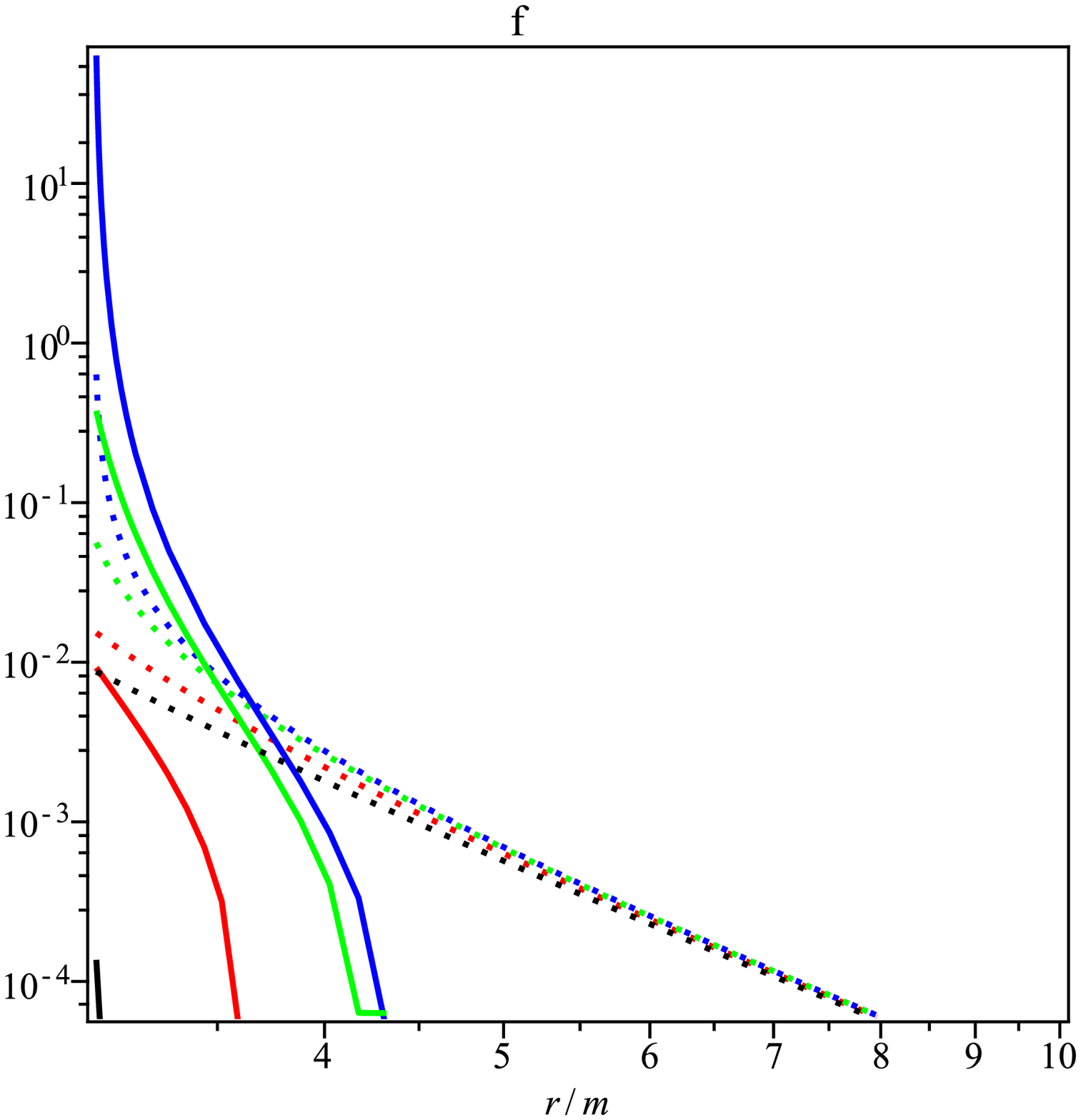}}
\centerline{\includegraphics[scale=.22]{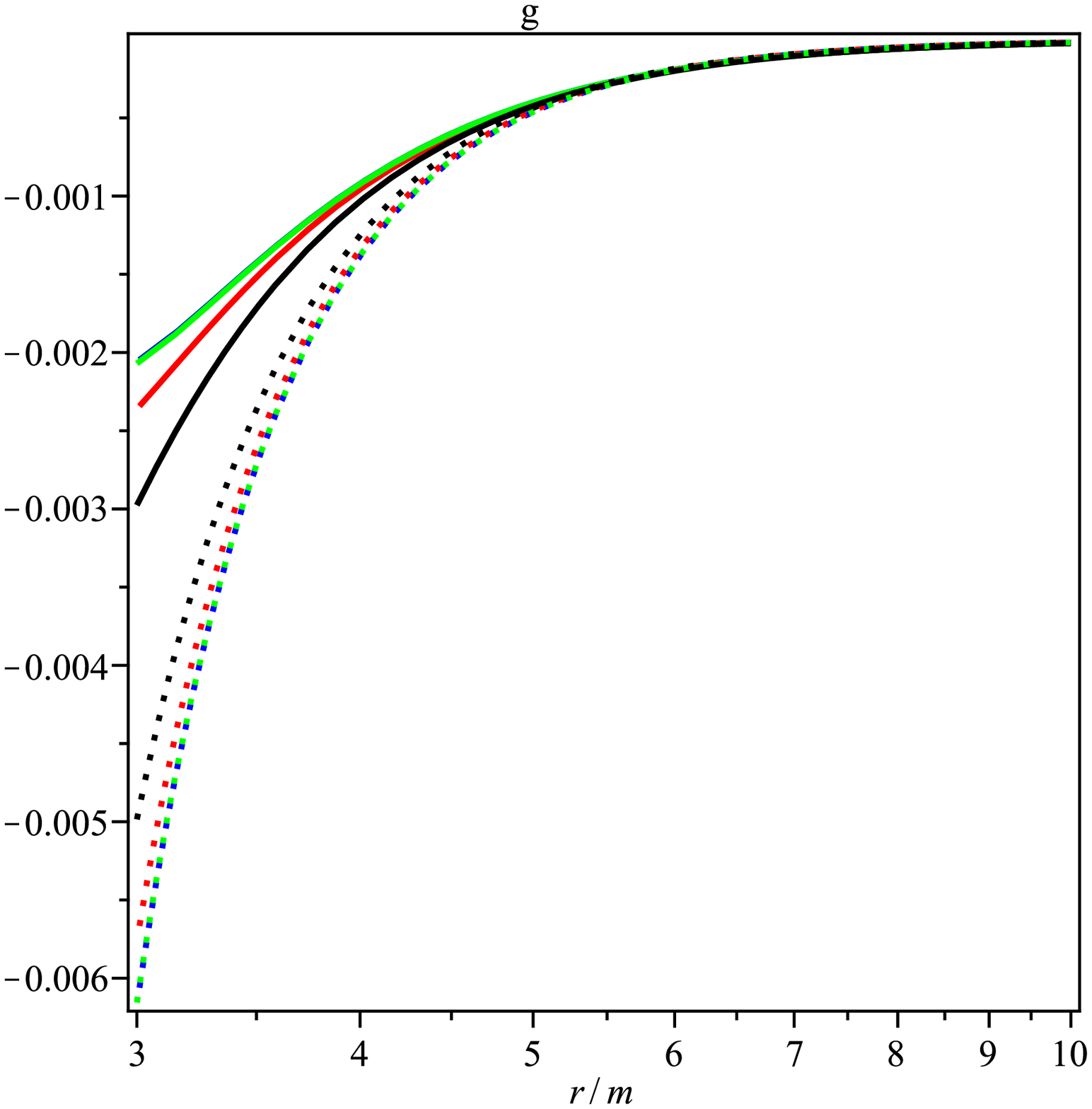}}
 \caption{The non-zero components of the shear and the bulk tensors with $\beta=1$ and $n=\frac{3}{2}$. $a=.9$ in black, $a=.5$ in red, $a=.1$ in green and $a=0$ in blue. a: Solid is $\sigma^{tt}$ and dotted is $-b^{tt}$, b: Solid is $\sigma^{rt}$ and dotted is $-b^{rt}$, c: Solid is $\sigma^{t\phi}$ and dotted is $-b^{t\phi}$, d: Solid is $\sigma^{rr}$ and dotted is $-b^{rr}$, e: Solid is $\sigma^{r\phi}$ and dotted is $-b^{r\phi}$, f: Solid is $\sigma^{\phi\phi}$ and dotted is $-b^{\phi\phi}$ and g: Solid is $\sigma^{\theta\theta}$ and dotted is $b^{\theta\theta}$.}
\label{figure4}
\end{figure}

From  equation (\ref{25}) the four velocity derives as
\begin{eqnarray}\label{28}
&&u^{\mu}=(\frac{\sqrt{r^{4}+r\beta^{2}}(r^{\frac{3}{2}}+a)}{r^{\frac{3}{2}}\sqrt{r^{4}+2ar^{\frac{5}{2}}-3r^{3}}},\nonumber\\&&-\frac{\beta\sqrt{r^{2}-2r+a^{2}}}{r^{\frac{5}{2}}},0,\frac{\sqrt{r^{4}+r\beta^{2}}}{r^{\frac{3}{2}}\sqrt{r^{4}+2ar^{\frac{5}{2}}-3r^{3}}}).
\end{eqnarray} 
In this state  $\Theta=\frac{\beta(-r^{2}+a^{2})}{2r^\frac{7}{2}\sqrt{r^{2}-2r+a^{2}}}$, also the figure \ref{figure4} shows the components of the shear and bulk tensors.
\subsection{$n=2$}          
\begin{figure}
\vspace{\fill}
\centerline{\includegraphics[scale=.22]{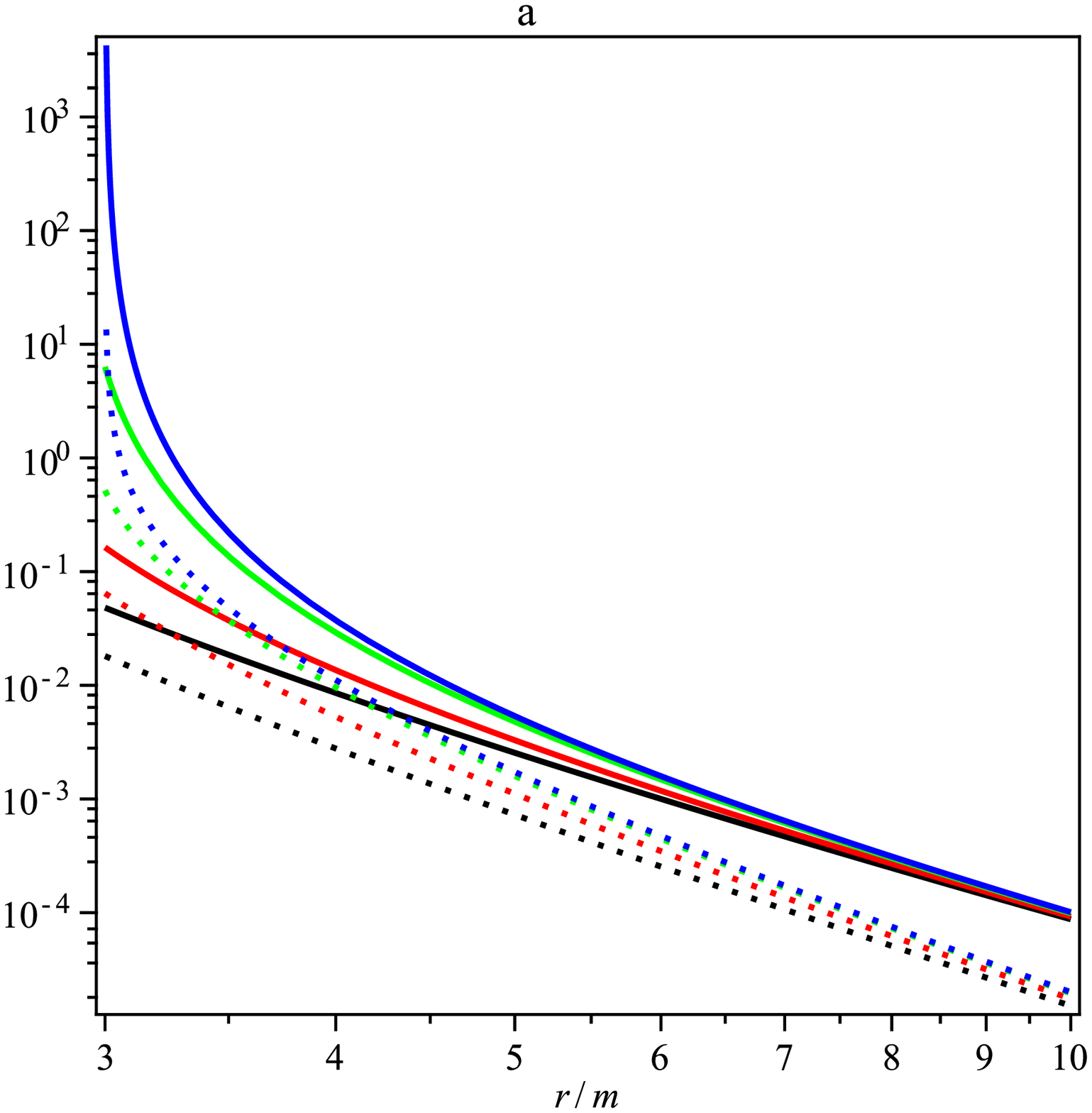}\includegraphics[scale=.22]{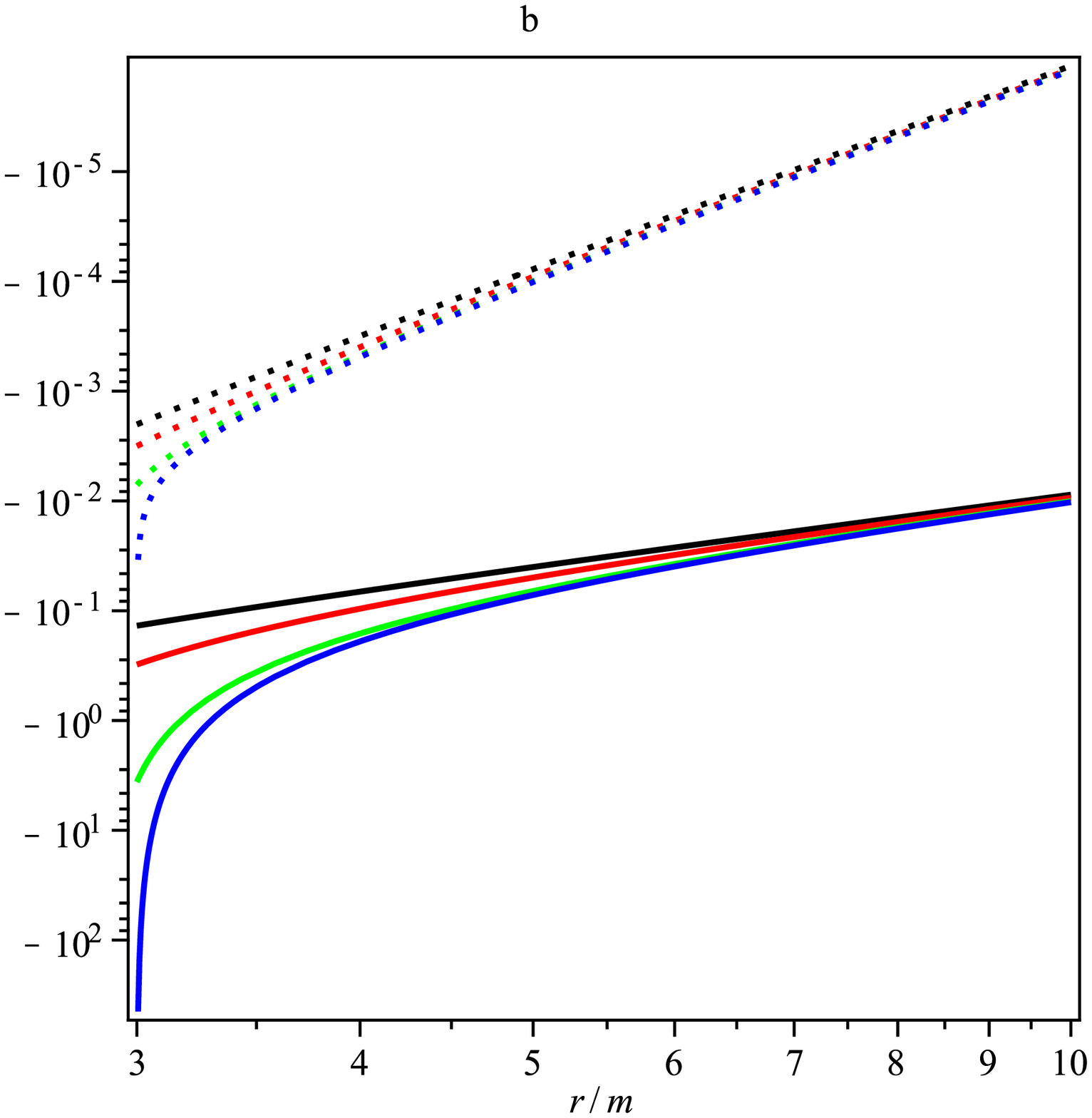}}
\centerline{\includegraphics[scale=.22]{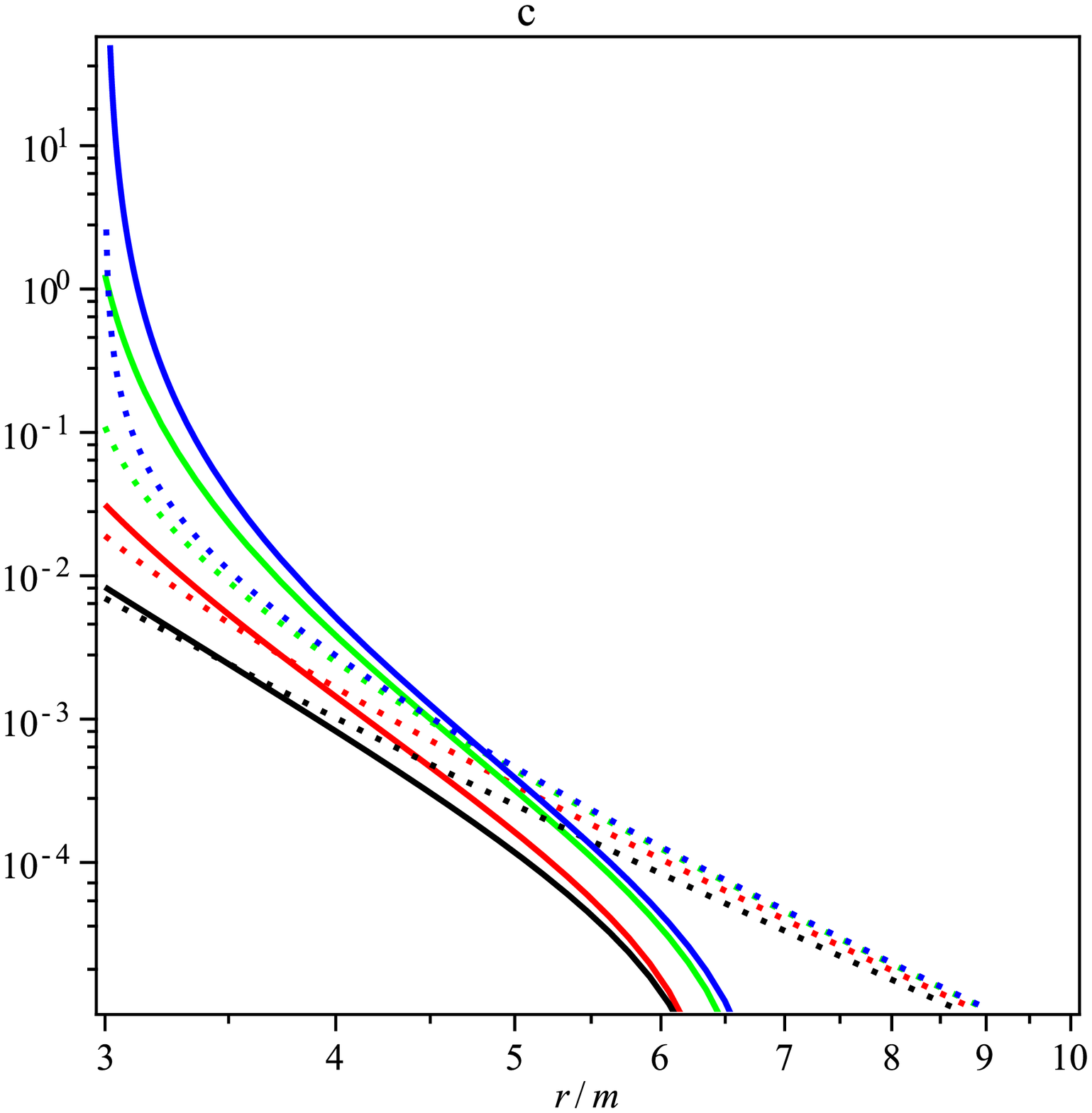}\includegraphics[scale=.22]{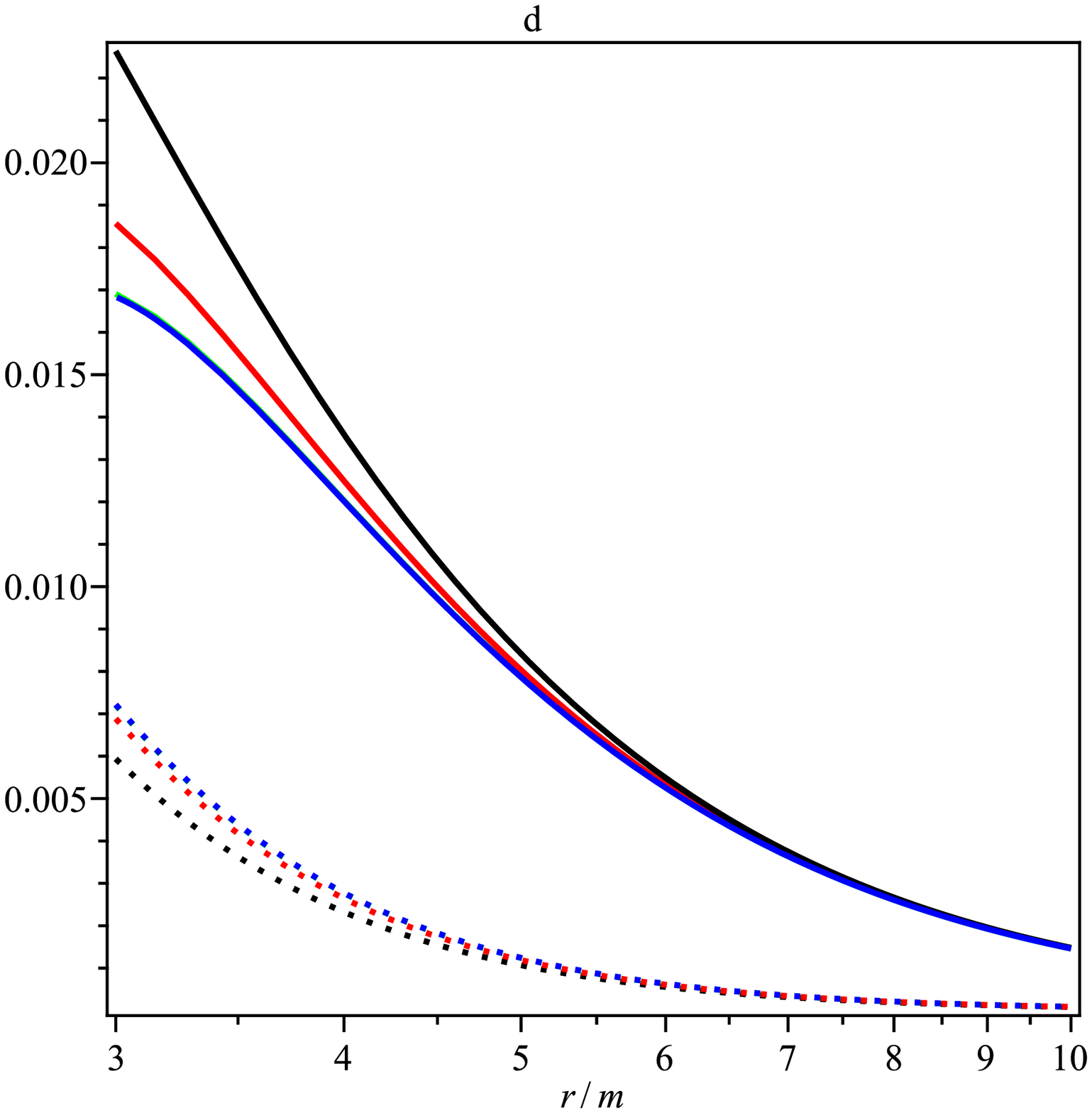}}
\centerline{\includegraphics[scale=.22]{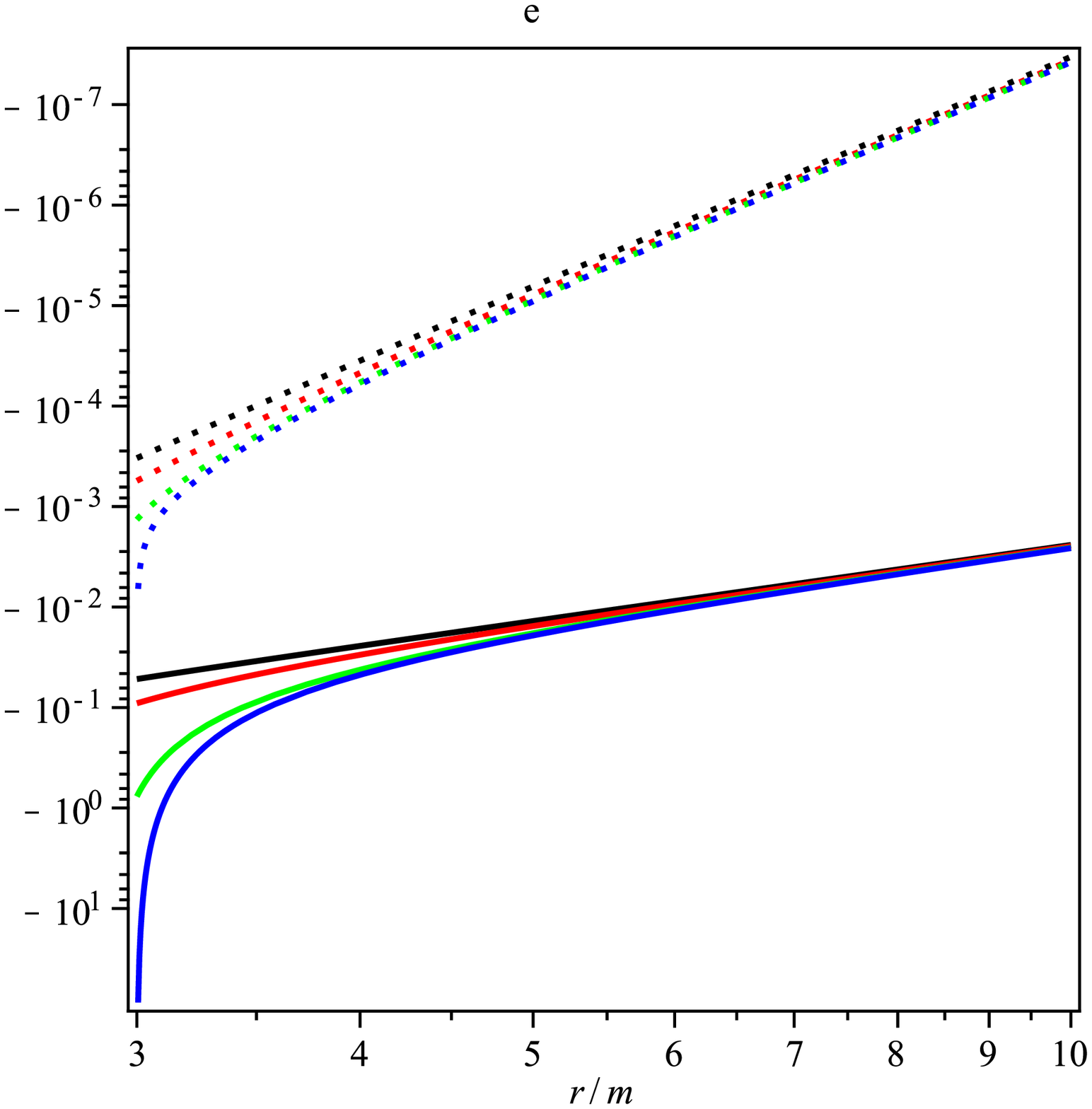} \includegraphics[scale=.22]{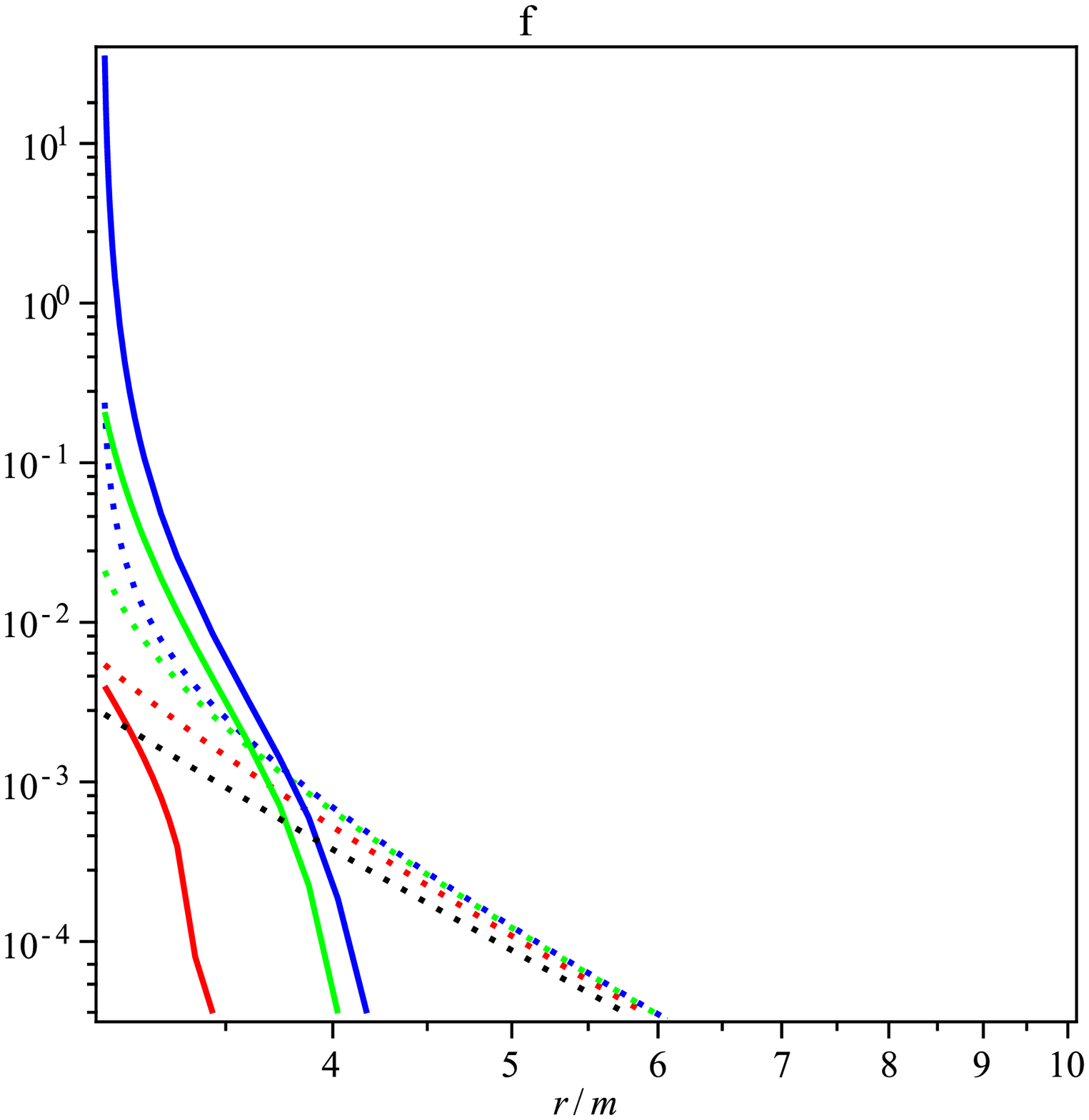}}
\centerline{\includegraphics[scale=.22]{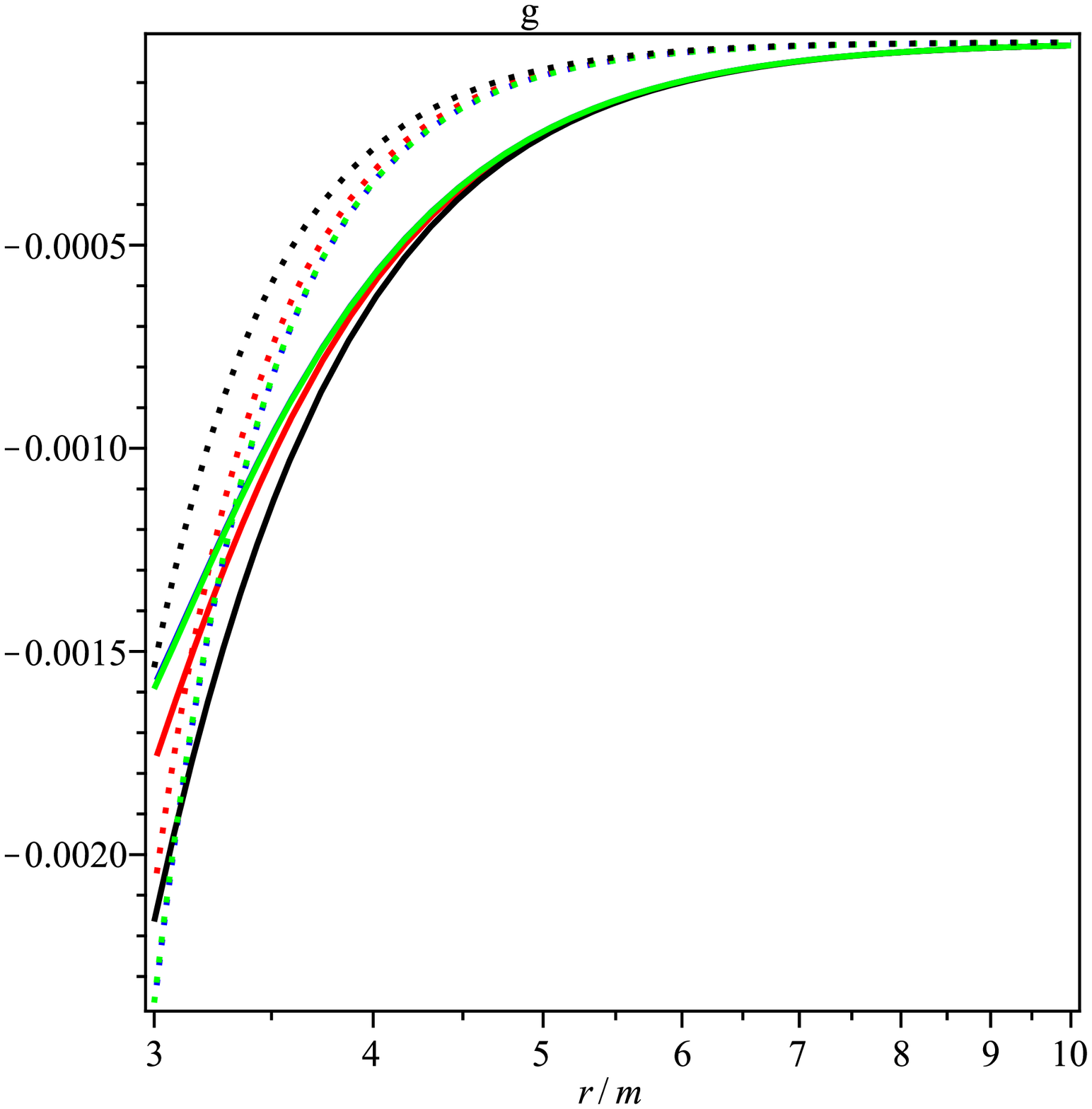}}
 \caption{The non-zero components of the shear and the bulk tensors with $\beta=1$ and $n=2$. $a=.9$ in black, $a=.5$ in red, $a=.1$ in green and $a=0$ in blue. a: Solid is $\sigma^{tt}$ and dotted is $-b^{tt}$, b: Solid is $\sigma^{rt}$ and dotted is $-b^{rt}$, c: Solid is $\sigma^{t\phi}$ and dotted is $-b^{t\phi}$, d: Solid is $\sigma^{rr}$ and dotted is $-b^{rr}$, e: Solid is $\sigma^{r\phi}$ and dotted is $-b^{r\phi}$, f: Solid is $\sigma^{\phi\phi}$ and dotted is $-b^{\phi\phi}$ and g: Solid is $\sigma^{\theta\theta}$ and dotted is $b^{\theta\theta}$.}
\label{figure5}
\end{figure}

In $n=2$ the four velocity is derived by equation (\ref{25}) as
\begin{eqnarray}\label{29}
&&u^{\mu}=(\frac{\sqrt{r^{5}+r\beta^{2}}(r^{\frac{3}{2}}+a)}{r^{2}\sqrt{r^{4}+2ar^{\frac{5}{2}}-3r^{3}}},\nonumber\\&& -\frac{\beta\sqrt{r^{2}-2r+a^{2}}}{r^{3}},0,\frac{\sqrt{r^{5}+r\beta^{2}}}{r^{2}\sqrt{r^{4}+2ar^{\frac{5}{2}}-3r^{3}}}).
\end{eqnarray}
In this state $\Theta=\frac{\beta(-r+a^{2})}{r^{4}\sqrt{r^{2}-2r+a^{2}}}$ and figure \ref{figure5} shows the components of shear and bulk tensors.
\subsection{$n=\frac{5}{2}$}
In this case the four velocity from equation (\ref{25}) calculates as:
\begin{eqnarray}\label{30}
&&u^{\mu}=(\frac{\sqrt{r^{6}+r\beta^{2}}(r^{\frac{3}{2}}+a)}{r^{\frac{5}{2}}\sqrt{r^{4}+2ar^{\frac{5}{2}}-3r^{3}}},\nonumber\\&& -\frac{\beta\sqrt{r^{2}-2r+a^{2}}}{r^{\frac{7}{2}}},0,\frac{\sqrt{r^{6}+r\beta^{2}}}{r^{\frac{5}{2}}\sqrt{r^{4}+2ar^{\frac{5}{2}}-3r^{3}}}).
\end{eqnarray}
 In this state $\Theta=\frac{\beta(-4r+r^{2}+3a^{2})}{2r^\frac{9}{2}\sqrt{r^{2}-2r+a^{2}}}$ and non-zero components of the shear and bulk tensors are shown in figure \ref{figure6}:
\begin{figure}
\vspace{\fill}
\centerline{\includegraphics[scale=.22]{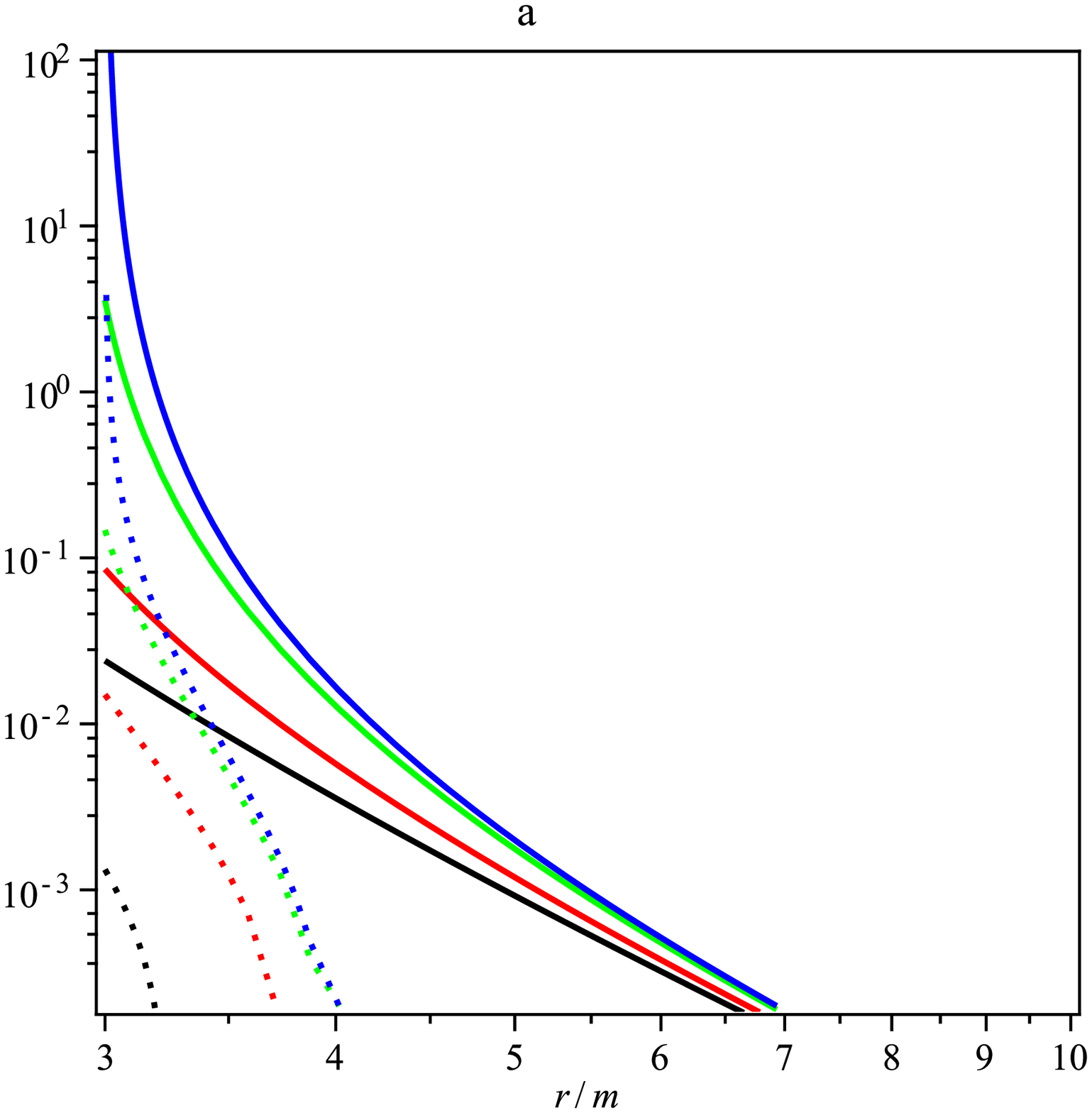}\includegraphics[scale=.22]{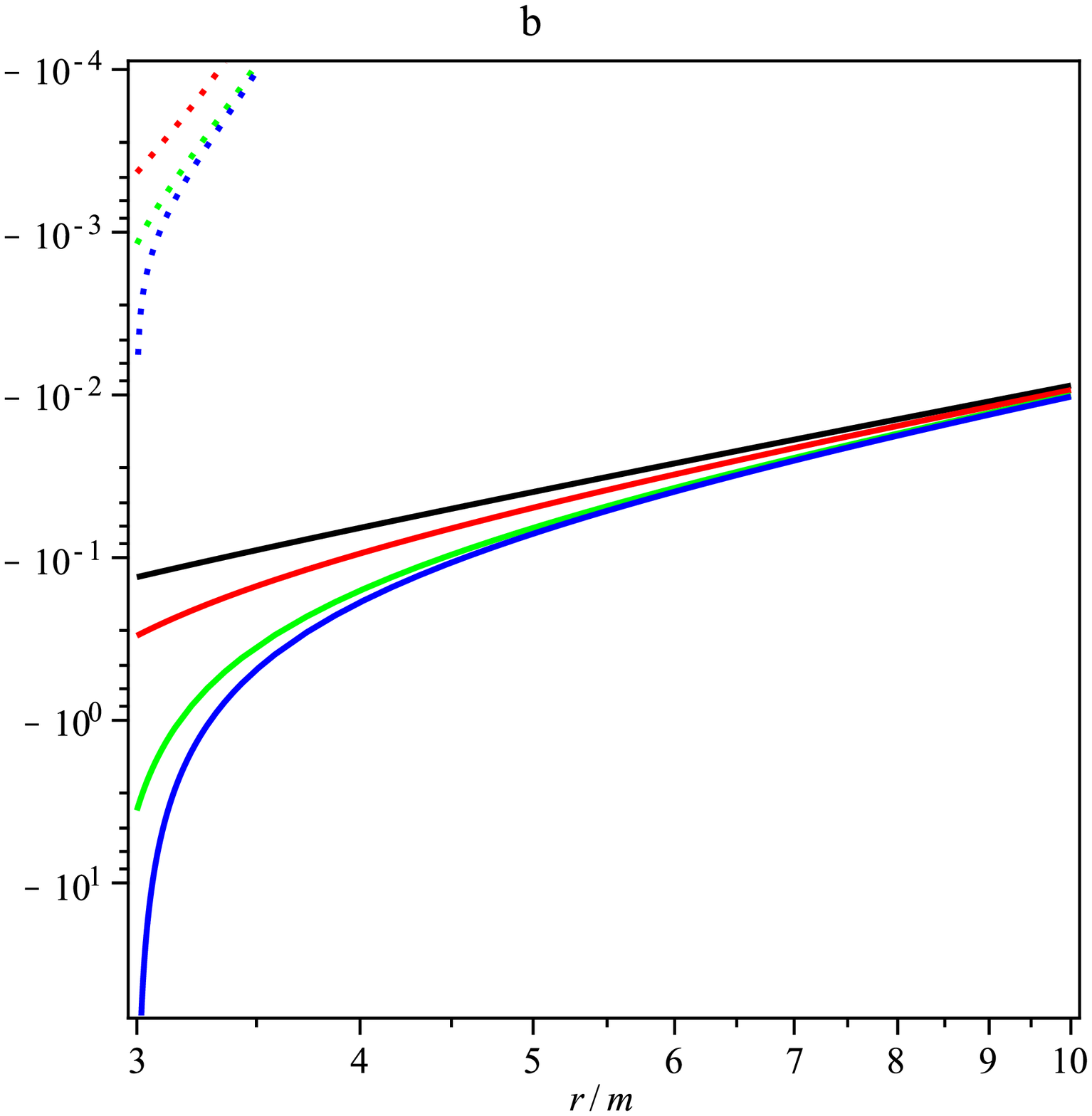}}
\centerline{\includegraphics[scale=.22]{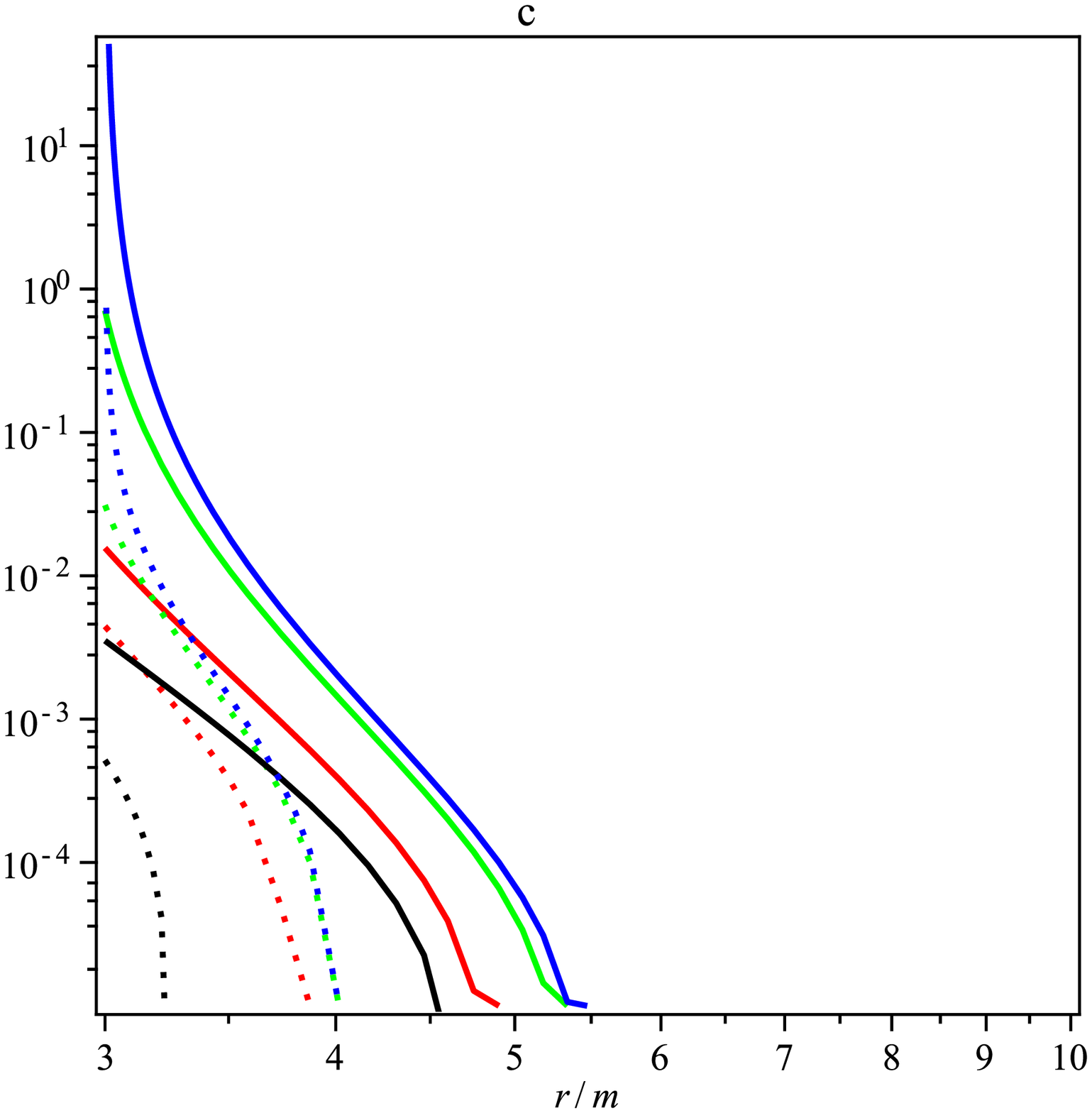}\includegraphics[scale=.22]{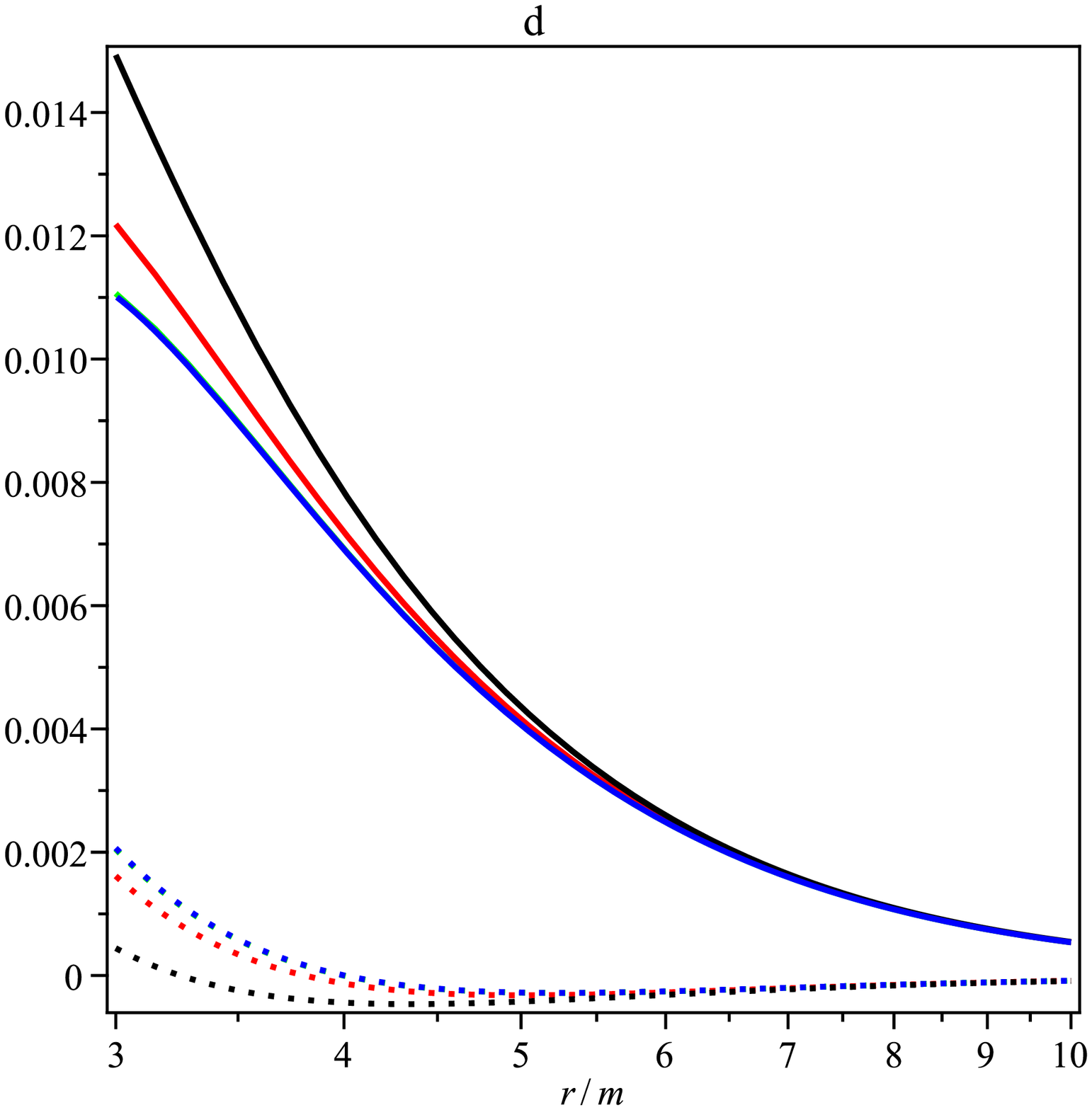}}
\centerline{\includegraphics[scale=.22]{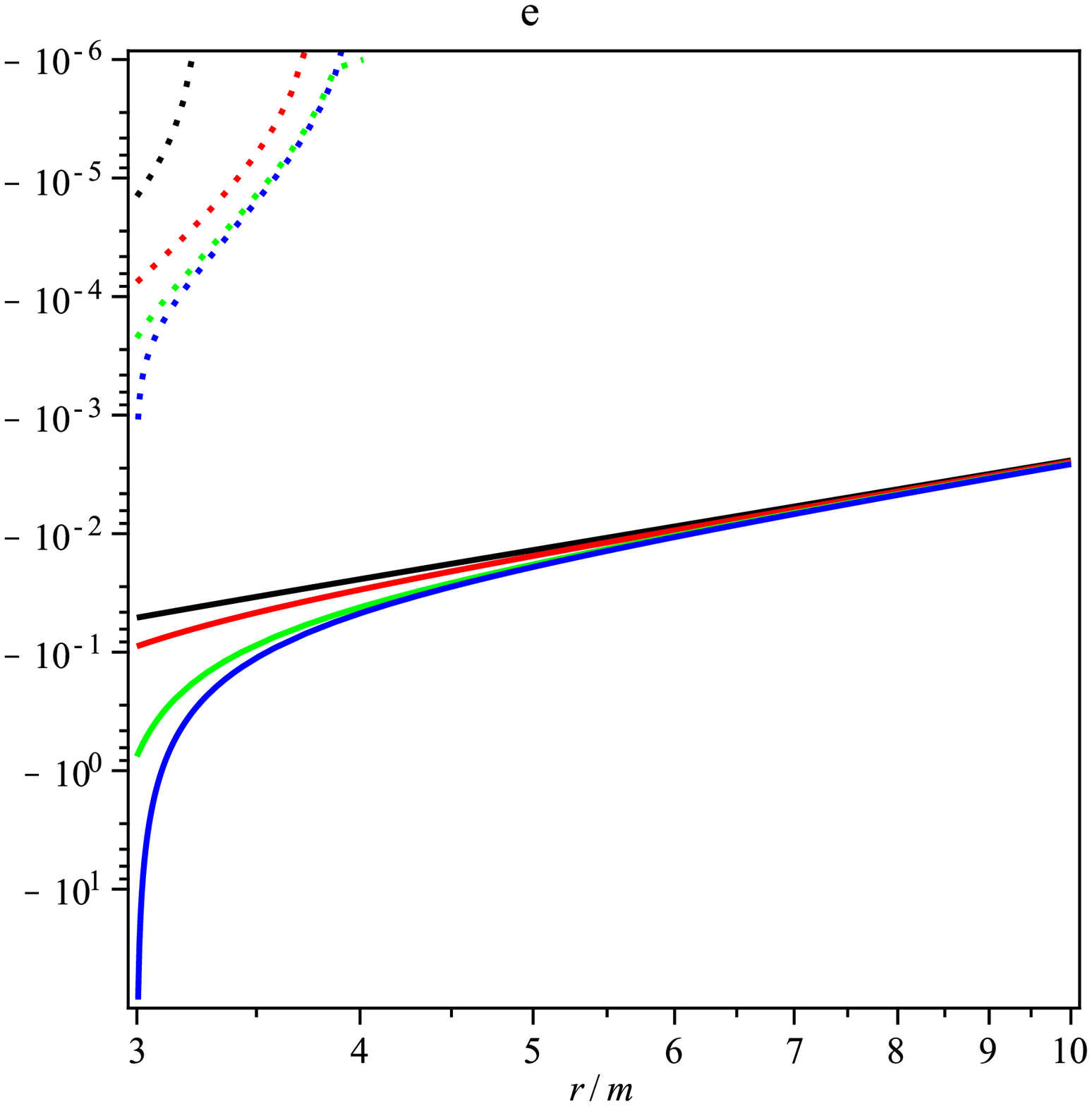} \includegraphics[scale=.22]{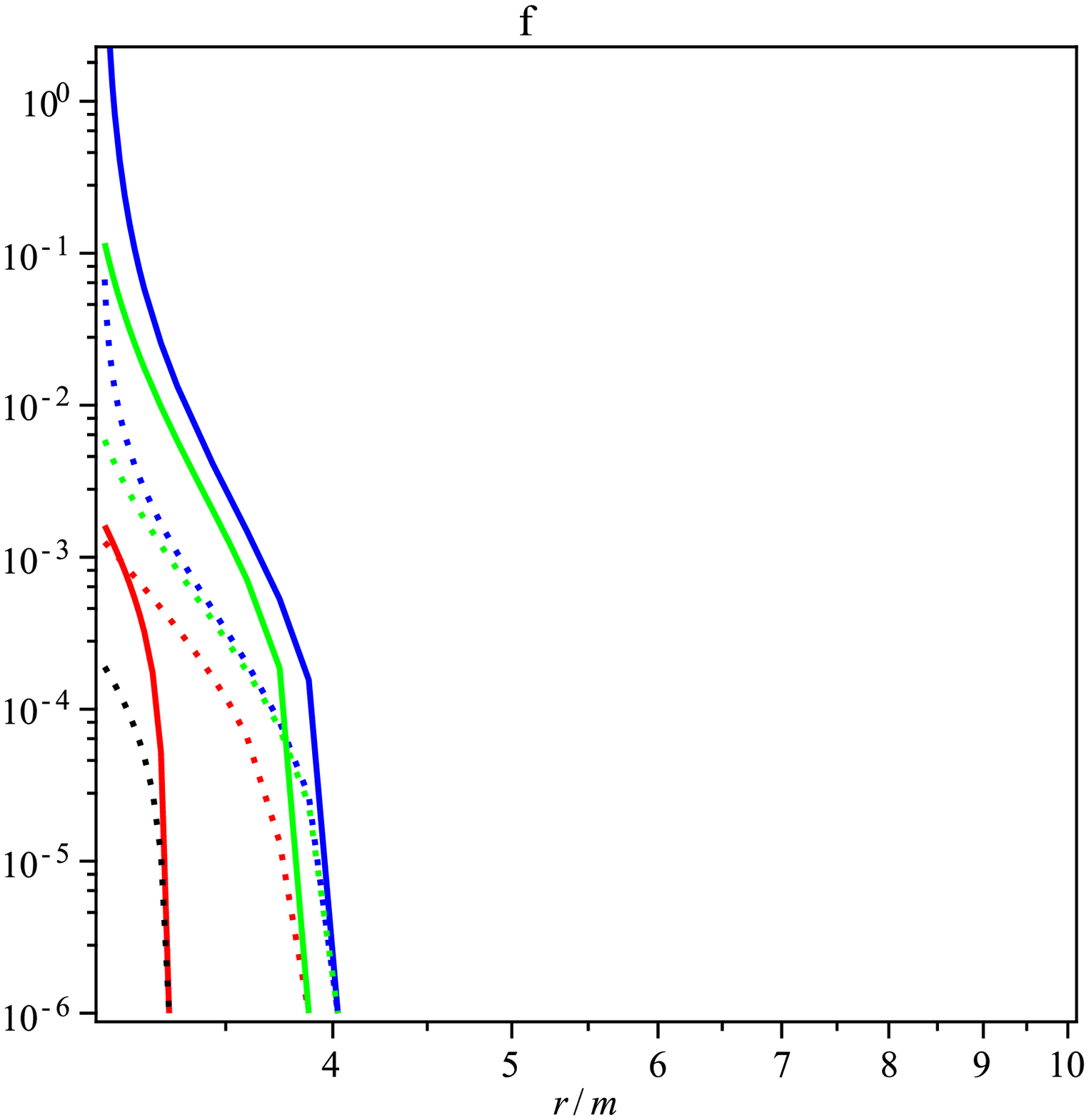}}
\centerline{\includegraphics[scale=.22]{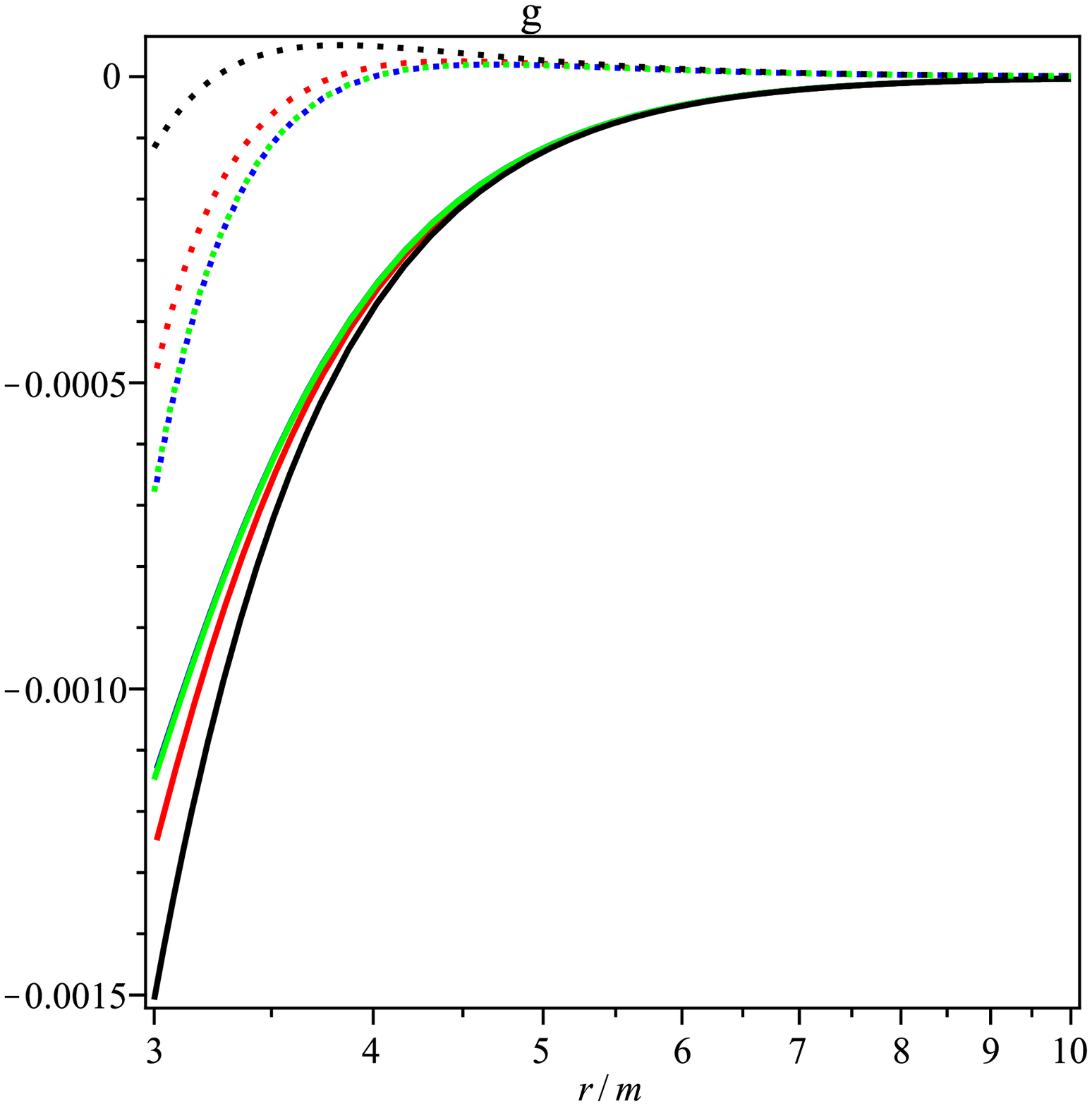}}
 \caption{The non-zero components of the shear and the bulk tensors with $\beta=1$ and $n=\frac{5}{2}$. $a=.9$ in black, $a=.5$ in red, $a=.1$ in green and $a=0$ in blue. a: Solid is $\sigma^{tt}$ and dotted is $-b^{tt}$, b: Solid is $\sigma^{rt}$ and dotted is $-b^{rt}$, c: Solid is $\sigma^{t\phi}$ and dotted is $-b^{t\phi}$, d: Solid is $\sigma^{rr}$ and dotted is $-b^{rr}$, e: Solid is $\sigma^{r\phi}$ and dotted is $-b^{r\phi}$, f: Solid is $\sigma^{\phi\phi}$ and dotted is $-b^{\phi\phi}$ and g: Solid is $\sigma^{\theta\theta}$ and dotted is $b^{\theta\theta}$.}
\label{figure6}
\end{figure}
\subsection{$n=3$}
\begin{figure}
\vspace{\fill}
\centerline{\includegraphics[scale=.22]{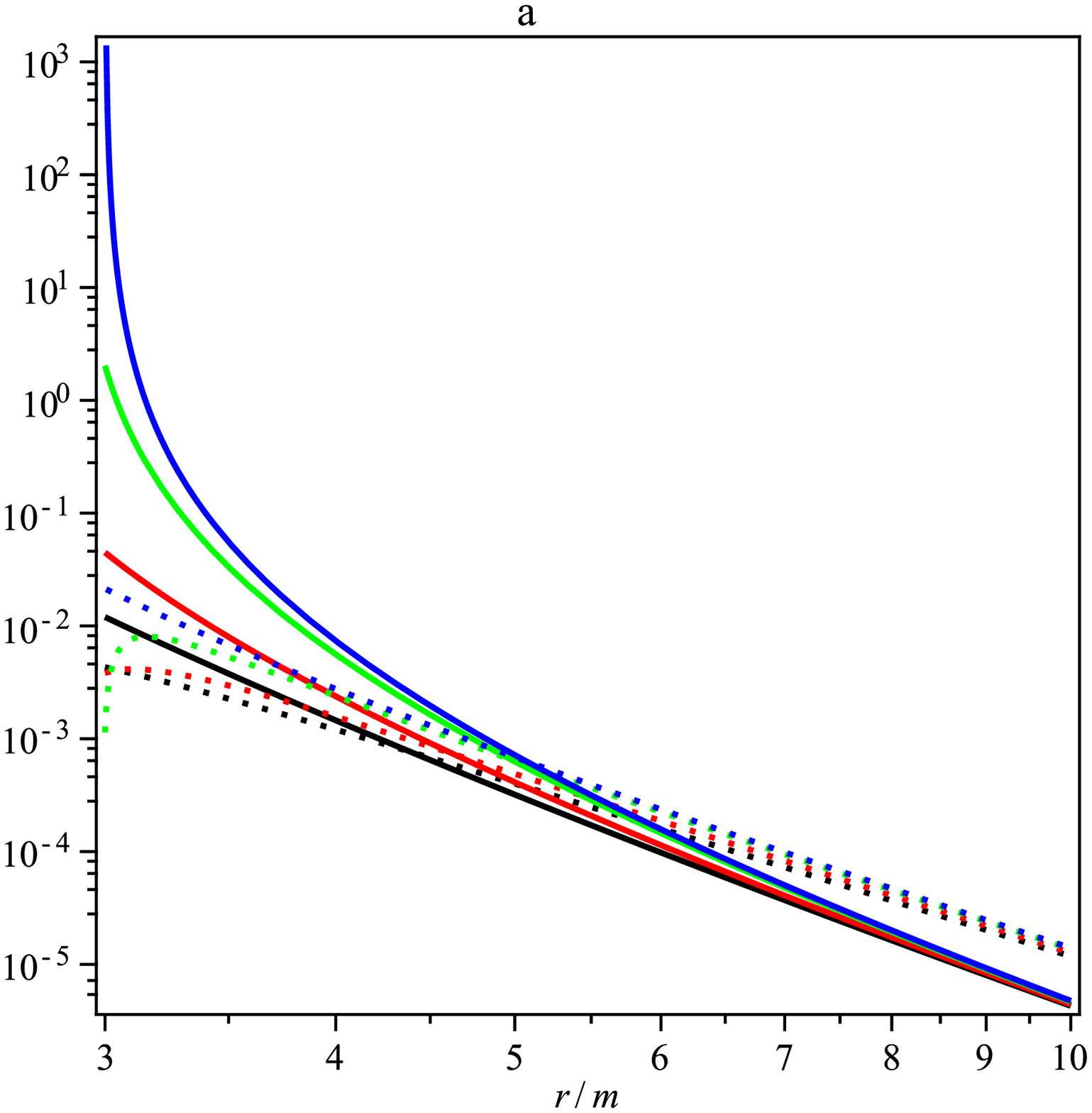}\includegraphics[scale=.22]{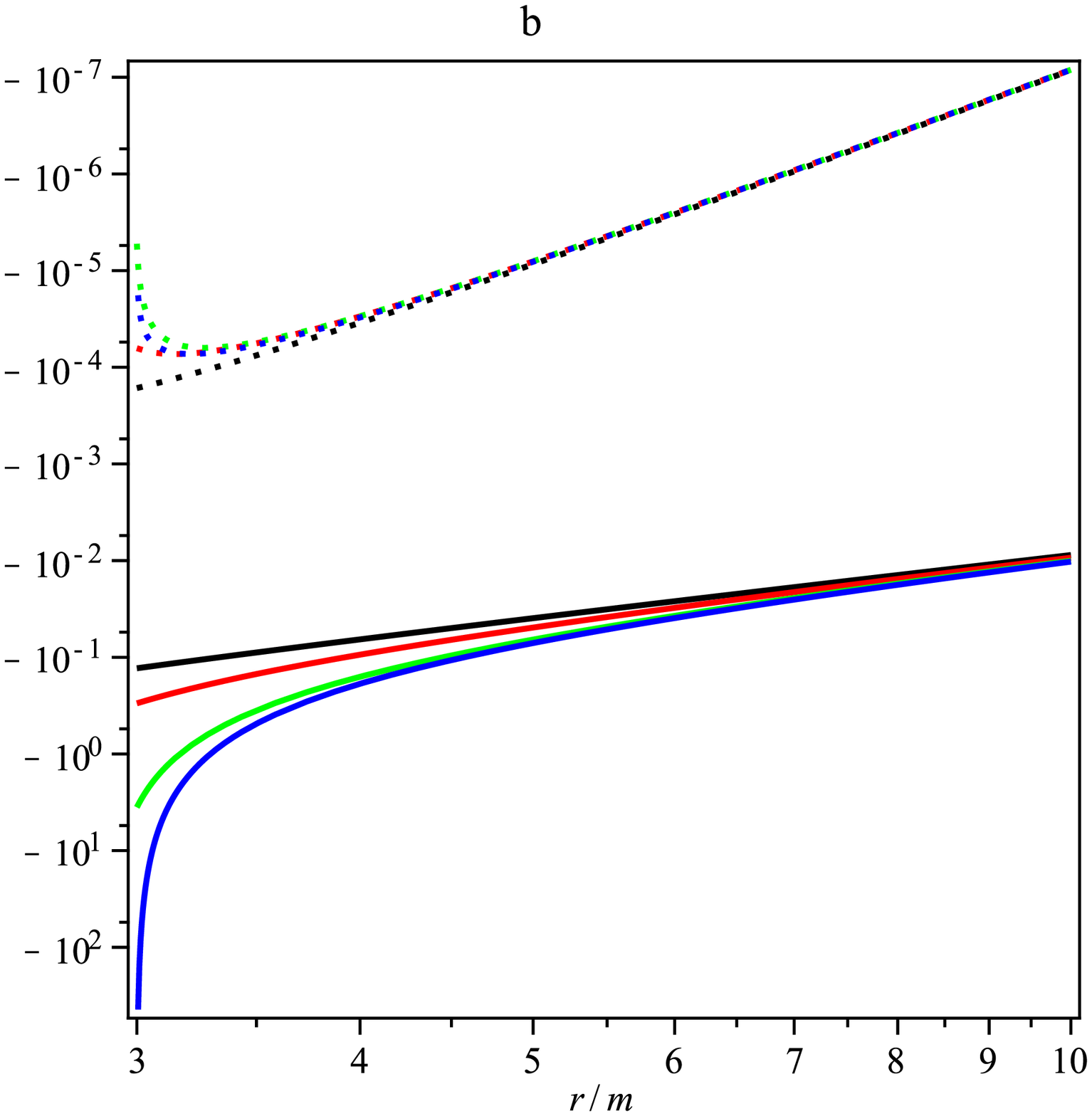}}
\centerline{\includegraphics[scale=.22]{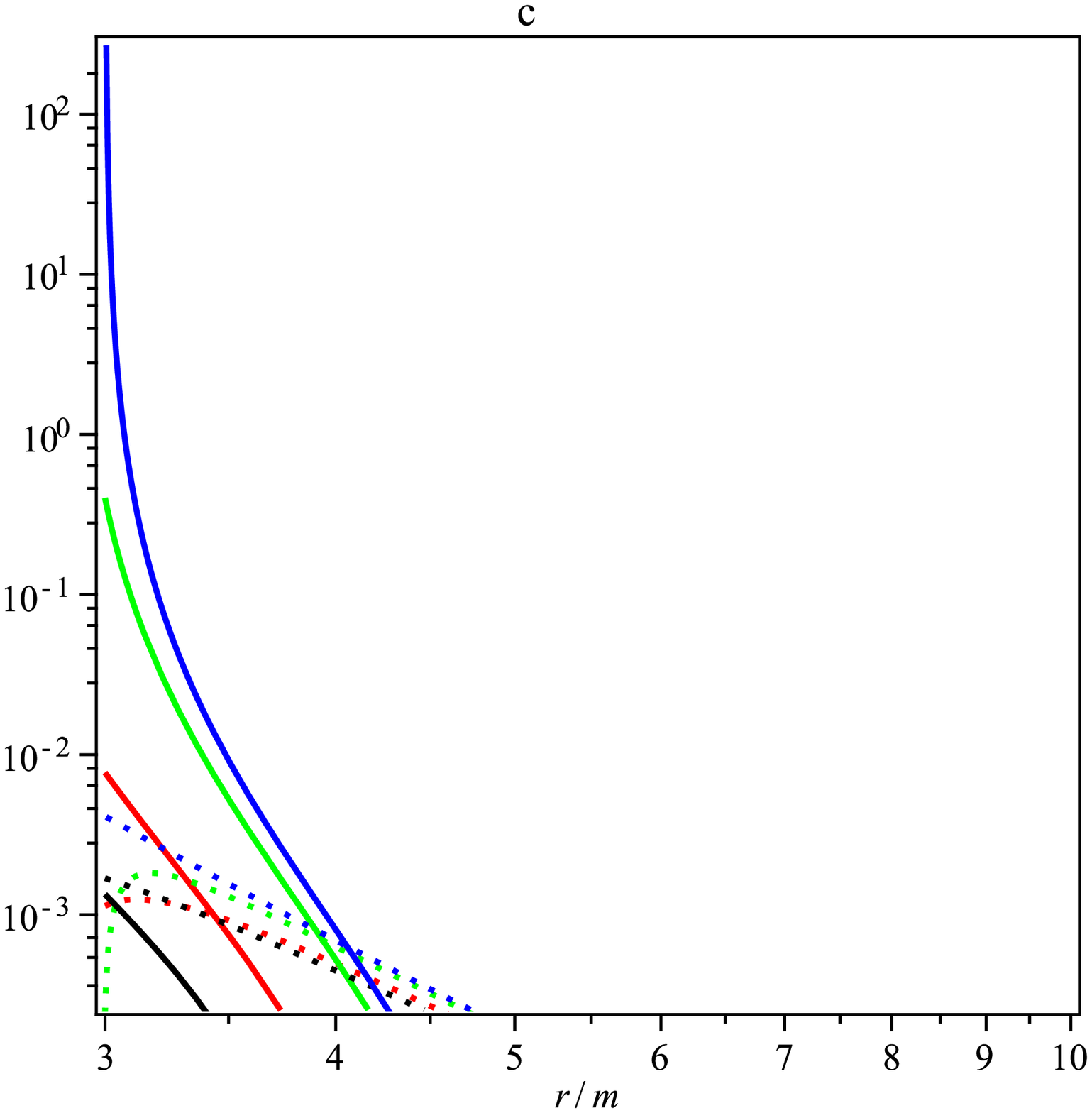}\includegraphics[scale=.22]{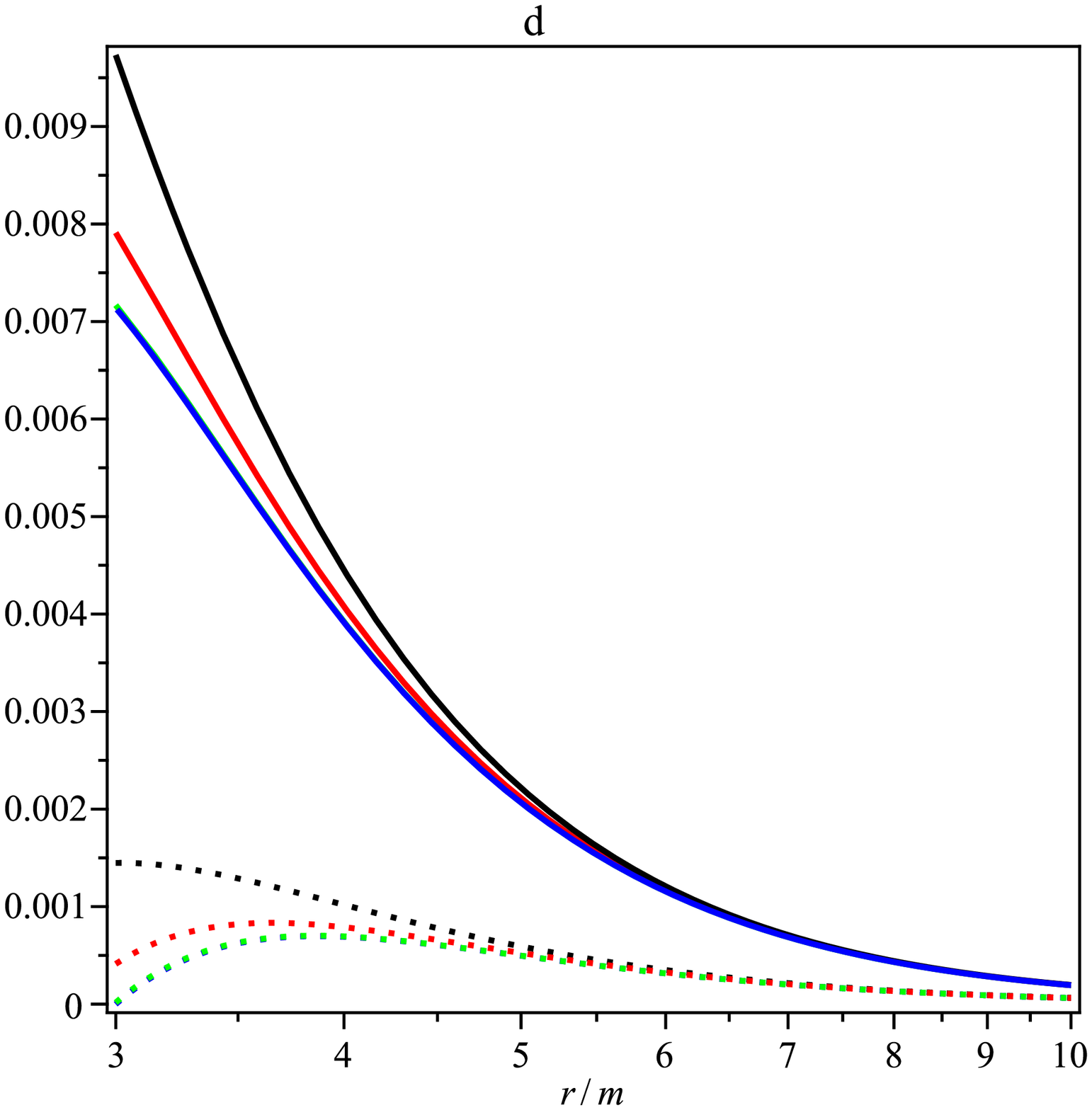}}
\centerline{\includegraphics[scale=.22]{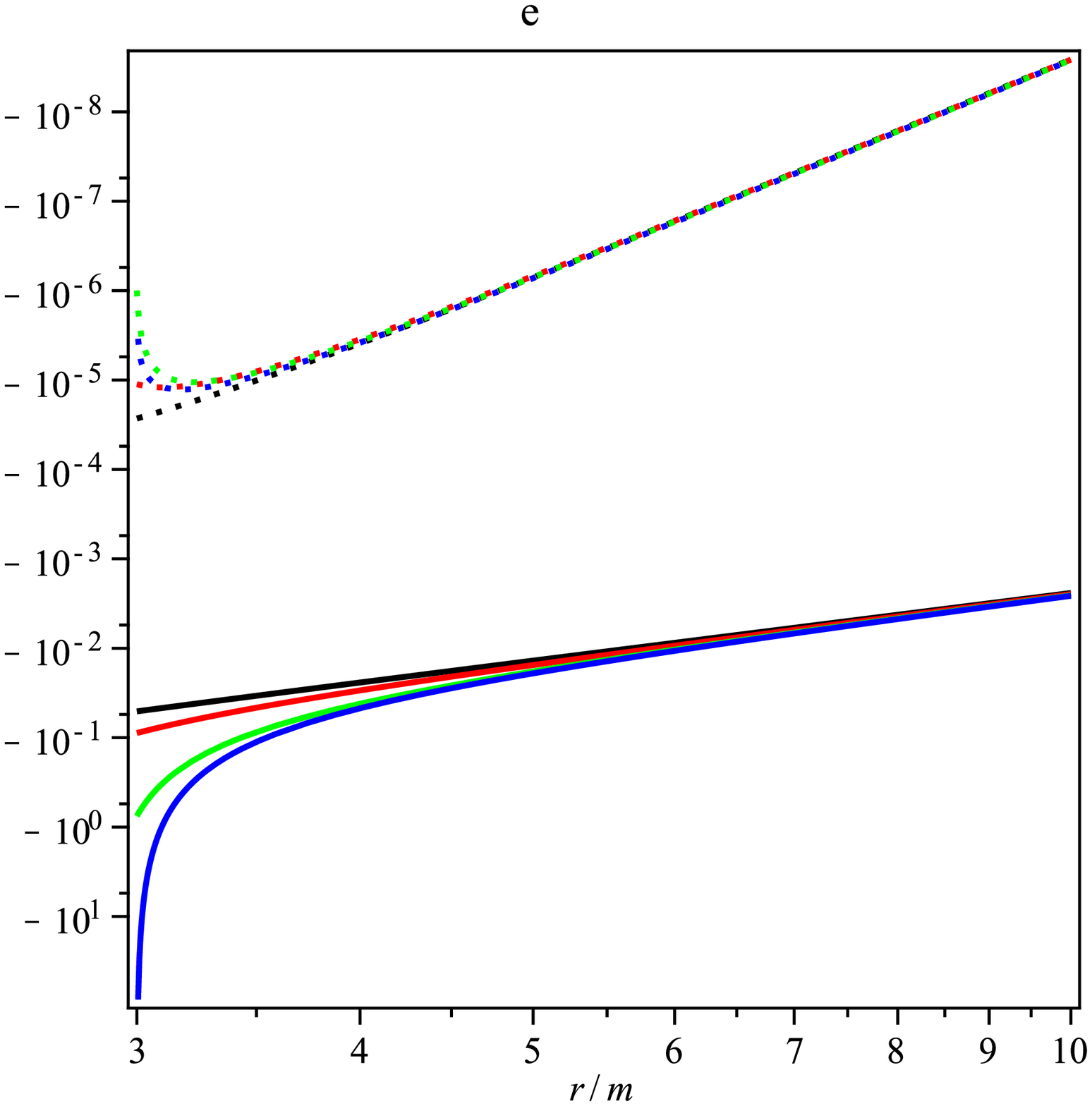} \includegraphics[scale=.22]{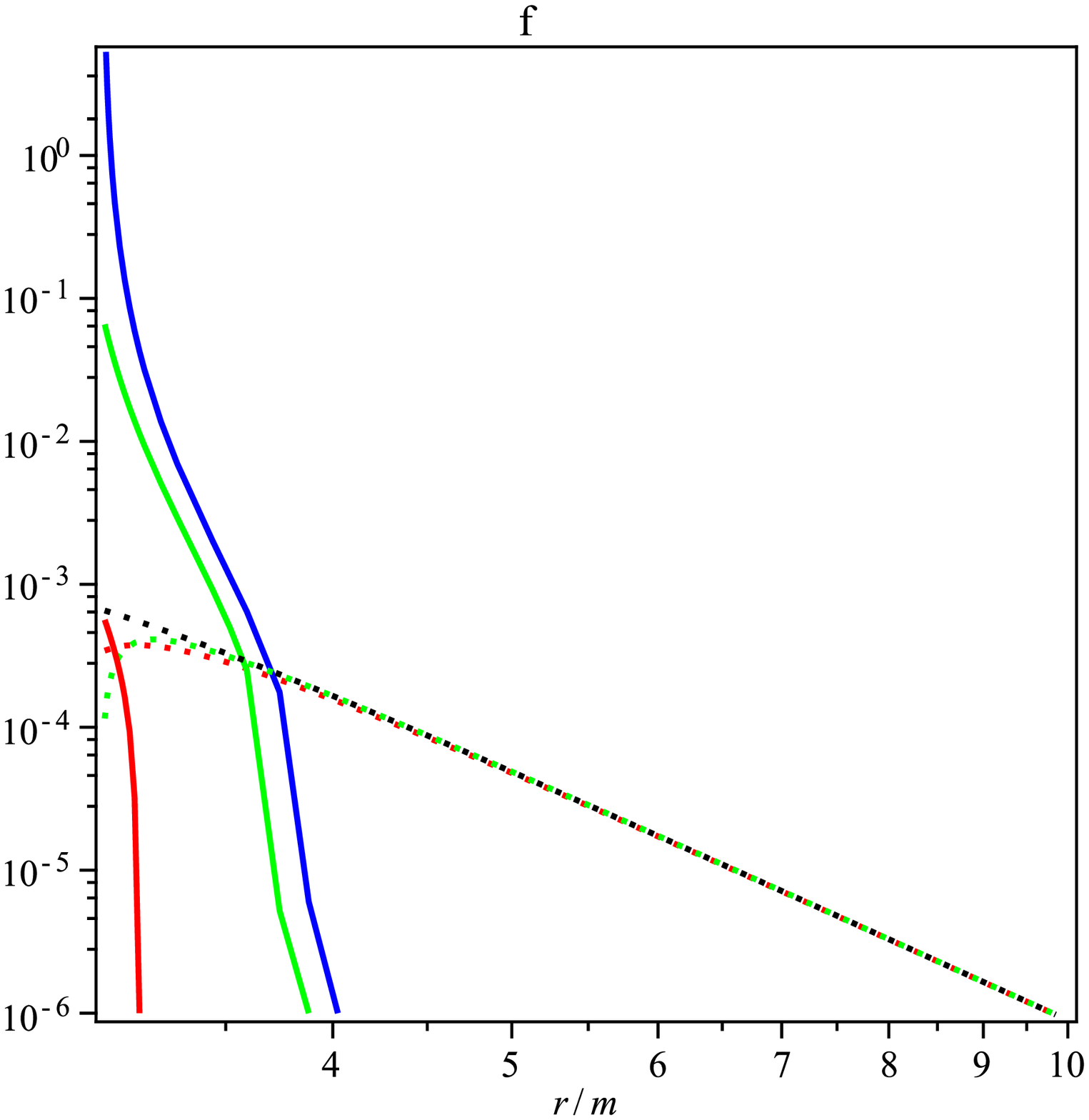}}
\centerline{\includegraphics[scale=.22]{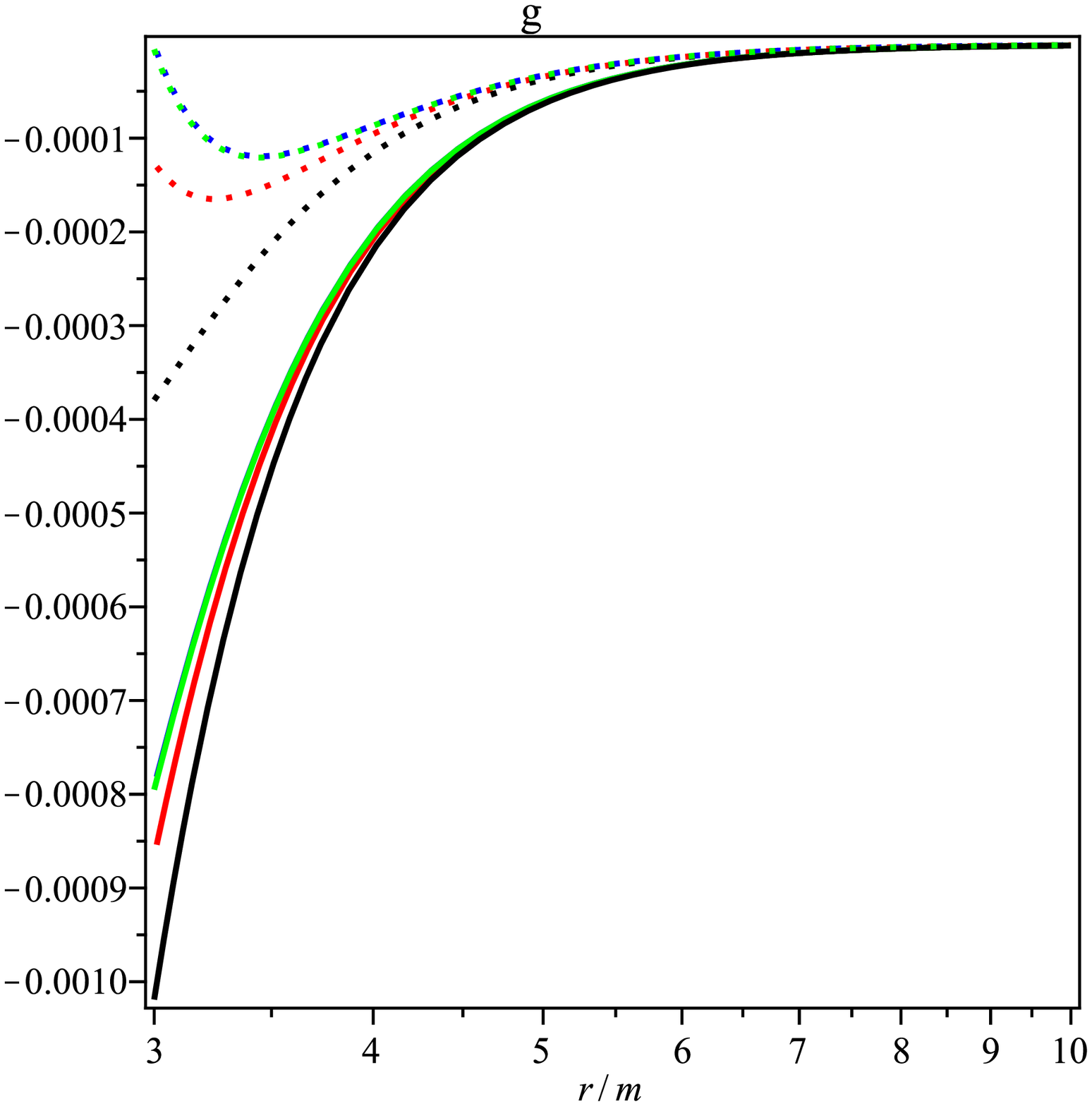}}
 \caption{The non-zero components of the shear and the bulk tensors with $\beta=1$ and $n=3$. $a=.9$ in black, $a=.5$ in red, $a=.1$ in green and $a=0$ in blue. a:Solid is $\sigma^{tt}$ and dotted is $b^{tt}$, b: Solid is $\sigma^{rt}$ and dotted is $b^{rt}$, c: Solid is $\sigma^{t\phi}$ and dotted is $b^{t\phi}$, d: Solid is $\sigma^{rr}$ and dotted is $b^{rr}$, e: Solid is $\sigma^{r\phi}$ and dotted is $b^{r\phi}$, f: Solid is $\sigma^{\phi\phi}$ and dotted is $b^{\phi\phi}$ and g: Solid is $\sigma^{\theta\theta}$ and dotted is $-b^{\theta\theta}$.}
\label{figure7}
\end{figure}

In this state four velocity is
\begin{eqnarray}\label{31} 
&&u^{\mu}=(\frac{\sqrt{r^{7}+r\beta^{2}}(r^{\frac{3}{2}}+a)}{r^{3}\sqrt{r^{4}+2ar^{\frac{5}{2}}-3r^{3}}},\nonumber\\&& -\frac{\beta\sqrt{r^{2}-2r+a^{2}}}{r^{4}},0,\frac{\sqrt{r^{7}+r\beta^{2}}}{r^{3}\sqrt{r^{4}+2ar^{\frac{5}{2}}-3r^{3}}}).
\end{eqnarray}
In this stae $\Theta=\frac{\beta(-3r+r^{2}+2a^{2})}{r^{5}\sqrt{r^{2}-2r+a^{2}}}$ and figure \ref{figure7} shows the components of the shear and bulk tensors.

Figure \ref{figure1} shows that the expansion of fluid world line ($\Theta$) in $n=\frac{1}{2}, 1, \frac{3}{2}, 2, \frac{5}{2},3$ and in $a=0.9, 0.5, 0.1,0$. In this figure we see that increase of $n$ and decrease of $a$ are caused decreasing of the expansion of fluid world line.  
In $n=\frac{1}{2}$ in $a=0.9, 0.5, 0.1, 0$, except $r\phi $ component the amounts of the other components of the bulk tensor are greater than  or about the amounts of components of the shear tensor. By increasing $n$, the bulk tensor and also the shear tensor are been smaller. So we see that in $n=1, \frac{3}{2}$ in $a=0.9, 0.5, 0.1$ and $0$, $tt$, $t\phi$, $\phi\phi$, $\theta\theta$ components, the amount of the bulk tensor are greater or about of the shear tensor. Also in the bigger $n$($n=2, \frac{5}{2}, 3$) in $a=0.9, 0.5, 0.1,0$ the amounts of components of the shear tensor and bulk tensor are near zero and the shear tensor is greater than the bulk tensor in the most components.
 \section{shear stress viscosity}
 \label{sec:7}
\begin{figure}
\vspace{\fill}
\centerline{\includegraphics[scale=.22]{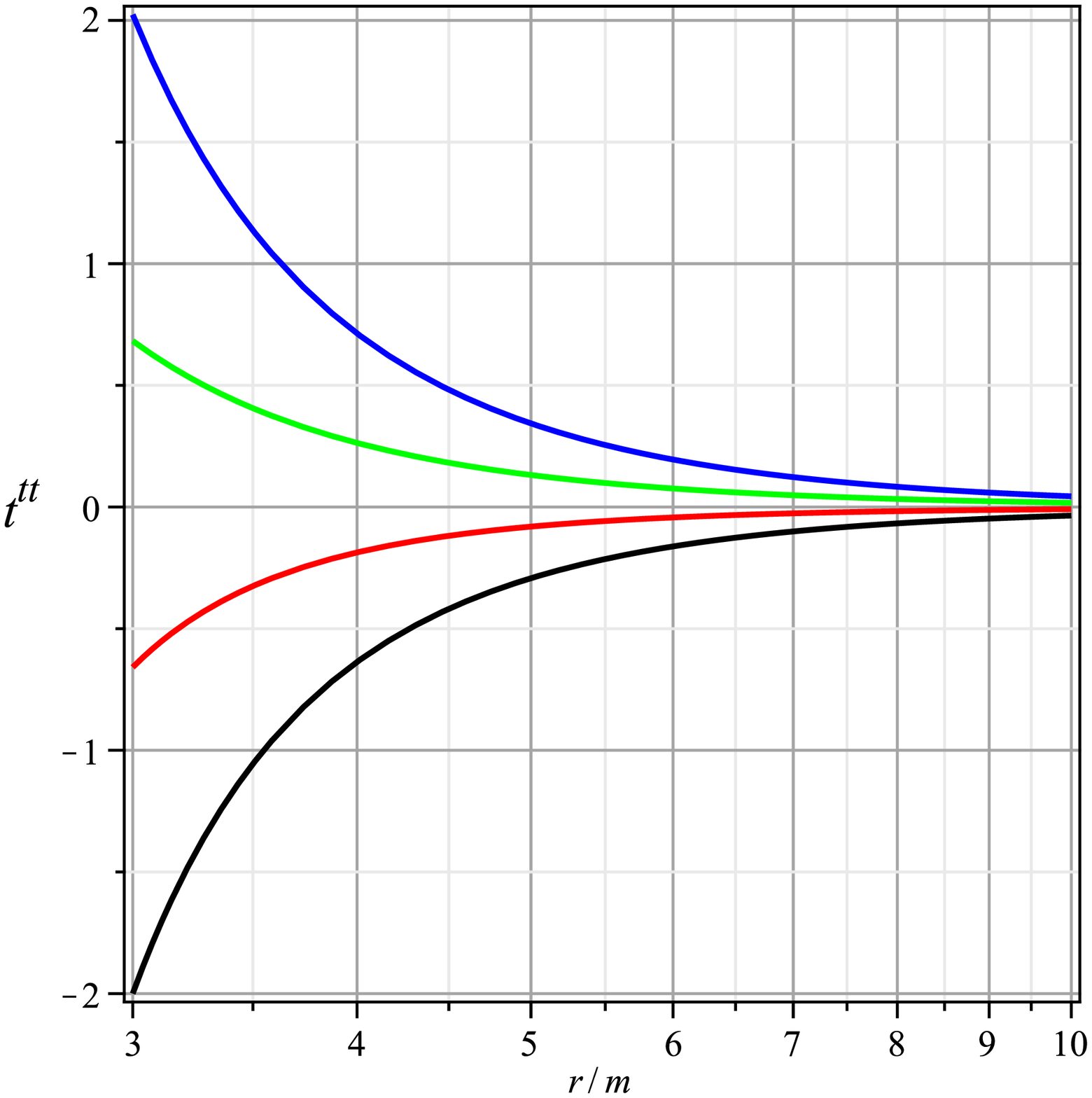}\includegraphics[scale=.22]{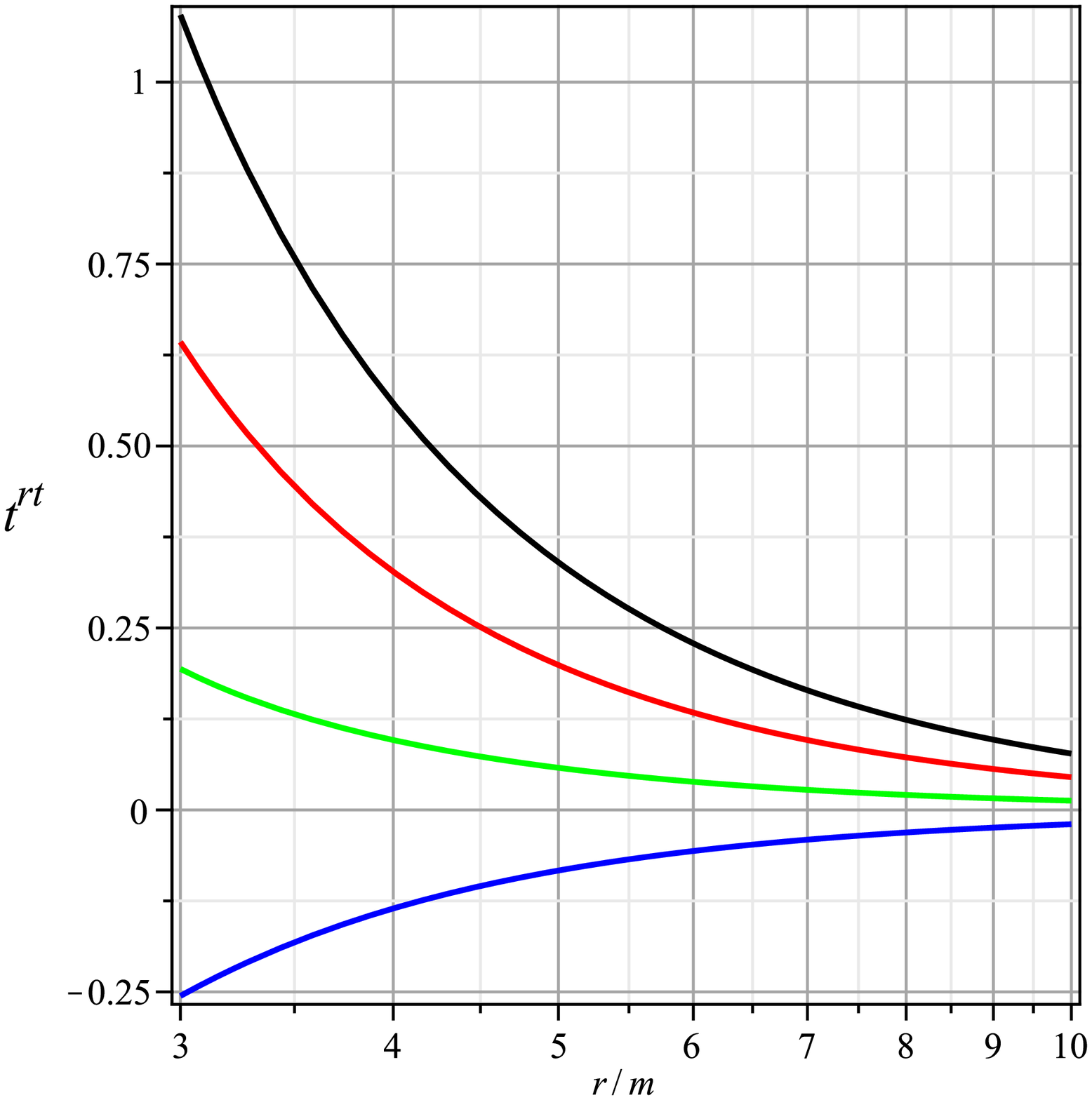}}
\centerline{\includegraphics[scale=.22]{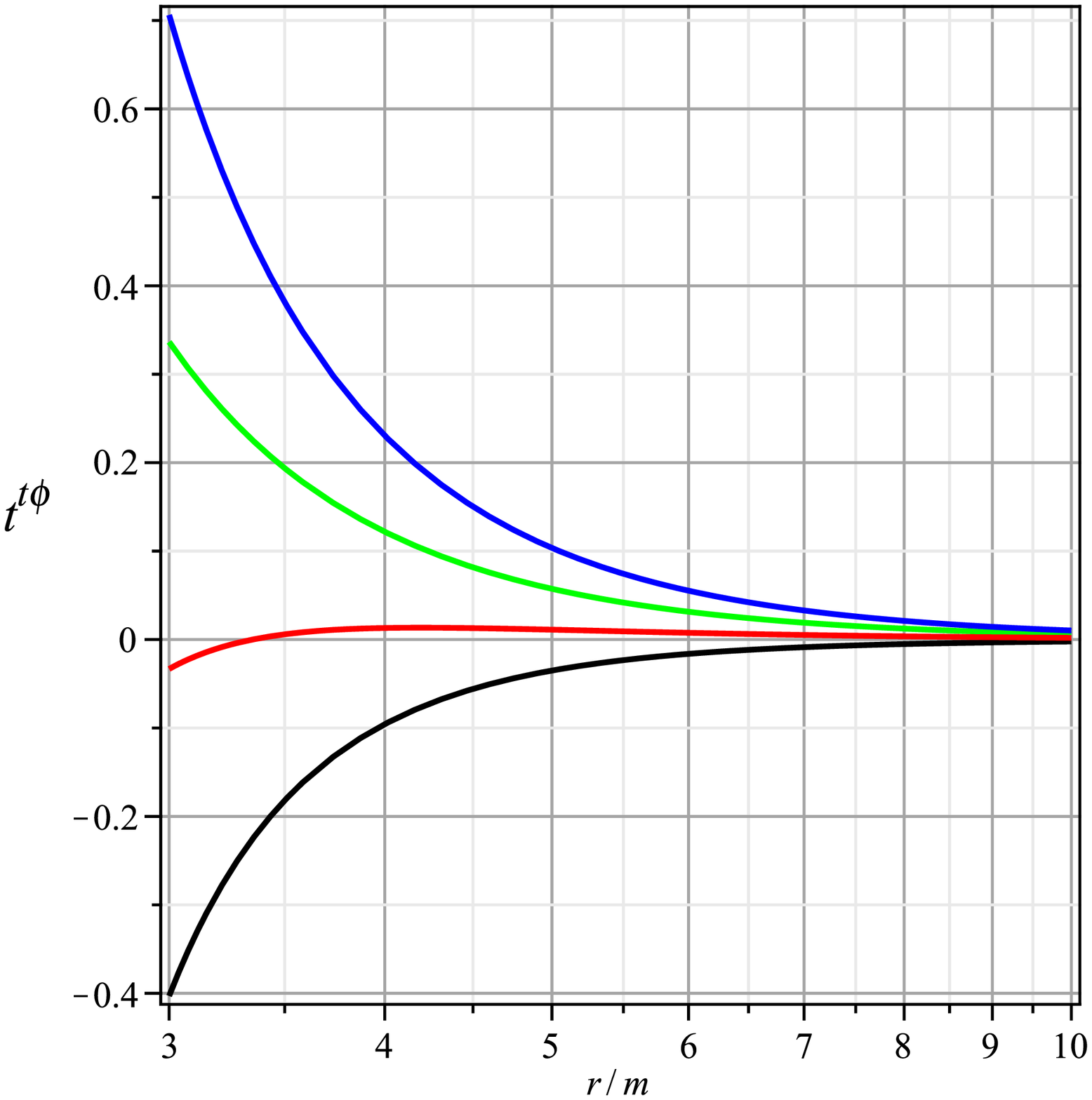}\includegraphics[scale=.22]{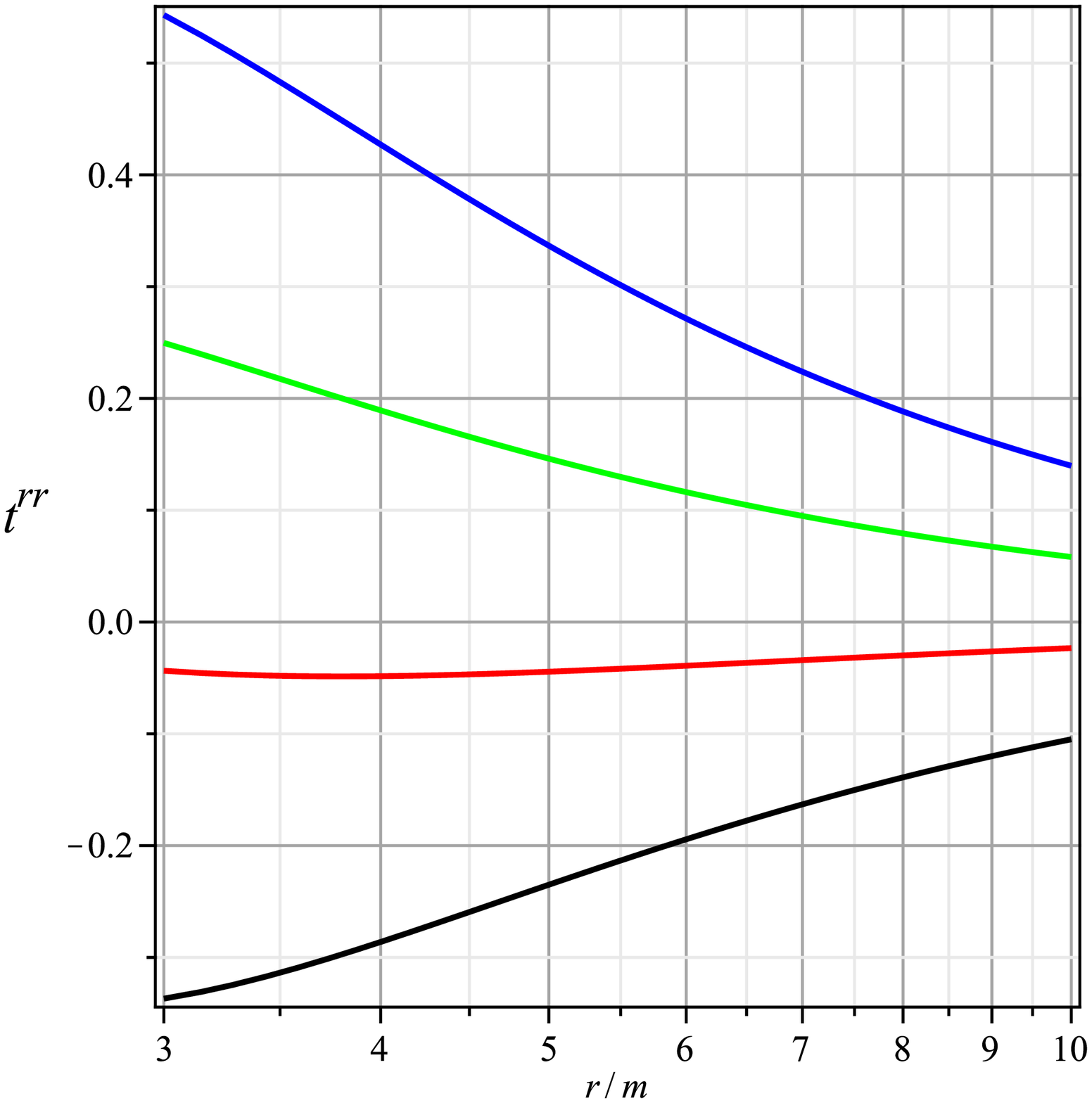}}
\centerline{\includegraphics[scale=.22]{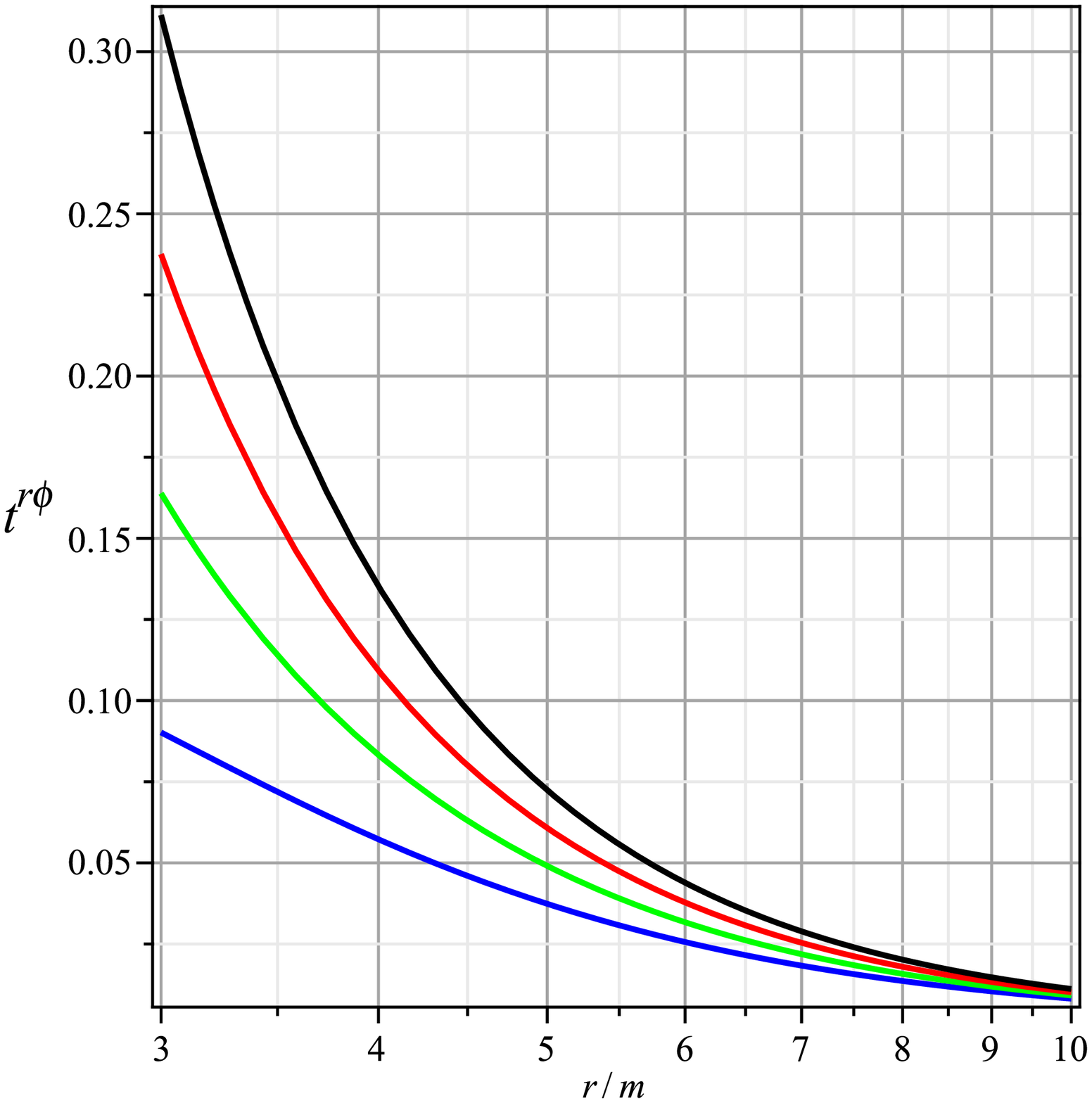} \includegraphics[scale=.22]{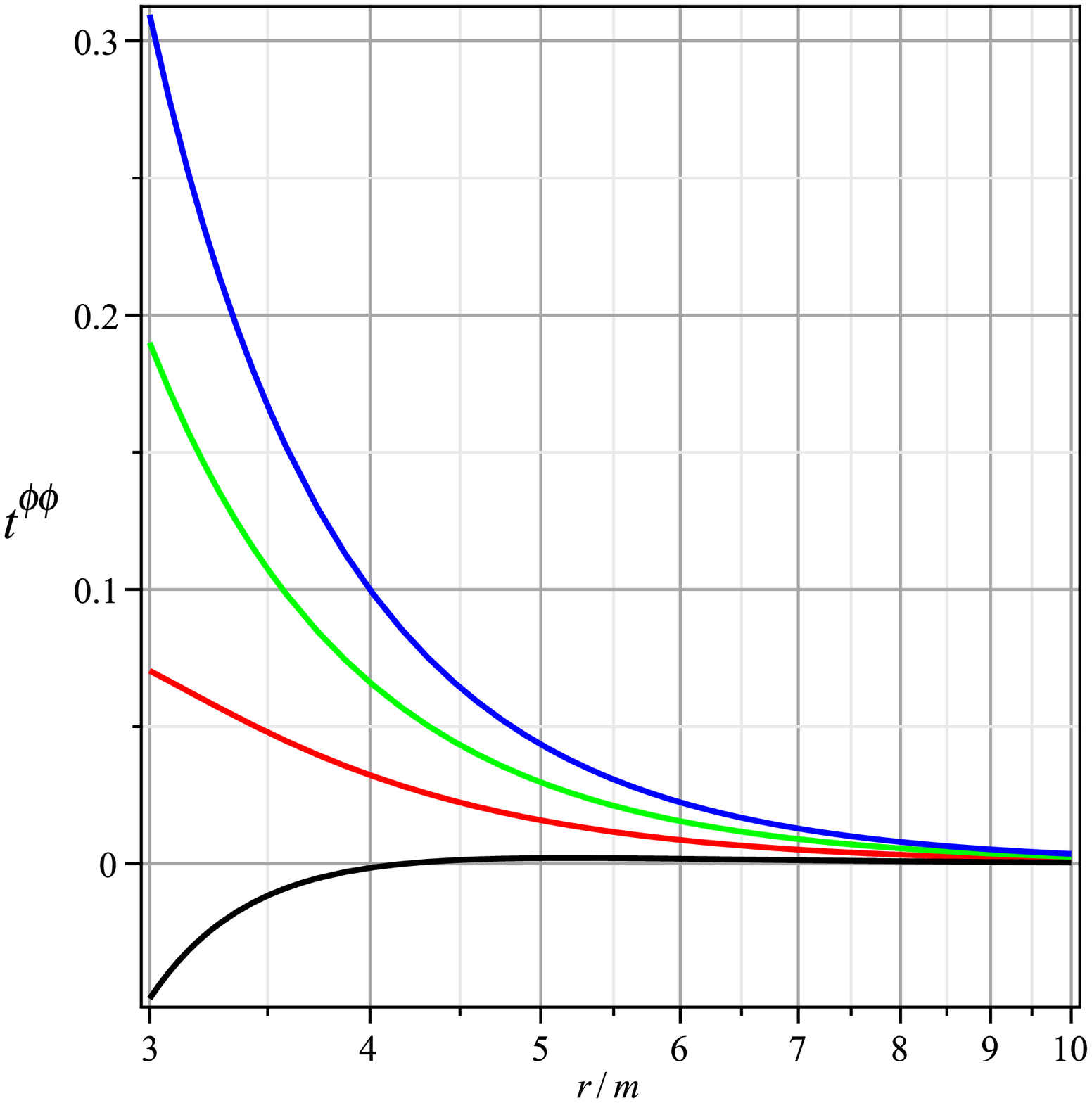}}
\centerline{\includegraphics[scale=.22]{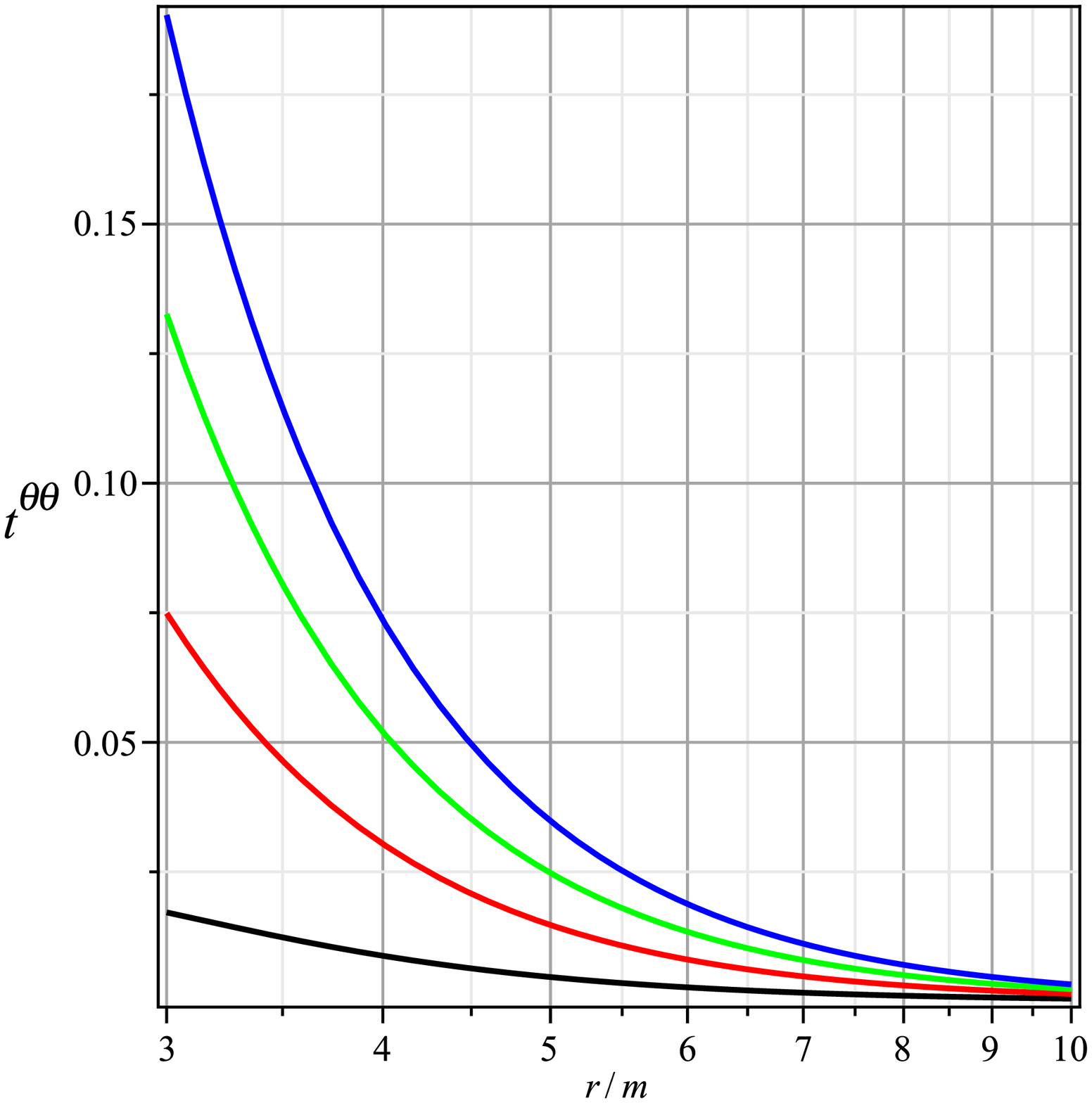}}
 \caption{Influence of the bulk viscosity in the non-zero components of the shear stress viscosity with $\beta=1$,  $n=\frac{1}{2}$, $a=.9$ and $\lambda=2$. With no bulk viscosity $\zeta=0$ in black, $\zeta=2$ in red, $\zeta=4$ in green and $\zeta=6$ in blue.}
\label{figure8}
\end{figure}

\begin{figure}
\vspace{\fill}
\centerline{\includegraphics[scale=.22]{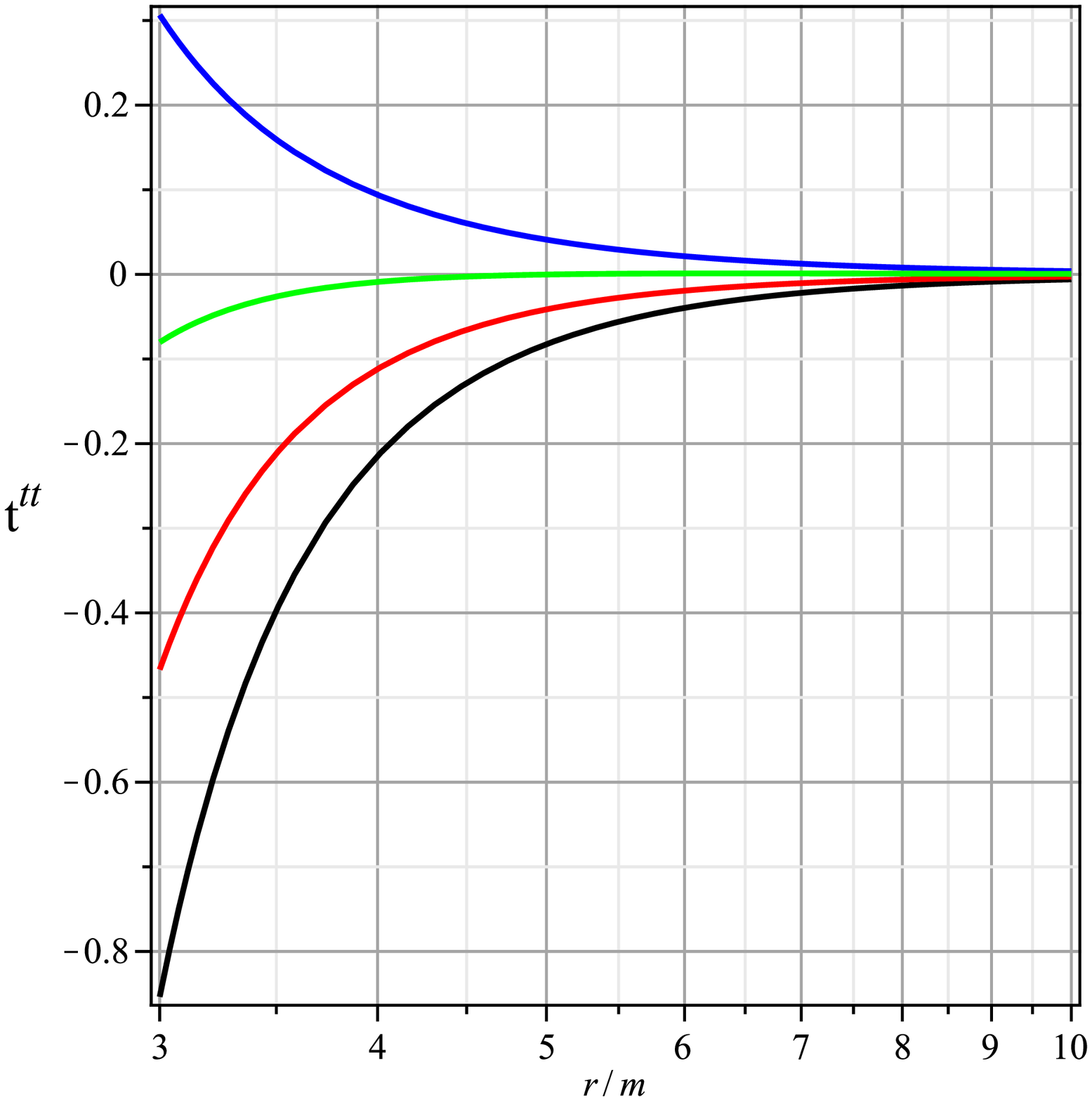}\includegraphics[scale=.22]{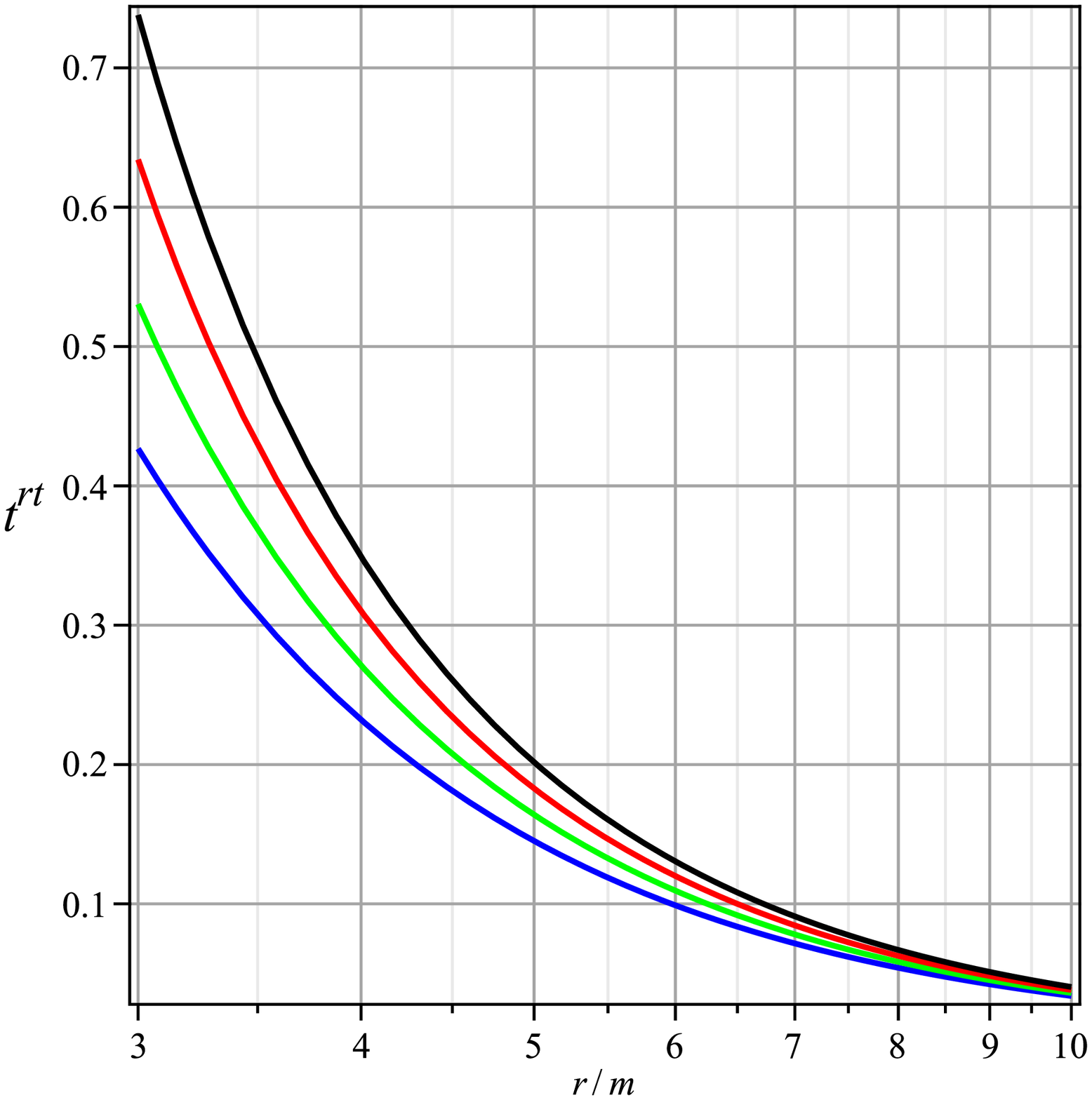}}
\centerline{\includegraphics[scale=.22]{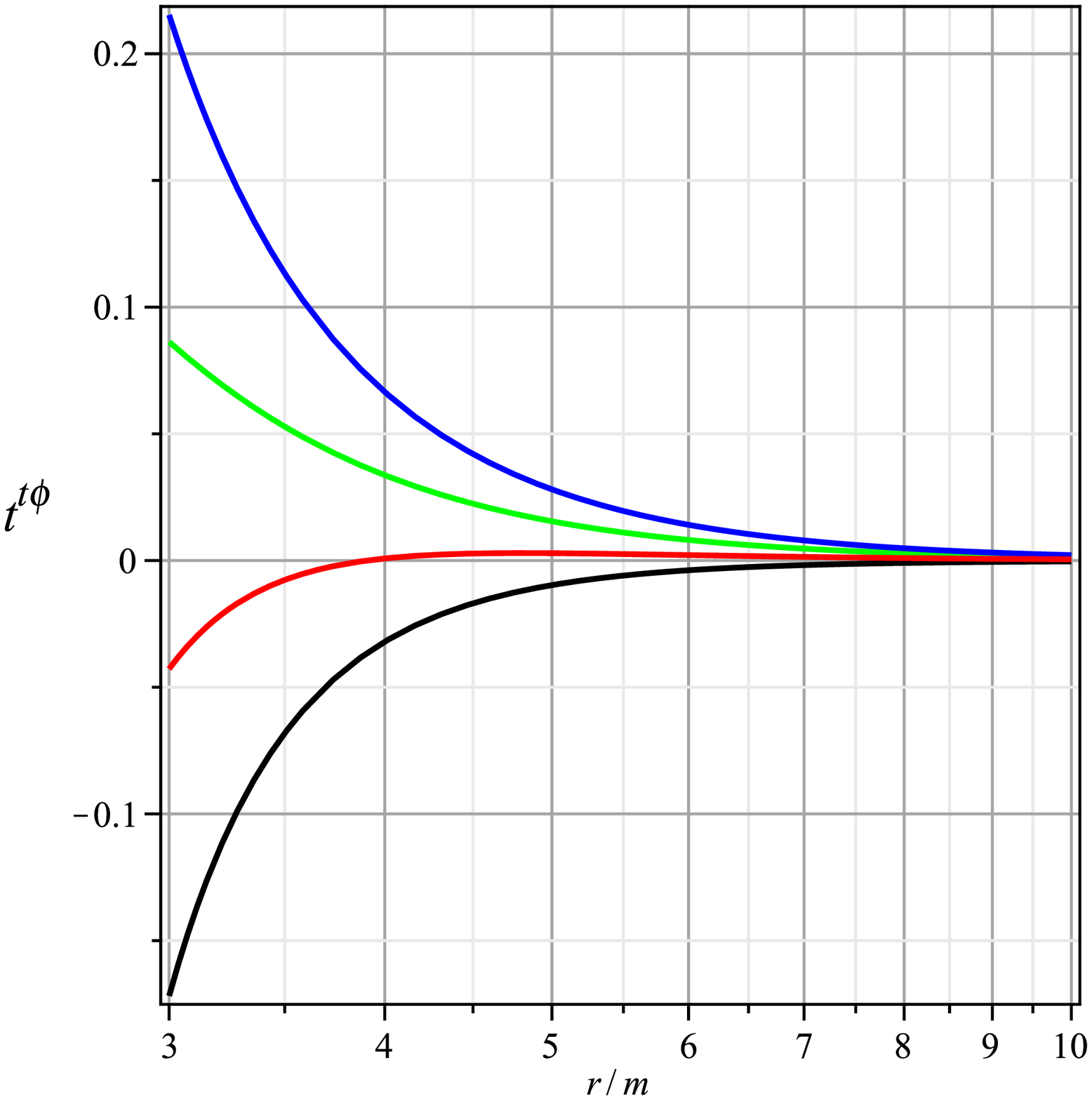}\includegraphics[scale=.22]{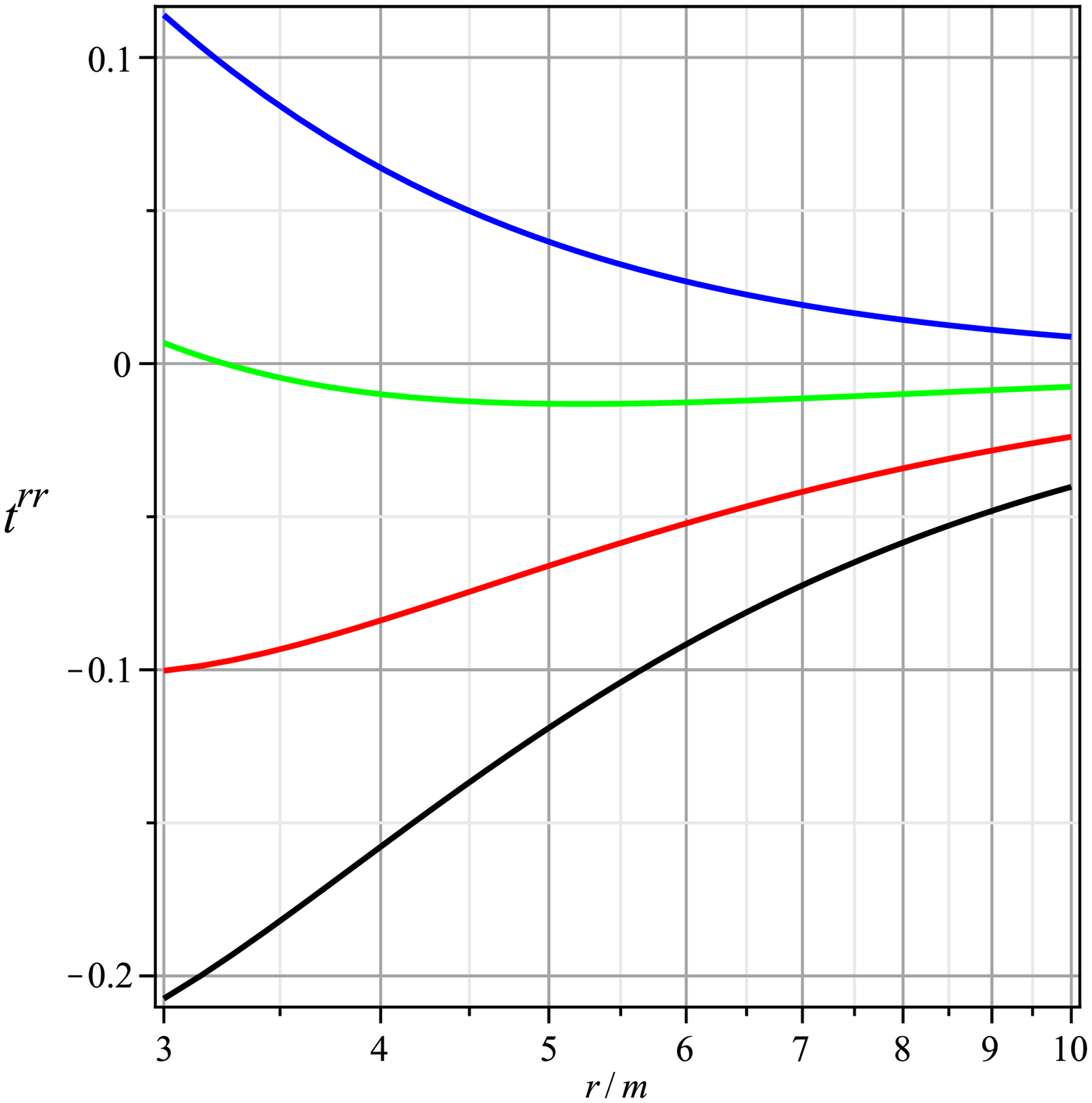}}
\centerline{\includegraphics[scale=.22]{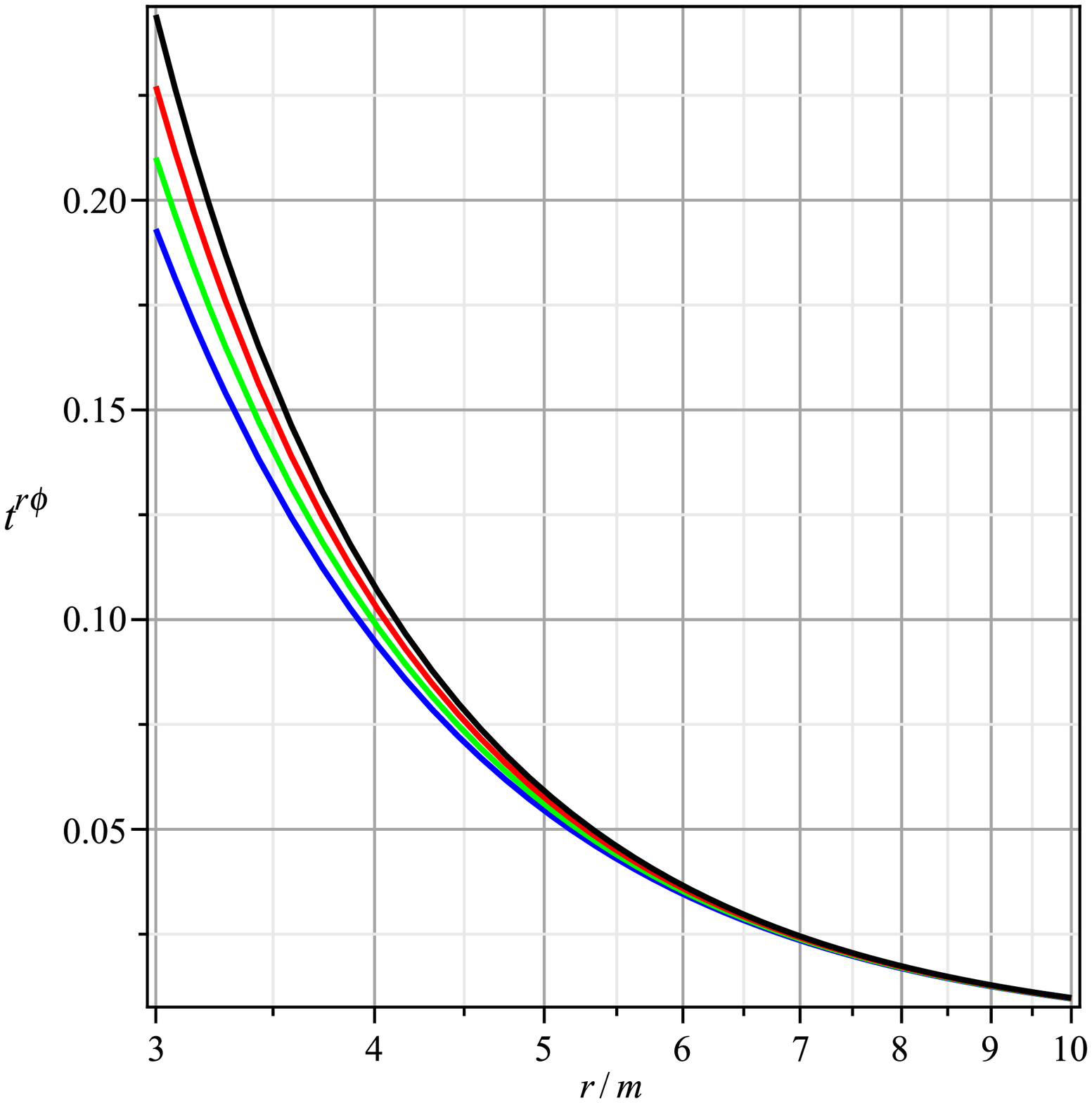} \includegraphics[scale=.22]{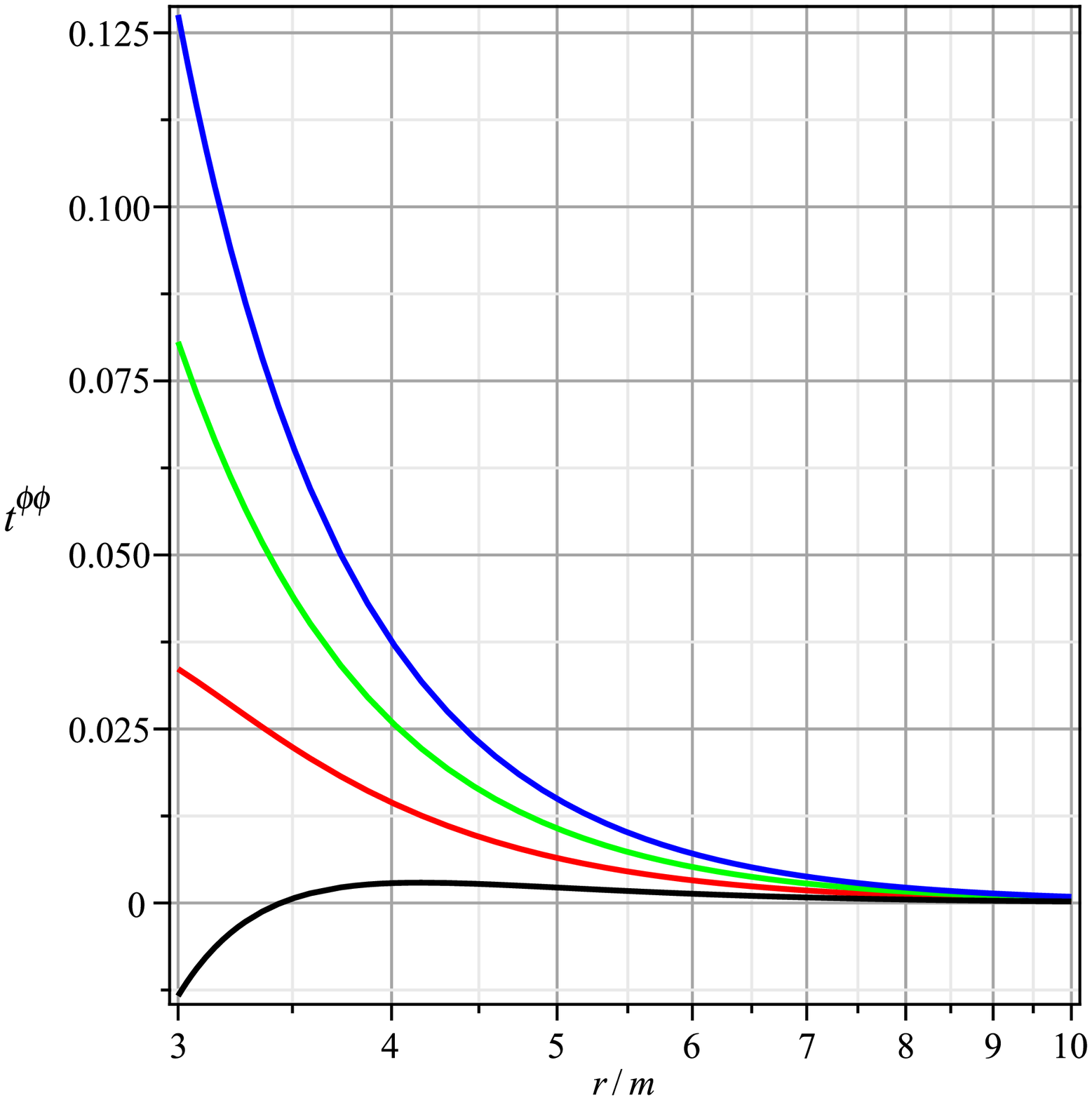}}
\centerline{\includegraphics[scale=.22]{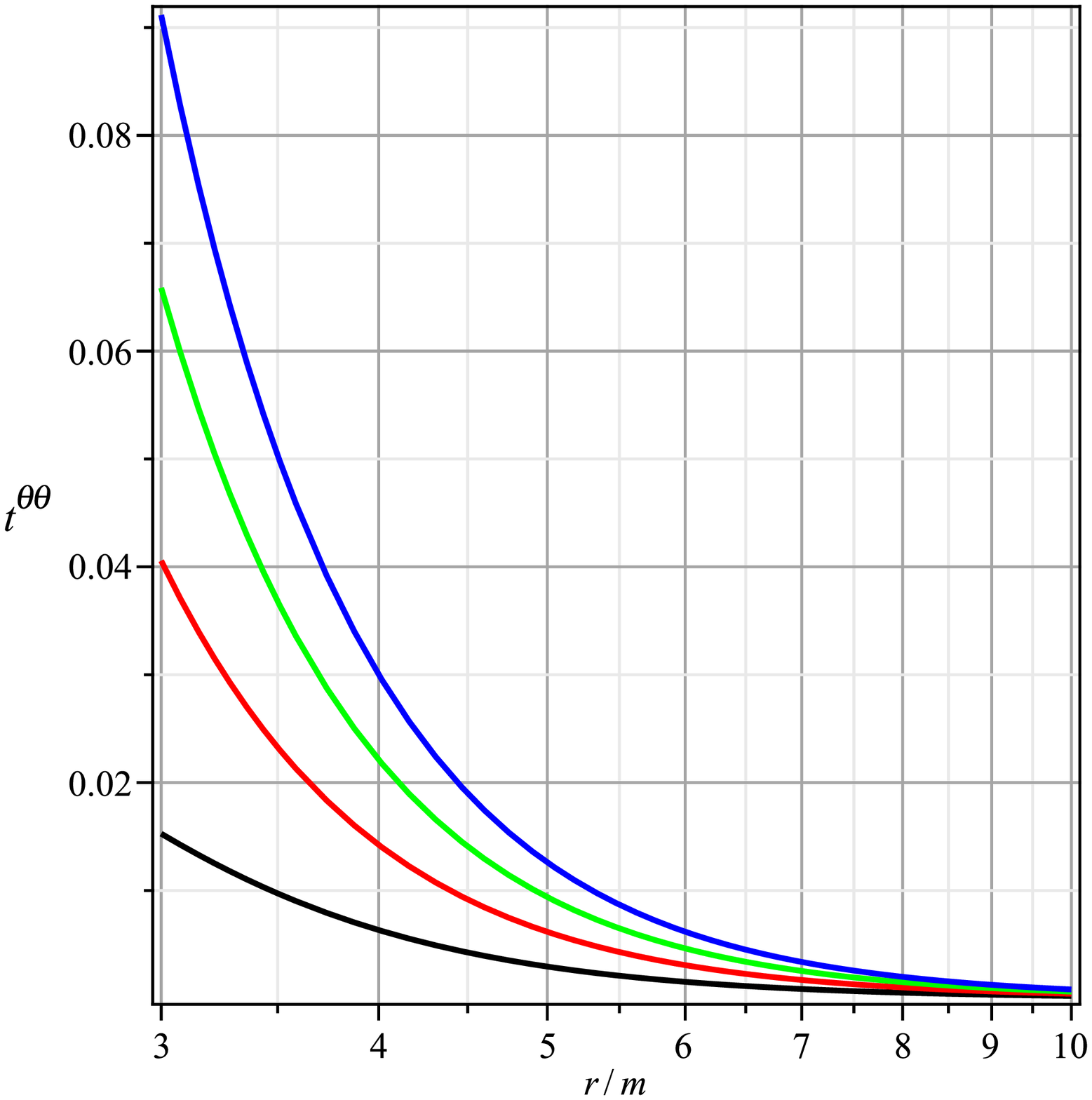}}
 \caption{Influence of the bulk viscosity in the non-zero components of the shear stress viscosity with $\beta=1$,  $n=1$, $a=.9$ and $\lambda=2$. With no bulk viscosity $\zeta=0$ in black, $\zeta=2$ in red, $\zeta=4$ in green and $\zeta=6$ in blue.}
\label{figure9}
\end{figure}

\begin{figure}
\vspace{\fill}
\centerline{\includegraphics[scale=.22]{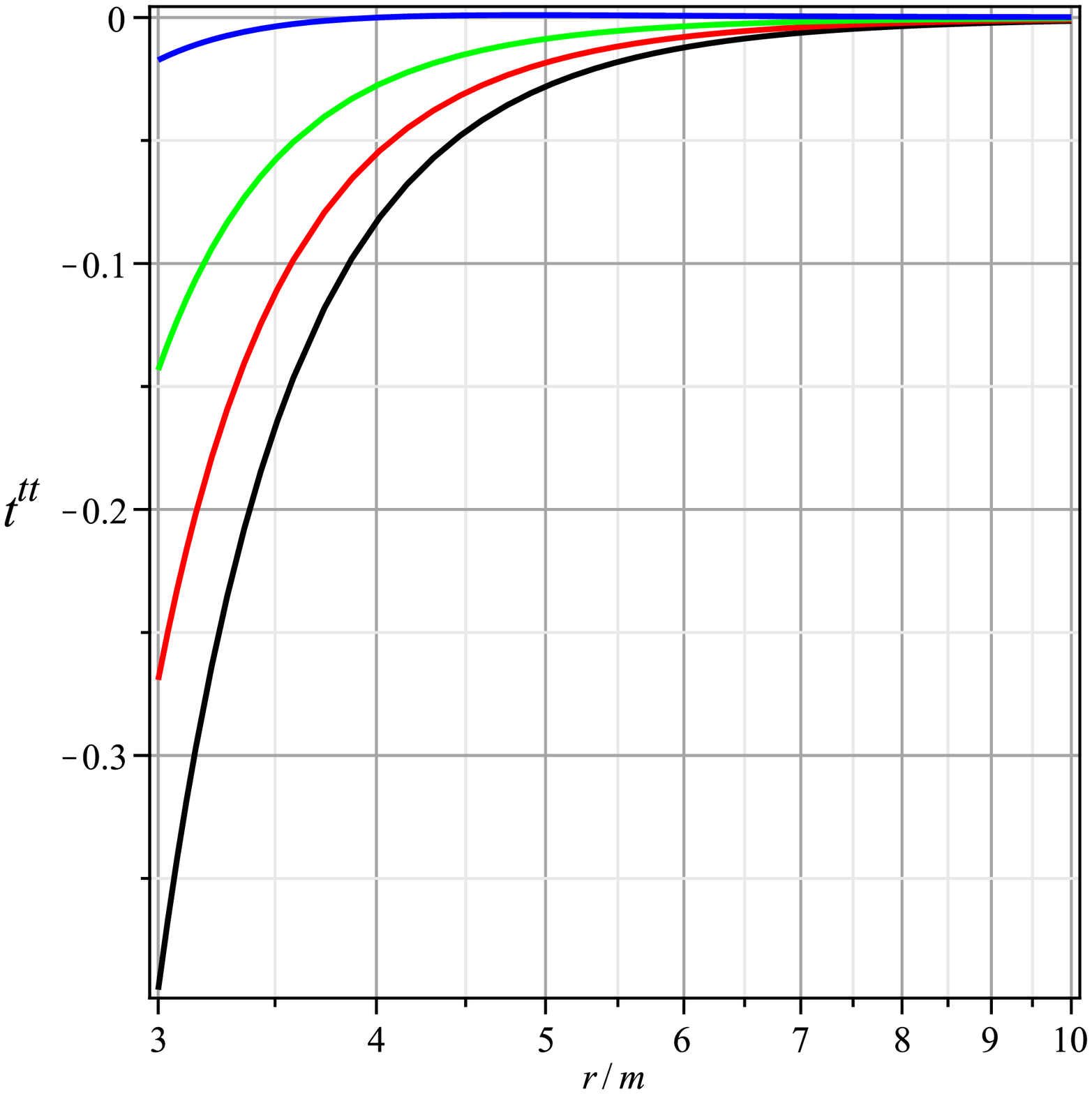}\includegraphics[scale=.22]{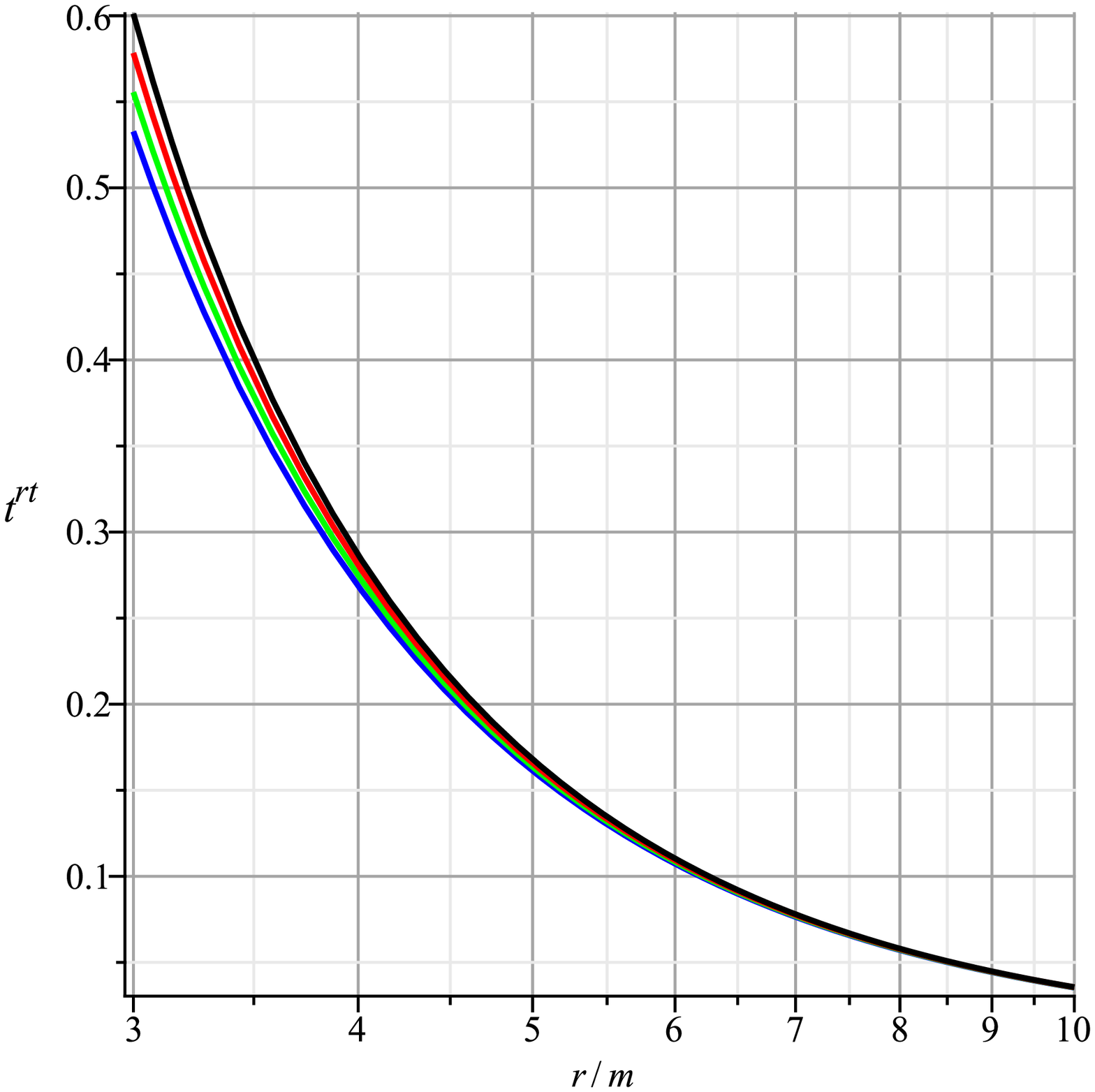}}
\centerline{\includegraphics[scale=.22]{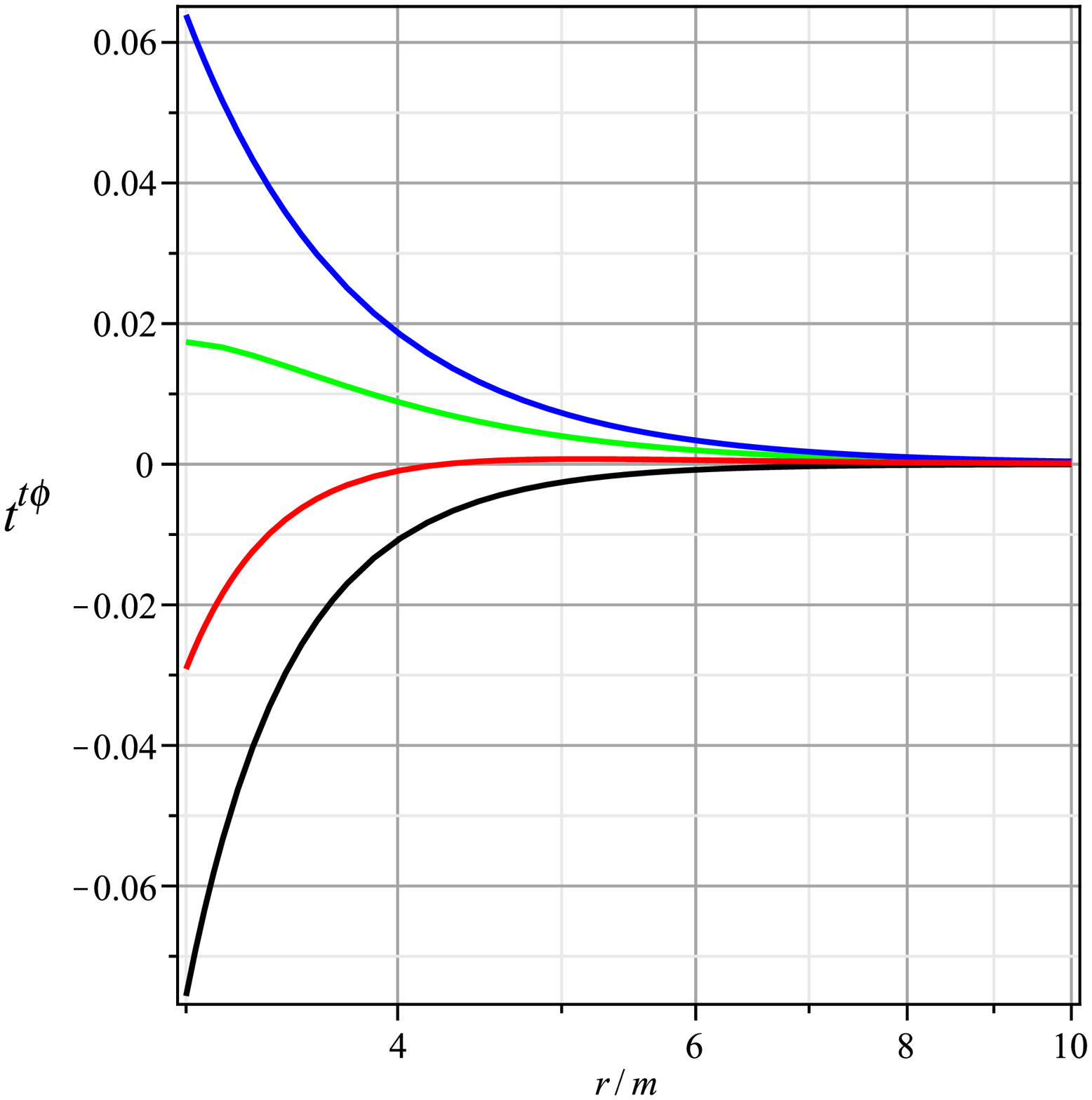}\includegraphics[scale=.22]{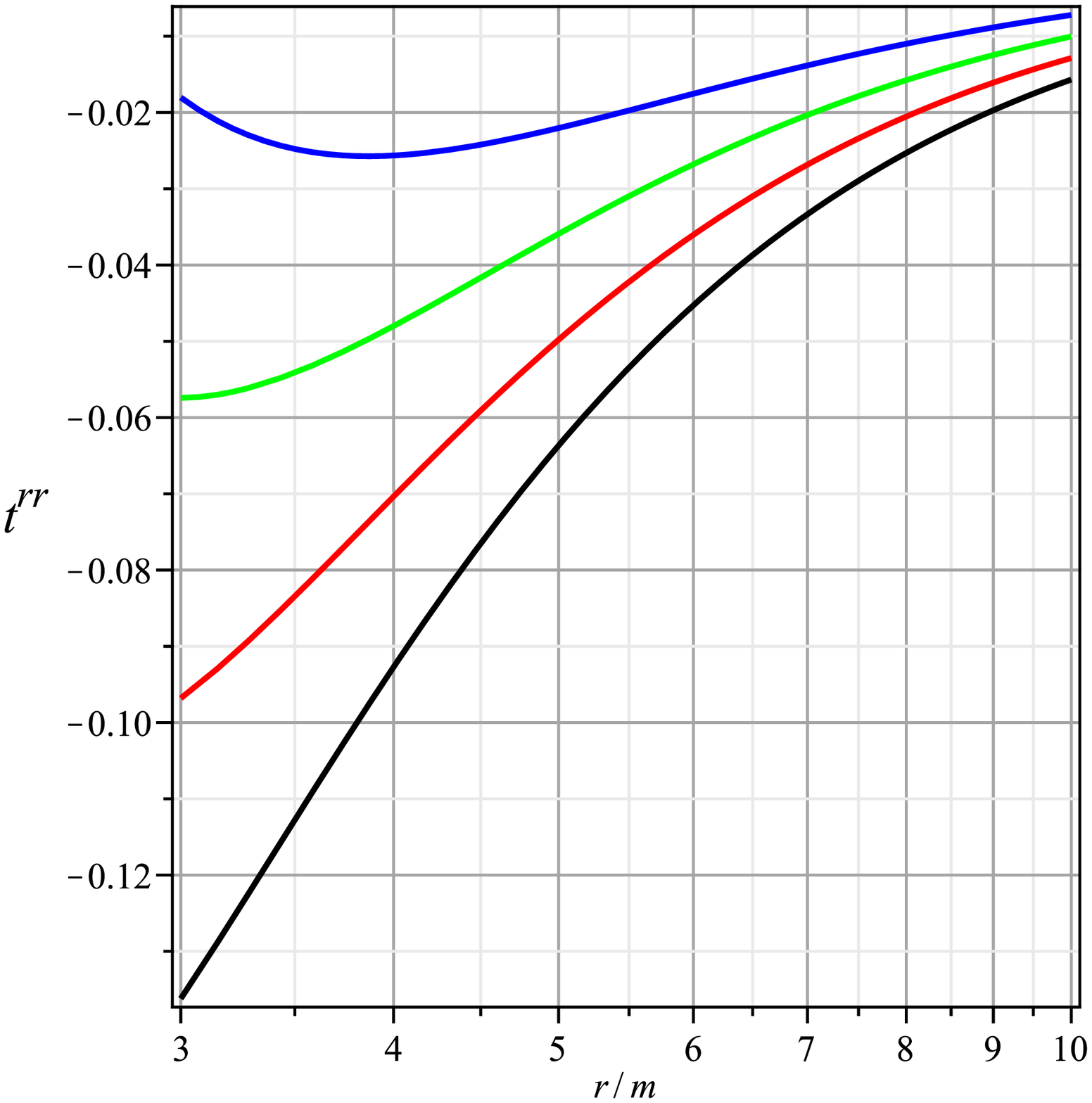}}
\centerline{\includegraphics[scale=.22]{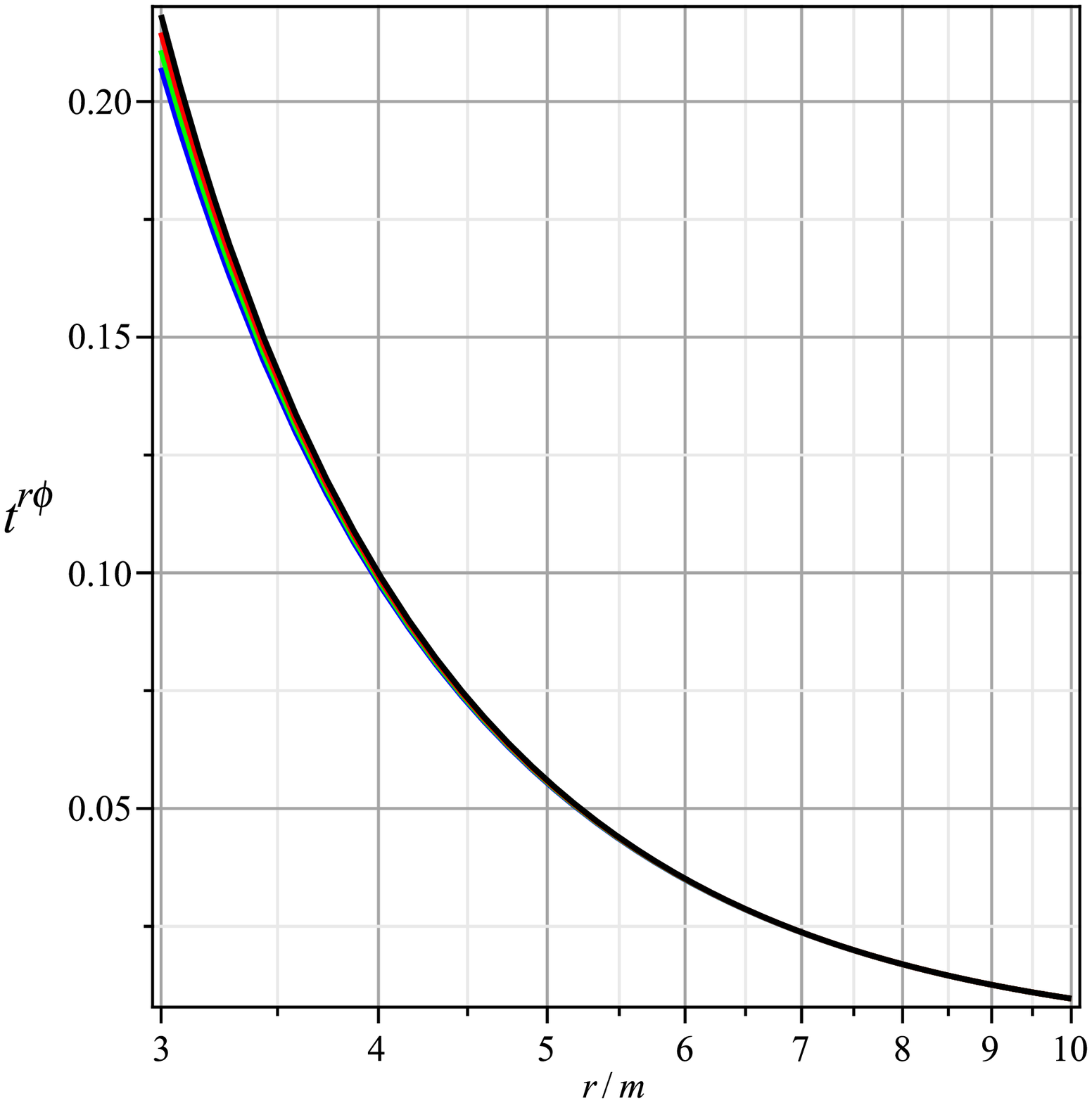} \includegraphics[scale=.22]{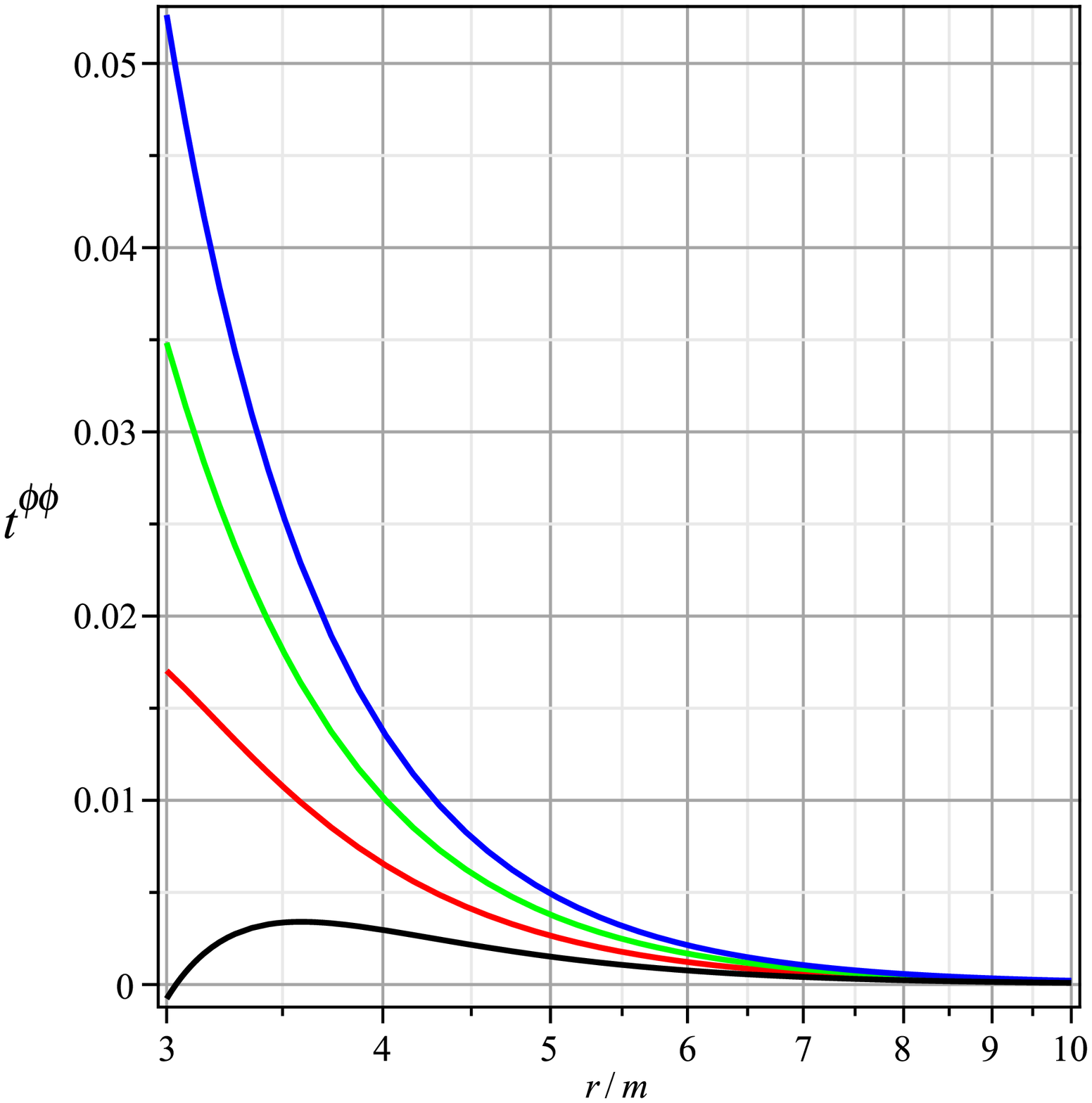}}
\centerline{\includegraphics[scale=.22]{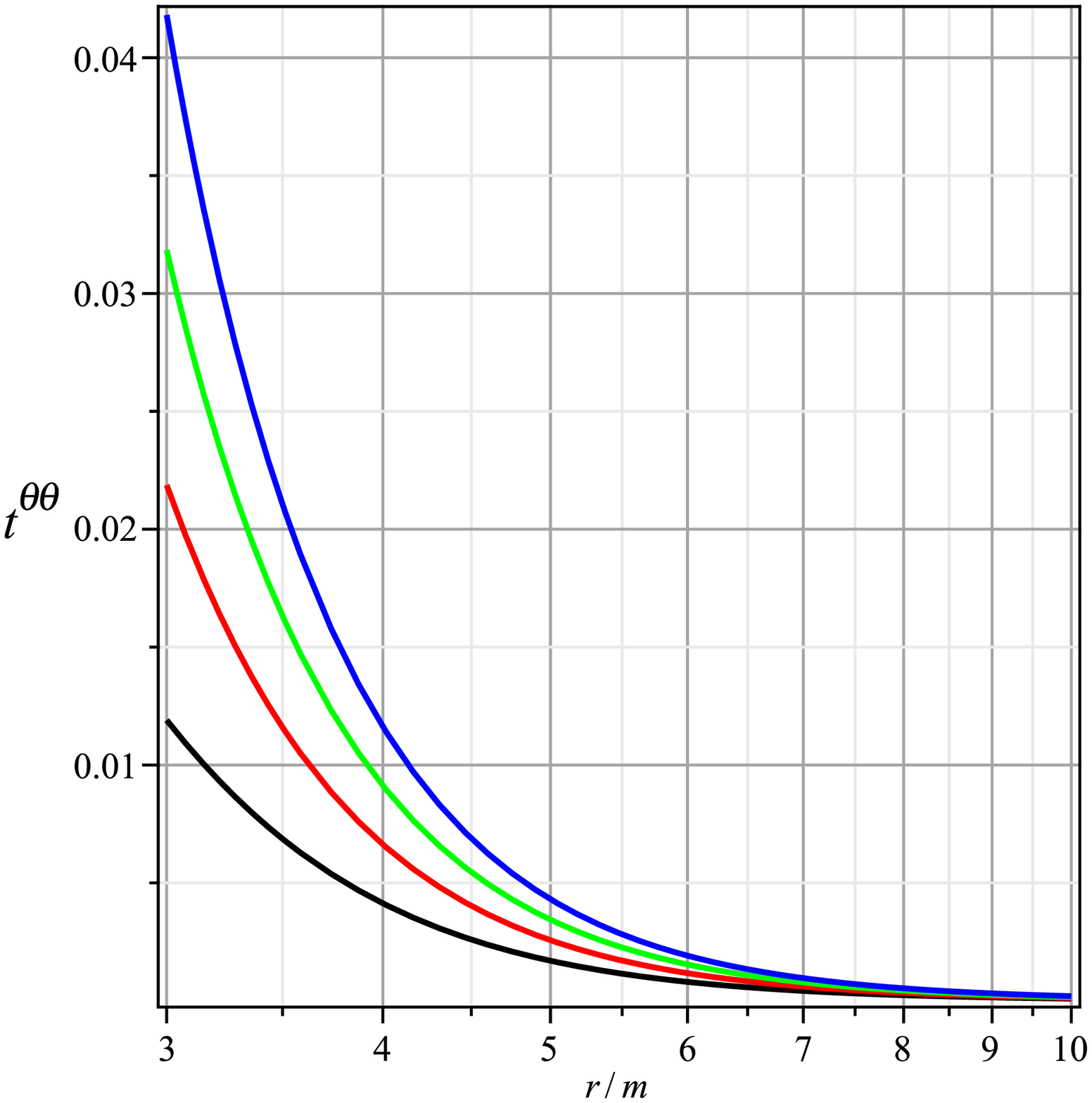}}
 \caption{Influence of the bulk viscosity in the non-zero components of the shear stress viscosity with $\beta=1$,  $n=\frac{3}{2}$, $a=.9$ and $\lambda=2$. With no bulk viscosity $\zeta=0$ in black, $\zeta=2$ in red, $\zeta=4$ in green and $\zeta=6$ in blue.}
\label{figure10}
\end{figure}

\begin{figure}
\vspace{\fill}
\centerline{\includegraphics[scale=.22]{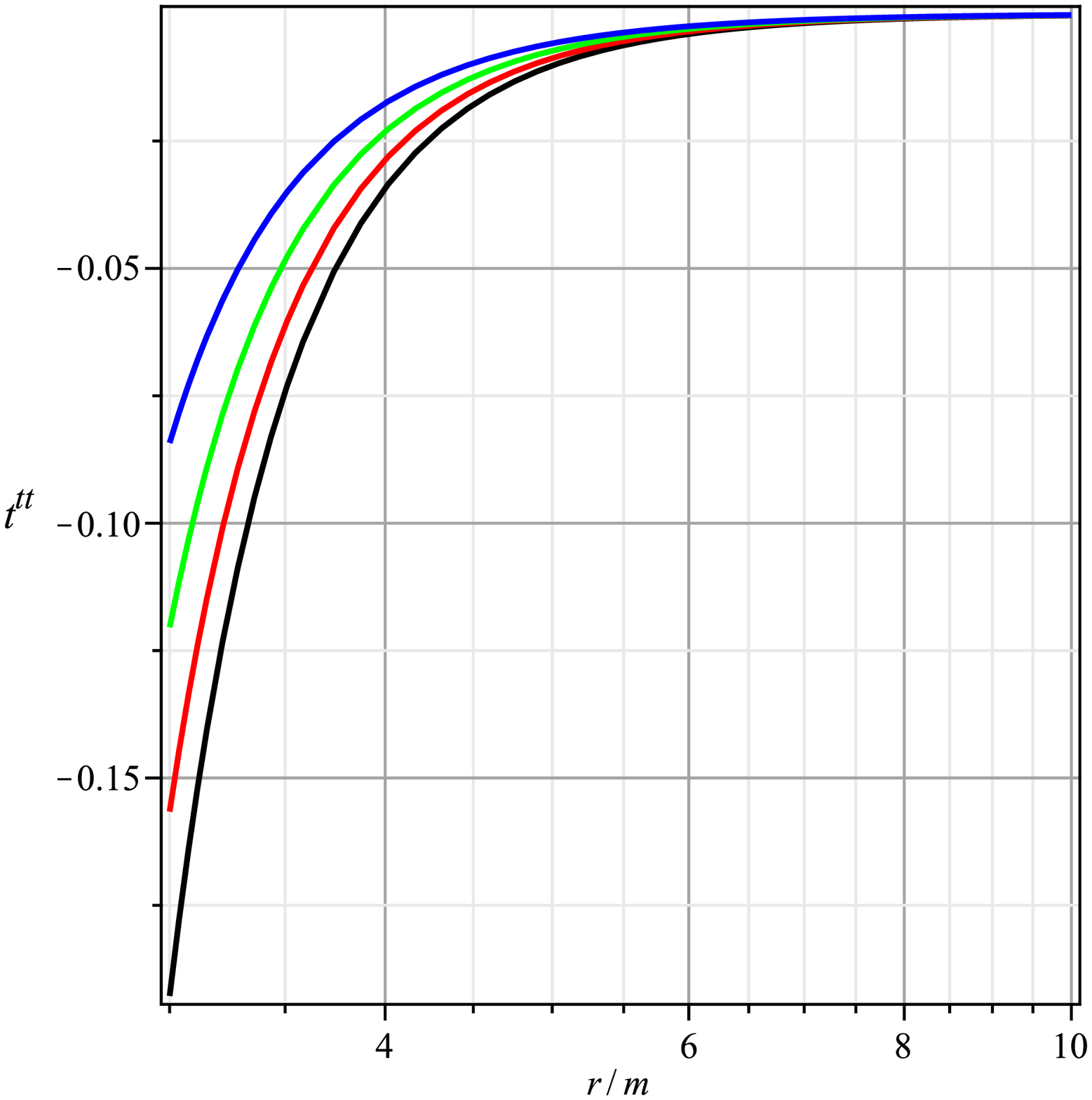}\includegraphics[scale=.22]{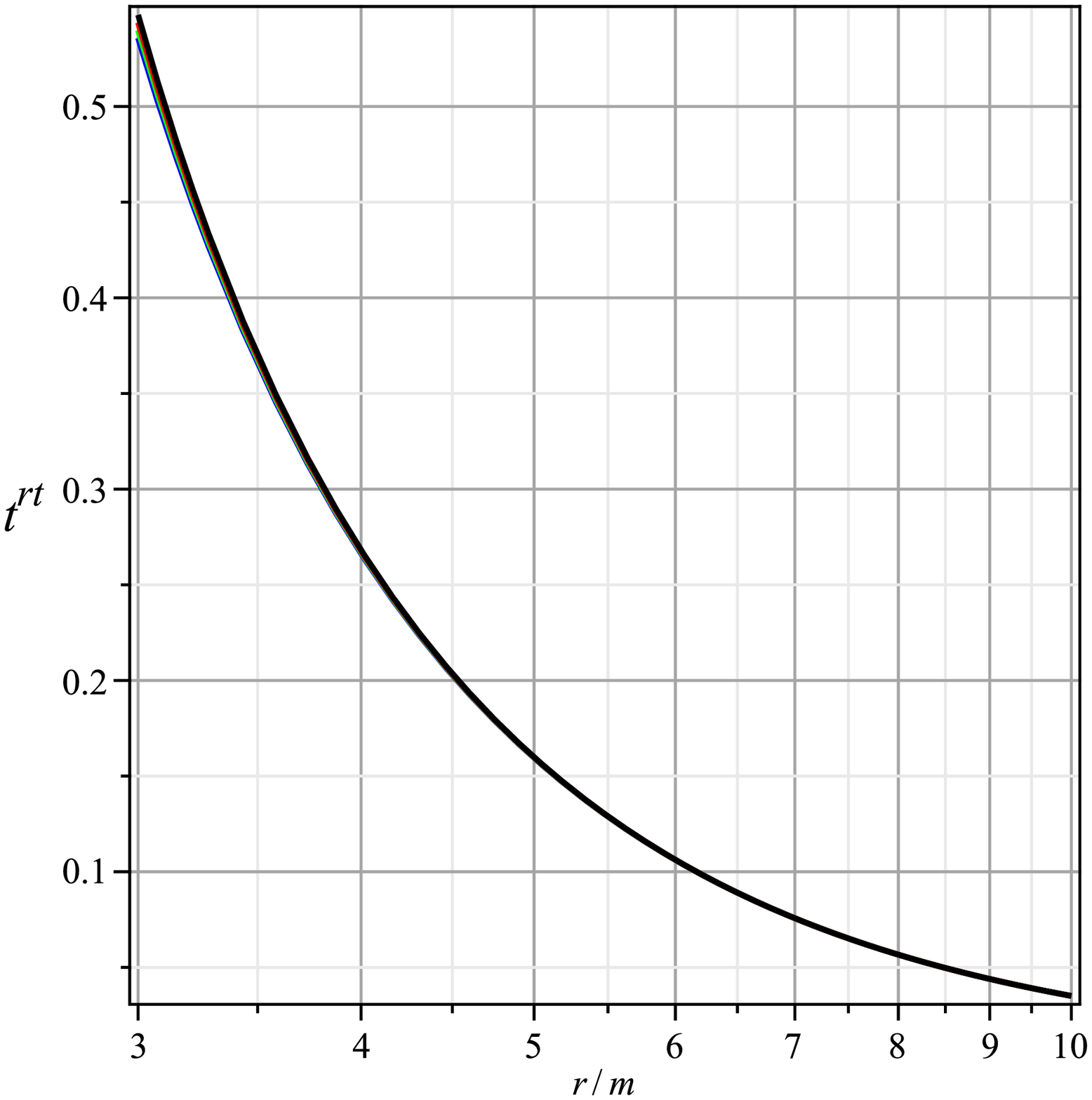}}
\centerline{\includegraphics[scale=.22]{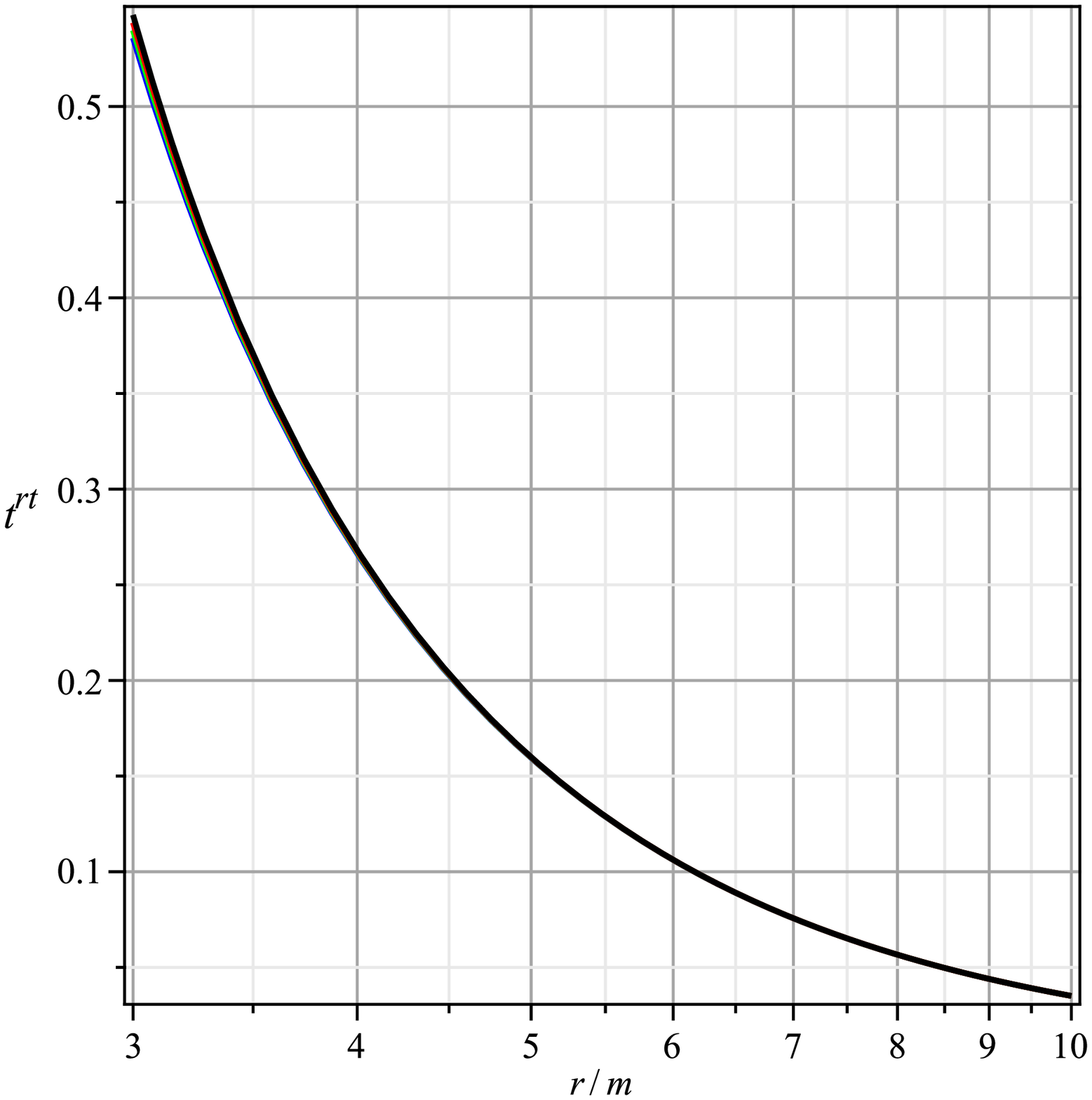}\includegraphics[scale=.22]{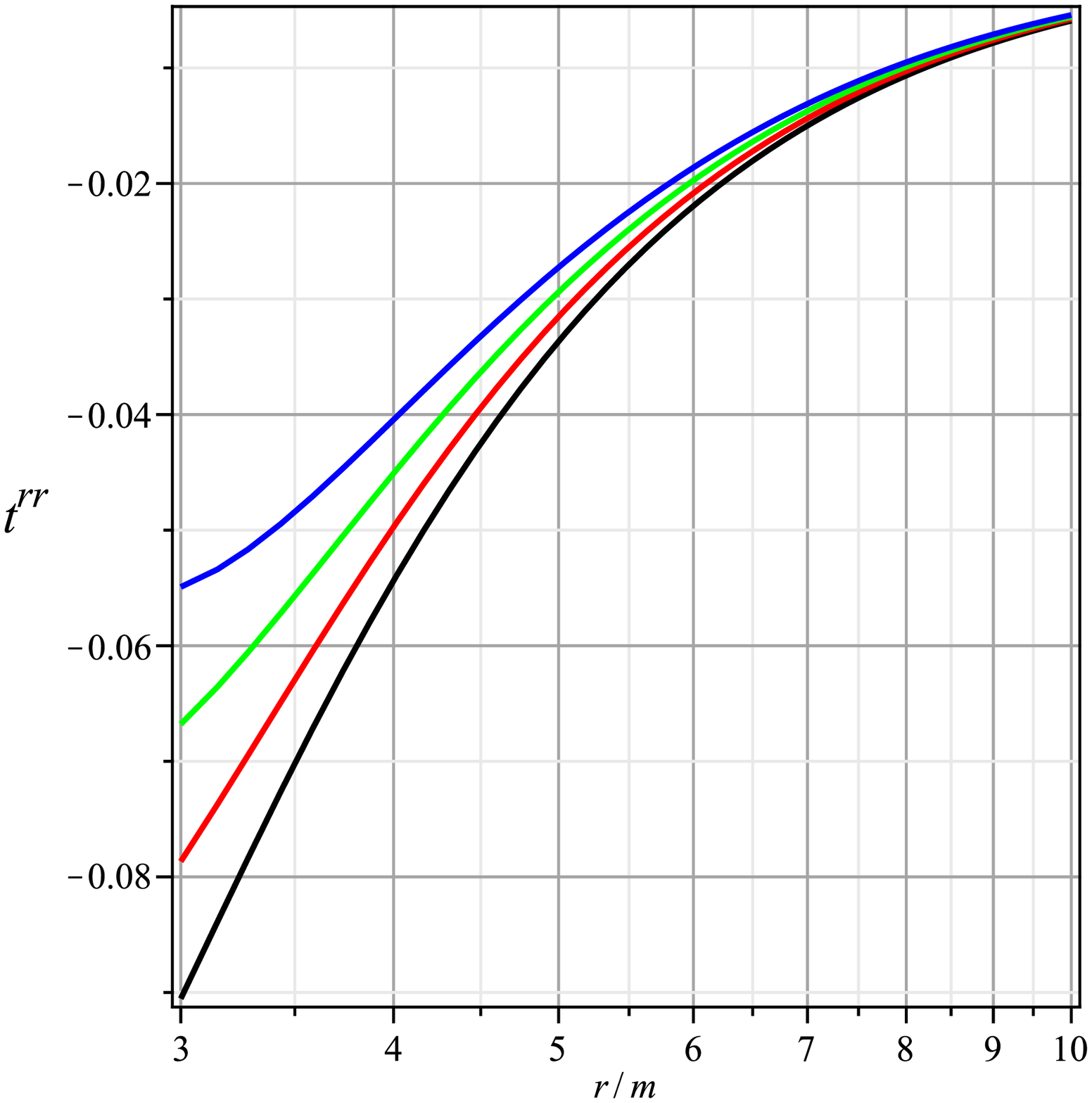}}
\centerline{\includegraphics[scale=.22]{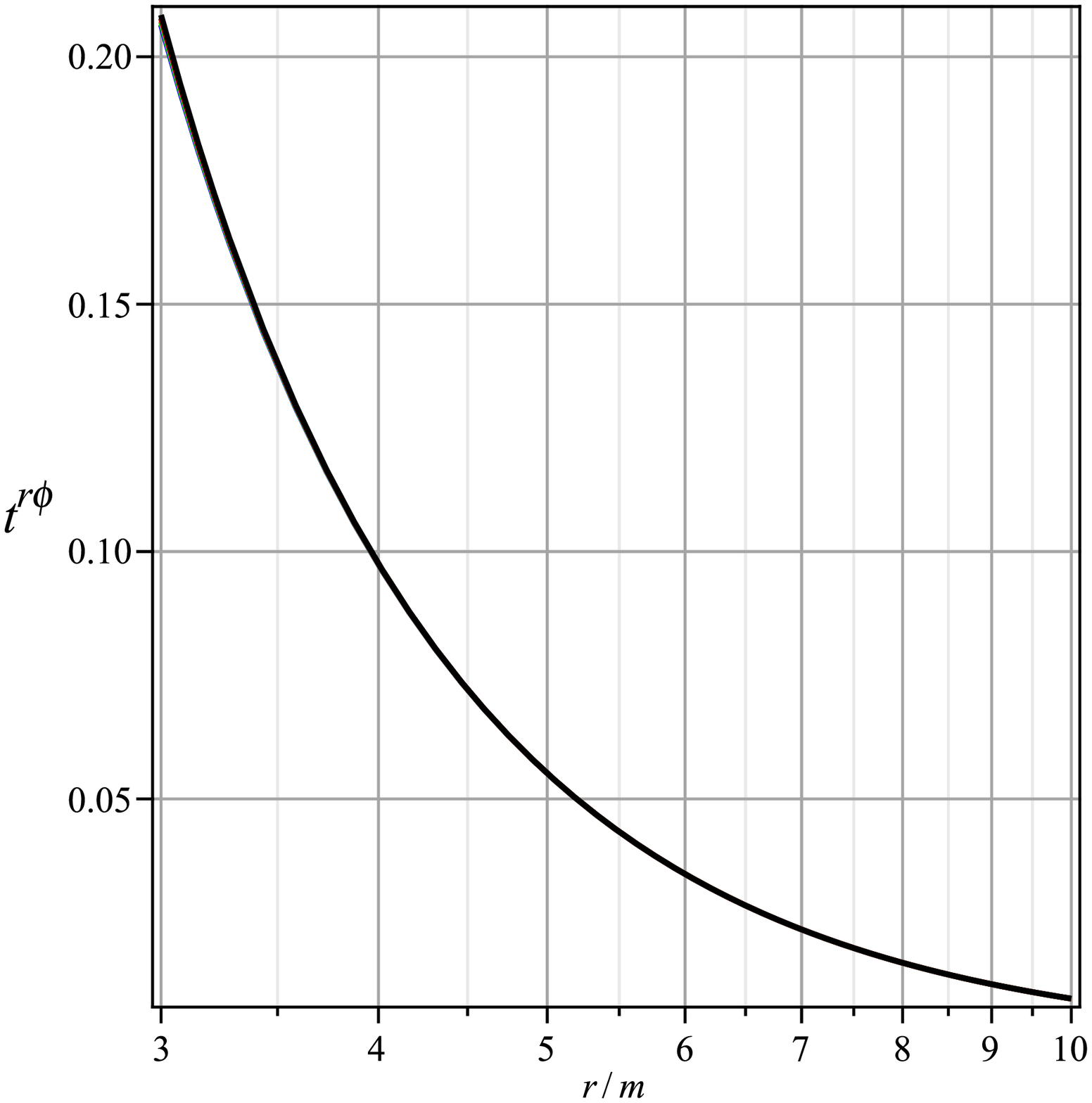}\includegraphics[scale=.22]{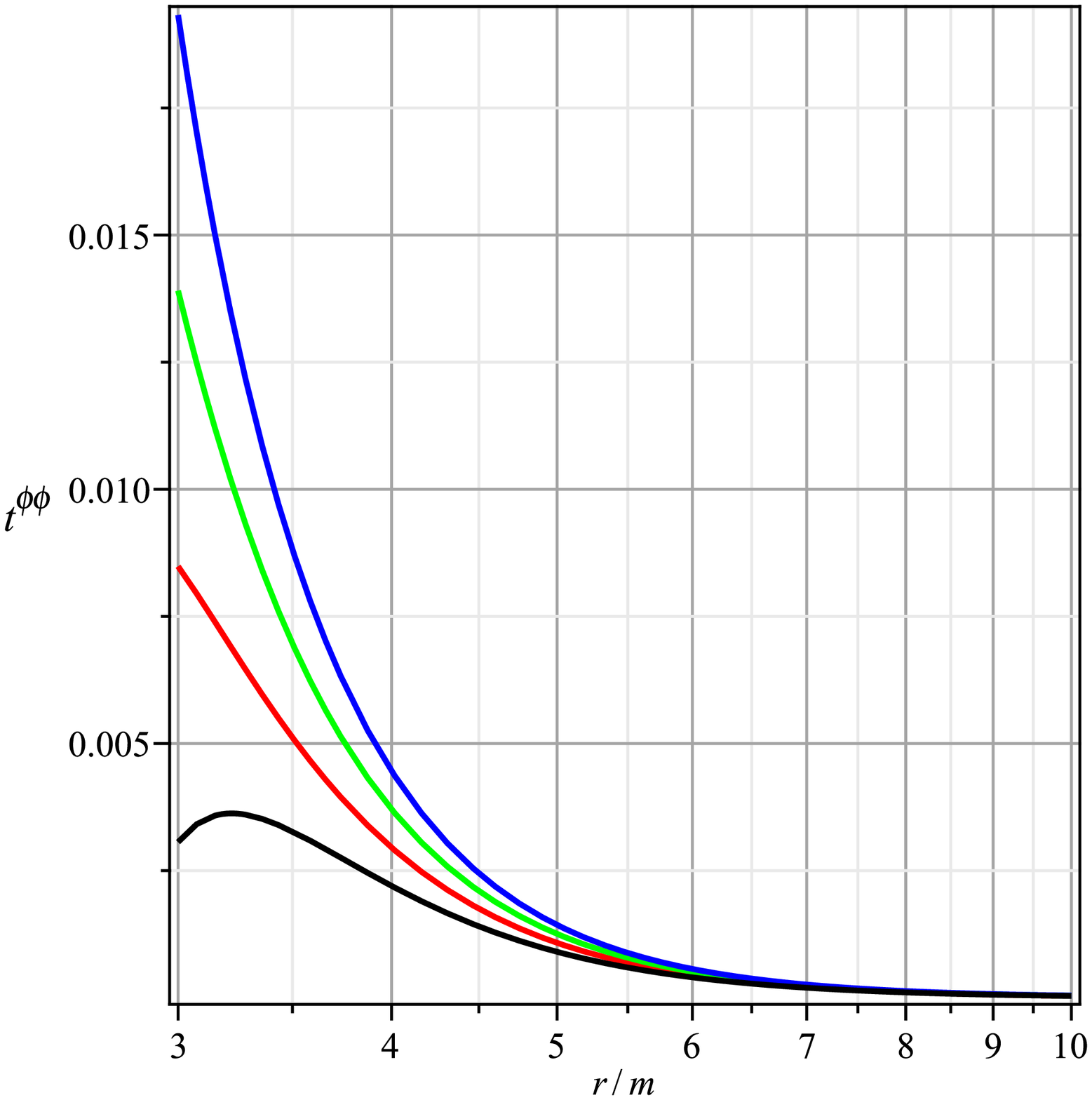}}
\centerline{\includegraphics[scale=.22]{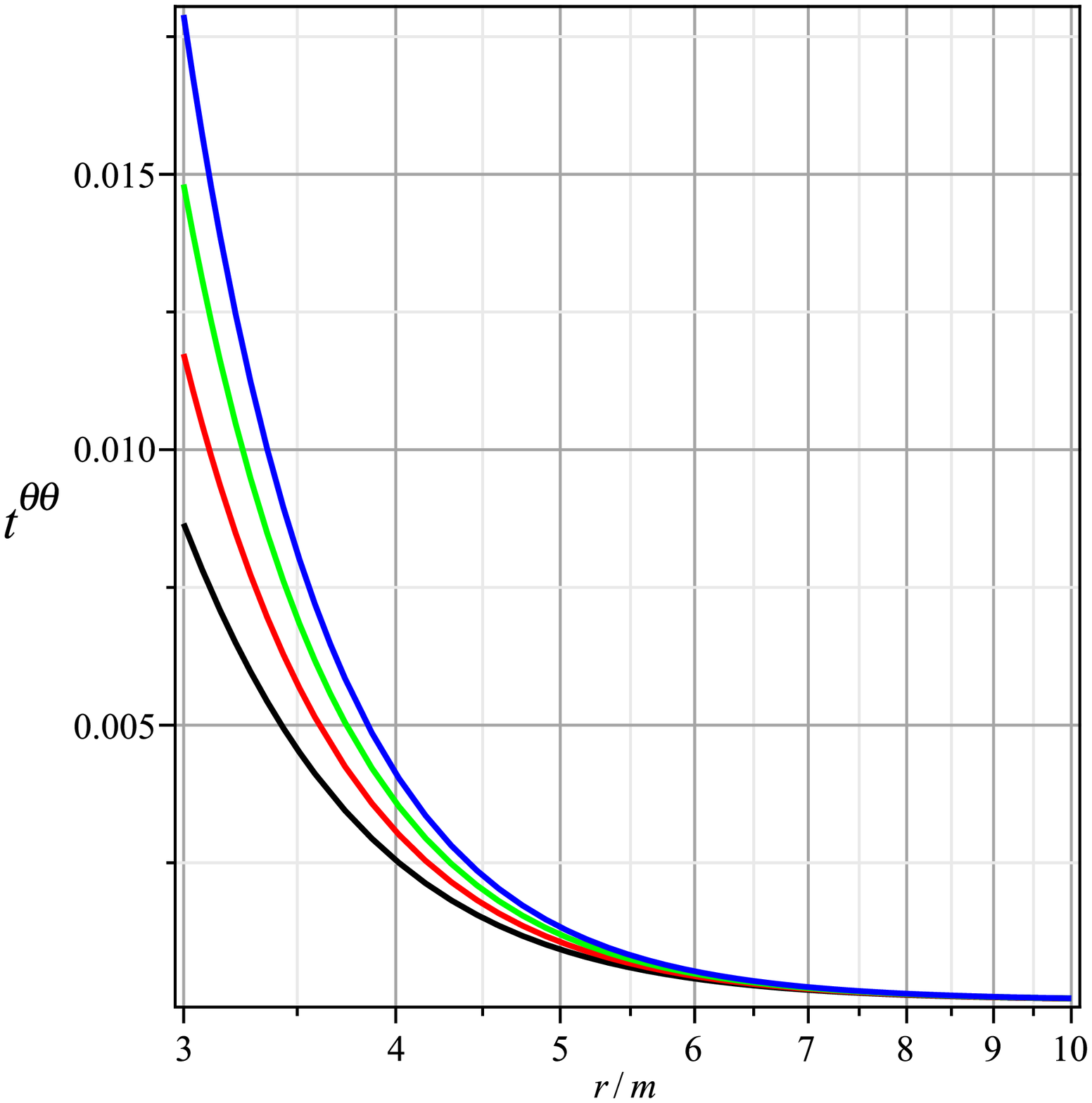}}
 \caption{Influence of the bulk viscosity in the non-zero components of the shear stress viscosity with $\beta=1$,  $n=2$, $a=.9$ and $\lambda=2$. With no bulk viscosity $\zeta=0$ in black, $\zeta=2$ in red, $\zeta=4$ in green and $\zeta=6$ in blue.}
\label{figure11}
\end{figure}

The relativistic Navier-Stokes shear stress viscosity, is written as (\cite{mt}):
\begin{equation}\label{32}
t^{\mu\nu}=-2\lambda\sigma^{\mu\nu}-\zeta\Theta h^{\mu\nu}=S^{\mu\nu}+B^{\mu\nu}.
\end{equation}
In this section the components of shear stress viscosity are calculated with the radial model with equations(\ref{6}), (\ref{11}), (\ref{12}) and (\ref{16}). Similar to shear tensor and bulk tensor, the shear stress viscosity have ten non-zero components as
\begin{eqnarray}\label{33}
t^{\mu\nu}=&&\left(
\begin{array}{cccc}
  t^{tt} & t^{tr} & 0 & t^{t\phi}\\
  t^{rt} & t^{rr} & 0 & t^{r\phi} \\
  0 & 0 & t^{\theta\theta} & 0 \\
  t^{\phi t} & t^{\phi r} & 0 & t^{\phi \phi} \\
\end{array}%
\right)
\end{eqnarray}
Figures \ref{figure8}, \ref{figure9}, \ref{figure10}  and \ref{figure11} show the influence of the bulk viscosity in non-zero components of shear stress viscosity in $n=\frac{1}{2},1,\frac{3}{2}$ and $n=2$ in $a=0.9$.
Figures 8-11 show that the influence of variation of the bulk viscosity in the shear stress viscosity are important in the smaller amounts of $n$ and in the inner radiuses. 
\section{Summery and Conclusions}
\label{sec:8}
In study of the accretion disks around the rotating black holes the bulk viscosity was assumed to be zero without any physical rasons therefore in this paper we tried to calculate the relativistic bulk viscosity and study influence of the relativistic bulk viscosity in the relativistic shear stress viscosity of accretion disks around the rotating black holes. In this paper the relations of all the components of the relativistic bulk tensor and the shear tensor for the accretion disk around the rotating black hole in the equatorial plan in the BLF with no approximation are derived. Ten non-zero components of the relativistic bulk tensor and shear tensor are $tt$, $tr$, $t\phi$, $rt$, $rr$, $r\phi$, $\theta\theta$, $\phi t$, $\phi r$ and $\phi\phi$ components.

 We see that the relativistic bulk tensor, $b^{\mu\nu}$ is caused by radial variation of the radial component of the four velocity and components of $u^{\mu}$ and $u^{\nu}$. But the relativistic shear tensor , $\sigma^{\mu\nu}$  can be create with radial variation of $u^{\mu}$, $u^{\nu}$, $u^{r}$ and with all the components of the four velocity.
Therefore we expect in the most cases the relativistic shear tensor is larger than the relativistic bulk tensor except in the big radial component of the four velocity.

We introduce a simple radial model for the radial component of the four velocity in the LNRF and with transformation we calculate the components of the four velocity in the BLF. In the Keplarian disks we calculate the four velocity in four cases. With the relativistic calculation $tt$, $tr$, $t\phi$, $rt$, $rr$, $r\phi$, $\theta\theta$, $\phi t$, $\phi r$ and $\phi\phi$ components of the bulk tensor, shear tensor and shear stress viscosity are derived. 

The non-zero components of the relativistic bulk tensor and shear tensor are in figures 2-7. In these figures we notice that in the smaller $n$ and bigger $a$ the amounts of the bulk tensor may be comparable with the shear tensor because the amounts of the radial componet of the four velocity and expansion of the fliud world line are bigger in those cases.

The none-zero components of the relativistic shear stress viscosity show in figures \ref{figure8}, \ref{figure9}, \ref{figure10} and \ref{figure11}, in those figures we see that except $t^{r\phi}$ and $t^{rt}$ which were used in the previous papers some other components are important especially in the inner radiuses. But in the outer radiuses these are been less important as we see in the outer radiuse of the disks $tt$, $t\phi$, $\phi t$, and $\phi\phi$ components are zero, so $tr$, $rt$, $rr$, $r\phi$, $\theta\theta$, and $\phi r$  components are non-zero components of the shear stress viscosity in the outer radiuse.

Comparison of figures show that the bulk tensor is important in bigger a, lower $n$  and in the inner radiuses, therefore if the radial component of the four velocity has the similar form of this model, in all cases the bulk tensor is not ignorable.



\appendix
\section{The Christoffel symboles in the BLF}
\label{apb}
After some calculation the Christoffel symboles ($\Gamma ^{\alpha}_{\beta\gamma}$) in the equatorial plan and in our scaling of the BLF are
\begin{eqnarray}\label{ap35}
&& \Gamma ^{t}_{tt}=0,\qquad \Gamma ^{t}_{rt}=\Gamma ^{t}_{tr}=\frac{a^{2}+r^{2}}{r^{2}(a^{2}+r^{2}-2r)},
\qquad \Gamma ^{t}_{t\theta}=\Gamma ^{t}_{\theta t}=0,\qquad \Gamma ^{t}_{t\phi}=\Gamma ^{t}_{\phi t}=0,\qquad\Gamma ^{t}_{rr}=0,\qquad\Gamma ^{t}_{r\theta}=\Gamma ^{t}_{\theta r}=0, 
\nonumber\\ &&\Gamma ^{t}_{r\phi}=\Gamma ^{t}_{\phi r}=-\frac{a(a^{2}+3r^{2})}{r^{2}(a^{2}+r^{2}-2r)}, \qquad\Gamma ^{t}_{\theta\theta}=0,\qquad  \Gamma ^{t}_{\theta\phi}=\Gamma ^{t}_{\phi \theta}=0, \qquad \Gamma ^{t}_{\phi\phi}=0, \qquad\Gamma ^{r}_{tt}=\frac{a^{2}+r^{2}-2r}{r^{4}},
\nonumber\\ &&\Gamma ^{r}_{tr}=\Gamma ^{r}_{rt}=0, \qquad \Gamma ^{r}_{t\theta}=\Gamma ^{r}_{\theta t}=0,\qquad\Gamma ^{r}_{t\phi}=\Gamma ^{r}_{\phi t}=-\frac{a(a^{2}+r^{2}-2r)}{r^{4}},\qquad \Gamma ^{r}_{rr}=\frac{a^{2}-r}{r(a^{2}+r^{2}-2r)},\qquad\Gamma ^{r}_{r\theta}=\Gamma ^{r}_{\theta r}=0,
 \qquad \Gamma ^{r}_{r\phi}=\Gamma ^{r}_{\phi r}=0, 
\nonumber\\ &&\Gamma ^{r}_{\theta\theta}=-\frac{a^{2}+r^{2}-2r}{r},\qquad \Gamma ^{r}_{\theta\phi}=\Gamma ^{r}_{\phi\theta}=0,
 \qquad\Gamma ^{r}_{\phi\phi}=\frac{(a^{2}+r^{2}-2r)(-r^{3}+a^{2})}{r^{4}},\qquad\Gamma ^{\theta}_{tt}=0,\qquad \Gamma ^{\theta}_{tr}=\Gamma ^{\theta}_{rt}=0,
\nonumber\\ &&\Gamma ^{\theta}_{t\theta}=\Gamma ^{\theta}_{\theta t}=0,\qquad\Gamma ^{\theta}_{t\phi}=\Gamma ^{\theta}_{\phi t}=0,\qquad \Gamma ^{\theta}_{rr}=0,\qquad \Gamma ^{\theta}_{r\theta}=\Gamma ^{\theta}_{\theta r}=\frac{1}{r},\qquad \Gamma ^{\theta}_{r\phi}=\Gamma ^{\theta}_{\phi r}=0,\qquad \Gamma ^{\theta}_{\theta\theta}=0, \qquad\Gamma ^{\theta}_{\theta\phi}=\Gamma ^{\theta}_{\phi\theta}=0,
\nonumber\\ && \Gamma ^{\theta}_{\phi\phi}=0\Gamma ^{\phi}_{tt}=0,\qquad\Gamma ^{\phi}_{tr}=\Gamma ^{\phi}_{rt}=\frac{a}{r^{2}(a^{2}+r^{2}-2r)},\qquad\Gamma ^{\phi}_{t\theta}=\Gamma ^{\phi}_{\theta t}=0,\qquad\Gamma ^{\phi}_{t\phi}=\Gamma ^{\phi}_{\phi t}=0,\qquad\Gamma ^{\phi}_{rr}=0,
\nonumber\\ && \Gamma ^{\phi}_{r\theta}=\Gamma ^{\phi}_{\theta r}=0,\qquad\Gamma ^{\phi}_{r\phi}=\Gamma ^{\phi}_{\phi r}=\frac{r^{3}-a^{2}-2r^{2}}{r^{2}(r^{2}(a^{2}+r^{2}-2r)},\qquad \Gamma ^{\phi}_{\theta\theta}=0,\qquad\Gamma ^{\phi}_{\theta\phi}=\Gamma ^{\phi}_{\phi\theta}=0,\qquad\Gamma ^{\phi}_{\phi\phi}=0
\end{eqnarray}
\section{components of shear tensor}
\label{apc}
All non-zero components of shaer tensor are calculated from equation (\ref{15}) as
\begin{eqnarray}\label{ap36}
&&\sigma^{tt}= (u^{t}_{,r}+\Gamma ^{t}_{rt}u^{t}+\Gamma ^{t}_{rr}u^{r}+\Gamma ^{t}_{r\phi}u^{\phi})h^{rt}+(\Gamma ^{t}_{tt}u^{t}+\Gamma ^{t}_{tr}u^{r}
 +\Gamma ^{t}_{t\phi}u^{\phi})h^{tt}+(\Gamma ^{t}_{\phi t}u^{t}+\Gamma ^{t}_{\phi r}u^{r}+\Gamma ^{t}_{\phi\phi}u^{\phi})h^{t\phi} -\frac{1}{3}(u^{r}_{,r}+\frac{2u^{r}}{r})h^{tt}
\nonumber\\&&\sigma^{tr}=\sigma^{rt}= \frac{1}{2}[(u^{t}_{,r}+\Gamma ^{t}_{rt}u^{t}+\Gamma ^{t}_{rr}u^{r}+\Gamma ^{t}_{r\phi}u^{\phi})h^{rr}+(\Gamma ^{t}_{tt}u^{t}+\Gamma ^{t}_{tr}u^{r}+\Gamma ^{t}_{t\phi}u^{\phi})h^{rt}+(\Gamma ^{t}_{\phi t}u^{t}+\Gamma ^{t}_{\phi r}u^{r}+\Gamma ^{t}_{\phi\phi}u^{\phi})h^{r\phi}+(u^{r}_{,r}
\nonumber\\ &&\qquad +\Gamma ^{r}_{rt}u^{t}+\Gamma ^{r}_{rr}u^{r}+\Gamma ^{r}_{r\phi}u^{\phi})h^{rt}+(\Gamma ^{r}_{tt}u^{t}+\Gamma ^{r}_{tr}u^{r}+\Gamma ^{r}_{t\phi}u^{\phi})h^{tt}+(\Gamma ^{r}_{\phi t}u^{t}+\Gamma ^{r}_{\phi r}u^{r}+\Gamma ^{r}_{\phi\phi}u^{\phi})h^{t\phi}]-\frac{1}{3}(u^{r}_{,r}+\frac{2u^{r}}{r})h^{rt}
\nonumber\\&&\sigma^{t\phi}=\sigma^{\phi t}= \frac{1}{2}[(u^{t}_{,r}+\Gamma ^{t}_{rt}u^{t}+\Gamma ^{t}_{rr}u^{r}+\Gamma ^{t}_{r\phi}u^{\phi})h^{r\phi}+(\Gamma ^{t}_{tt}u^{t}+\Gamma ^{t}_{tr}u^{r}+\Gamma ^{t}_{t\phi}u^{\phi})h^{t\phi}+(\Gamma ^{t}_{\phi t}u^{t}+\Gamma ^{t}_{\phi r}u^{r}+\Gamma ^{t}_{\phi\phi}u^{\phi})h^{\phi\phi}+(u^{\phi}_{,r}
\nonumber\\ &&\qquad +\Gamma ^{\phi}_{rt}u^{t}+\Gamma ^{\phi}_{rr}u^{r}+\Gamma ^{\phi}_{r\phi}u^{\phi})h^{rt}+(\Gamma ^{\phi}_{tt}u^{t}+\Gamma ^{\phi}_{tr}u^{r}
+\Gamma ^{\phi}_{t\phi}u^{\phi})h^{tt}+(\Gamma ^{\phi}_{\phi t}u^{t}+\Gamma ^{\phi}_{\phi r}u^{r}+\Gamma ^{\phi}_{\phi\phi}u^{\phi})h^{t\phi}]-\frac{1}{3}(u^{r}_{,r}+\frac{2u^{r}}{r})h^{t\phi}
\nonumber\\&&\sigma^{rr}= (u^{r}_{,r}+\Gamma ^{r}_{rt}u^{t}+\Gamma ^{r}_{rr}u^{r}+\Gamma ^{r}_{r\phi}u^{\phi})h^{rr}(\Gamma ^{r}_{tt}u^{t}+\Gamma ^{r}_{tr}u^{r}+\Gamma ^{r}_{t\phi}u^{\phi})h^{rt}+(\Gamma ^{r}_{\phi t}u^{t}+\Gamma ^{r}_{\phi r}u^{r}+\Gamma ^{r}_{\phi\phi}u^{\phi})h^{r\phi}-\frac{1}{3}(u^{r}_{,r}+\frac{2u^{r}}{r})h^{rr}
\nonumber\\&&\sigma^{r\phi}=\sigma^{\phi r}= \frac{1}{2}[(u^{r}_{,r}+\Gamma ^{r}_{rt}u^{t}+\Gamma ^{r}_{rr}u^{r}+\Gamma ^{r}_{r\phi}u^{\phi})h^{r\phi}+(\Gamma ^{r}_{tt}u^{t}
+\Gamma ^{r}_{tr}u^{r}+\Gamma ^{r}_{t\phi}u^{\phi})h^{t\phi}+(\Gamma ^{r}_{\phi t}u^{t}+\Gamma ^{r}_{\phi r}u^{r}+\Gamma ^{r}_{\phi\phi}u^{\phi})h^{\phi\phi}+(u^{\phi}_{,r}
\nonumber\\ &&\qquad +\Gamma ^{\phi}_{rt}u^{t}+\Gamma ^{\phi}_{rr}u^{r}+\Gamma ^{\phi}_{r\phi}u^{\phi})h^{rr}+(\Gamma ^{\phi}_{tt}u^{t}+\Gamma ^{\phi}_{tr}u^{r}
+\Gamma ^{\phi}_{t\phi}u^{\phi})h^{rt}+(\Gamma ^{\phi}_{\phi t}u^{t}+\Gamma ^{\phi}_{\phi r}u^{r}+\Gamma ^{\phi}_{\phi\phi}u^{\phi})h^{r\phi}]
-\frac{1}{3}(u^{r}_{,r}+\frac{2u^{r}}{r})h^{r\phi}
\nonumber\\&&\sigma^{\theta\theta}= (u^{\theta}_{,r}+\Gamma ^{\theta}_{rt}u^{t}+\Gamma ^{\theta}_{rr}u^{r}+\Gamma ^{\theta}_{r\phi}u^{\phi})h^{r\theta}+(\Gamma ^{\theta}_{tt}u^{t}+\Gamma ^{\theta}_{tr}u^{r}
+\Gamma ^{\theta}_{t\phi}u^{\phi})h^{t\theta}+(\Gamma ^{\theta}_{\phi t}u^{t}+\Gamma ^{\theta}_{\phi r}u^{r}+\Gamma ^{\theta}_{\phi\phi}u^{\phi})h^{\theta\phi}
-\frac{1}{3}(u^{r}_{,r}+\frac{2u^{r}}{r})h^{\theta\theta}
\nonumber\\&&\sigma^{\phi\phi}= (u^{\phi}_{,r}+\Gamma ^{\phi}_{rt}u^{t}+\Gamma ^{\phi}_{rr}u^{r}+\Gamma ^{\phi}_{r\phi}u^{\phi})h^{r\phi}+(\Gamma ^{\phi}_{tt}u^{t}+\Gamma ^{\phi}_{tr}u^{r}
+\Gamma ^{\phi}_{t\phi}u^{\phi})h^{t\phi}+(\Gamma ^{\phi}_{\phi t}u^{t}+\Gamma ^{\phi}_{\phi r}u^{r}+\Gamma ^{\phi}_{\phi\phi}u^{\phi})h^{\phi\phi}
-\frac{1}{3}(u^{r}_{,r}+\frac{2u^{r}}{r})h^{\phi\phi}\nonumber\\
\end{eqnarray}


\label{lastpage}

\end{document}